\newcommand\anon[2]{{#2}} 

\documentclass[twoside,leqno,twocolumn]{article}

\usepackage[letterpaper]{geometry}

\usepackage{ltexpprt}
\usepackage{hyperref}
\usepackage{subfig}
\usepackage{graphicx}
\usepackage{stackengine}
\usepackage{amsfonts}
\usepackage{amsmath}

\usepackage[utf8]{inputenc}
\usepackage{graphicx}
\usepackage{marginnote}

\usepackage{cleveref}

\usepackage{color}

\definecolor{forestgreen}{cmyk}{0.76,0,0.76,0.45}

\newcommand{\briancode}{Tusqh}

\begin{document}

\newcommand\relatedversion{}

\title{\Large \briancode: Topological Control of Volume-Fraction Meshes\\ Near Small Features and Dirty Geometry\relatedversion \anon{(anonymized)}{} }
\anon{
    \author{Submission ID 1003}  
}
{ 
    \author{Brian Shawcroft\thanks{Brigham Young University Department of Civil and Construction Engineering} \thanks{Sandia National Laboratories, Center for Computing Research.}
    \and Kendrick Shepherd\footnotemark[1]
    \and Scott Mitchell\footnotemark[2]}    
}

\date{}

\maketitle







\begin{abstract} \small\baselineskip=9pt 
This work develops a framework to create meshes with user-specified homology from potentially dirty geometry by coupling background grids, persistent homology, and a generalization of volume fractions.
For a mesh with fixed grid size, the topology of the output mesh changes predictably and monotonically as its volume-fraction threshold decreases.
Topological anti-aliasing methods are introduced to resolve pinch points and disconnected regions that are artifacts of user choice of grid size and orientation, making the output meshes suitable for downstream processes including analysis.
The methodology is demonstrated on geographical, mechanical, and graphics models in 2D and 3D using a custom-made software called \briancode{.}
The work demonstrates that the proposed framework is viable for generating meshes on topologically invalid geometries and for automatic defeaturing of small geometric artifacts.
Finally, the work shows that although subdividing the background grid frequently improves the topological and geometrical fidelity of the output mesh, there are simple 2D examples for which the topology does not converge under refinement for volume-fraction codes.
\end{abstract}

\section{Introduction}
Getting geometry that is suitable for mesh generation is often more difficult and time consuming than creating a mesh from that geometry.
``Ugly'' geometry is ubiquitous ``in the wild.'' 
Industrial and commercial CAD data are often hand-designed ``blueprints'' to guide assembly, and do not represent the as-built part. 
Data from imaging and segmentation may have topological inconsistencies.
Even in cases with valid geometry and topology, analysts must carefully review, modify, and defeature models based on the intended purpose of the mesh because typical techniques  generate meshes whose topology and geometry match the input models.
Common issues in ``ugly'' geometry are gaps and overlaps,  features smaller than the desired mesh size, topological complexities such as small holes, and small angles and thin regions that would produce poor-quality elements.

However, the mesh is a discrete approximation to the geometry. Why require a higher fidelity in the input than is aspired to in the output?  Indeed, the community is developing tools to mesh ugly geometry, robustly producing meshes that are topologically correct and have high-quality elements despite topological and geometrical defects and small features in the input.

Sculpt~\cite{owen2012parallel,owen2015evaluation,sculpt} is one such tool, achieving a hexahedral mesh of reasonable quality, but reconstructing an approximation of the input geometry and topology.
Inexact reconstruction is a \emph{benefit} in the case of gaps, overlaps, and small features. 
The Sculpt algorithm starts with a background grid overlaying the input geometry. 
The fraction of each grid cell that lies inside the geometry of an input material is its \emph{volume fraction}.
Cells with volume fractions above a threshold (e.g., one-half) are retained; the rest are discarded. 
Heuristics remove undesirable  topology such as pinch points and components consisting of only a few cells. Retained cells are then snapped to the geometry, and mesh quality is achieved through pillowing~\cite{mitchel1995pillowing,10.1007/978-3-540-87921-3_28,ZHANG2010405}, smoothing~\cite{Knupp:2012:TMOP}, and other changes to mesh topology and node positions.
Similarly, Morph~\cite{morph_noble_2,morph_staten_1} is a parallel tet mesher using a background grid that snaps nodes to geometry based on dimension, proximity, and how other nodes are snapped.
When no suitable node snap is found, Morph adds new nodes at the intersections with the geometry to produce nodes on the geometry boundary. 
In both Sculpt and Morph, the size of the background grid indirectly determines the geometric fidelity of the output to the input.
TetWild~\cite{10.1145/3386569.3392385,10.1145/3197517.3201353} uses a Delaunay triangulation rather than a background grid. Triangles representing the geometry are incrementally inserted and the mesh is refined. Edge length and geometric proximity parameters control the mesh resolution and geometric fidelity.

Though these tools always produce meshes with valid topology, there is no a priori knowledge of what the homology of the output will be, nor how it will compare to the input topology or the desired topology.
For some downstream operations, small holes and features play a critical role in final results, while for others, these same holes and features are extraneous, may lead to overly dense meshes, and must be (manually) removed.
In sum, a single model may require multiple representations depending on its intended purpose, each with varying mesh sizes and topological needs.
However, tools with control over how to select the appropriate topology of the mesh for its intended purpose are in short supply, meaning that much of this work is deferred to time-consuming manual manipulation by engineers.

Herein, we explore how to robustly predict and achieve the desired mesh topology for algorithms based on background grids and volume fractions (including Sculpt) 
 through the use of persistent homology and generalized winding numbers.

To accomplish this goal, \briancode{}---a prototype mesher now available on GitHub \cite{tusqh_github:2025}---was developed in Rhinoceros 3D.
It is a testbed for research and demonstrates that our techniques are effective.
\briancode{} mimics the initial steps of Sculpt, using a background grid and volume fractions to decide which grid cells to retain.
As with TetWild, it uses generalized winding numbers~\cite{barill2018fast,Jacobson:2013}
to define the ``interior'' for valid and invalid geometries.
We explore the topological structure of potential meshes under different volume-fraction thresholds using persistent homology~\cite{Edelsbrunner:2002,Otter:2017}.
Local connectivity decisions based on sub-sampling volume-fractions mitigate the effects of the arbitrary orientation and offset of the background grid.
This enables the analyst to measure and select the desired mesh topology, which then informs which volume-fraction threshold to select. 
The user may also adjust the volume-fraction threshold on a sliding scale and visualize the choice of meshes. 
These meshes can serve as input for subsequent steps to improve geometric fidelity, such as Sculpt's snapping, pillowing, and smoothing.
Concluding theoretical results demonstrate that obvious applications of grid-based volume-fraction methods cannot guarantee consistent topological output.


\section{Background Material}
The proposed method, which selects which cells of a background grid to retain, is related to the computer graphics problem of rasterization~\cite{GPUGemsRaster}. 
Consequently, we first introduce background information about rasterization and anti-aliasing, after which fundamentals about tools employed in this work are introduced, including homology, persistent homology, and winding numbers.

\begin{figure}[!htb]
\centering
\includegraphics[width=0.47\columnwidth]{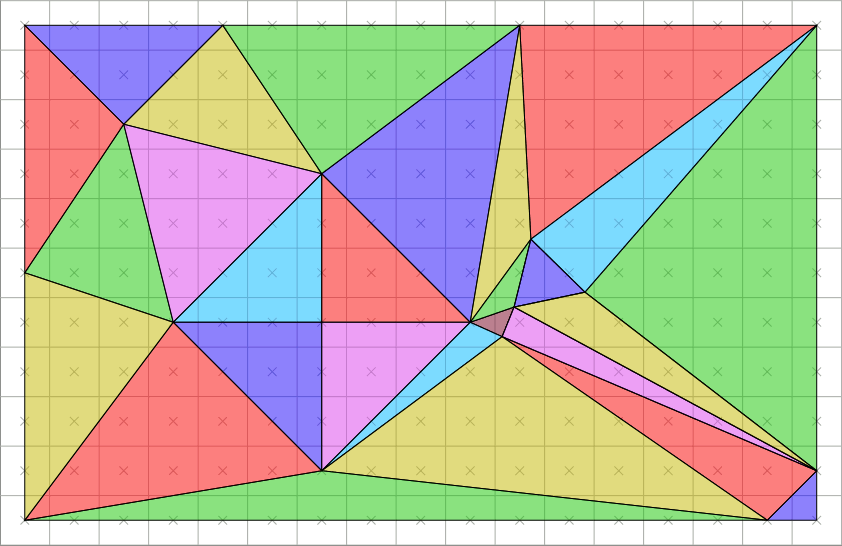}
\hspace{0.02\columnwidth}
\includegraphics[width=0.47\columnwidth]{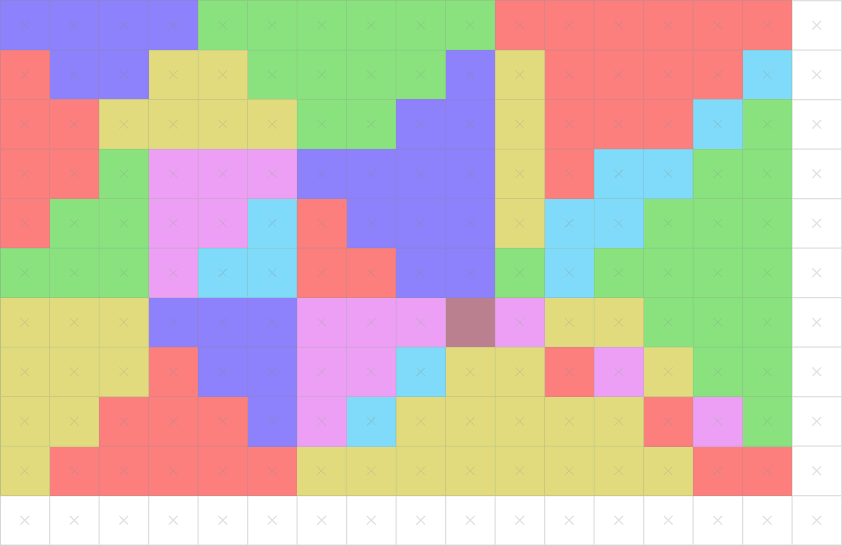}
\caption{Rasterization of triangles into pixels for computer graphics.  Note the pinches from the two left cyan triangles, the archipelago from the lower right pink triangle, and the multitude of additional topological errors in the lower right. Image courtesy Wikipedia \url{https://en.wikipedia.org/wiki/Rasterisation}}\label{fig:TopologicalErrorRaster}
\end{figure}

\paragraph{Rasterization.}\label{sec:rasterization}
 \emph{Rasterization} is the process of converting an arbitrary geometry into a grid-based representation.
In traditional computer graphics, the background grid is screen pixels, and the objects are triangles embedded in floating point $\mathbb{R}^2$. 
The problem is to select which color and intensity to display in each pixel. 
\emph{Aliasing}~\cite{crow1977aliasing,mitchell1990antialiasing} is a significant problem in rasterization: 
pixel values are sensitive to the offset, rotation, and size of the objects, as well as which locations within a pixel are sampled.
Consider a non axis-aligned edge shared by a blue and a red triangle.
For a given pixel, if we choose red or blue we get increased contrast but also stair-step patterns called ``jaggies.''
If we choose a purple mixture the image appears smoother, but shading can produce Moir\'e patterns.
Both patterns are glaring to human eyes. 
For small triangles, the pixel topology may not match the triangle topology; see \cref{fig:TopologicalErrorRaster}.
Such topological errors may be visually insignificant, but they can lead to significant errors in simulation results.

\begin{figure}[b]
\centering
\subfloat
{
\begin{minipage}{0.28\columnwidth}
\centering
\includegraphics[width=\columnwidth]{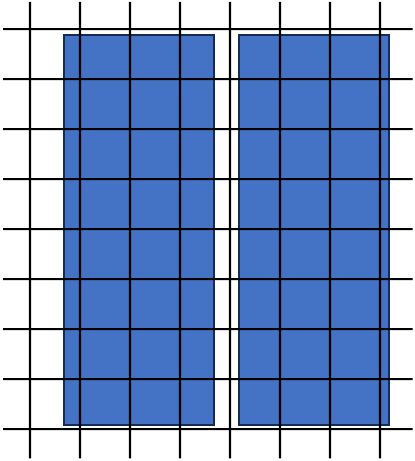}
\footnotesize gap$<$grid$/2$ \\ closed
\label{fig:parallel_gap_small}
\end{minipage}
}
\hspace{3pt}
\subfloat
{
\begin{minipage}{0.28\columnwidth}
\centering
\includegraphics[width=\columnwidth]{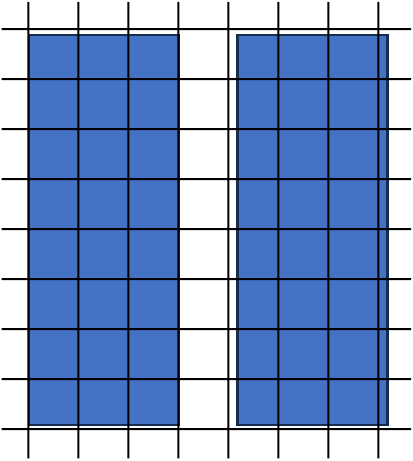}
\footnotesize gap$>$grid \\ open
\label{fig:parallel_gap_big}
\end{minipage}
}
\hspace{3pt}
\subfloat
{
\begin{minipage}{0.28\columnwidth}
\centering
\includegraphics[width=\columnwidth]{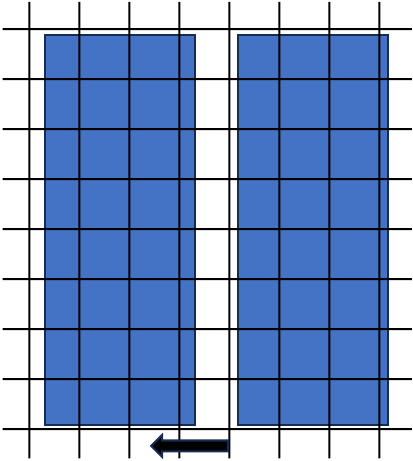}
\footnotesize gap $\in \lbrack\frac{1}{2},1\rbrack$ grid \\ depends
\label{fig:parallel_gap_medium}
\end{minipage}
}
\caption{Small grid-aligned gaps are closed, large gaps are open, and intermediate gaps depend on their offset.}\label{fig:parallel_axis_aligned_gap}
\end{figure}

\paragraph{Topological Anti-aliasing.}\label{sec:antialiasing}
As in graphics, volume-fraction meshing is sensitive to the offset and rotation of objects, as well as the grid size (analogous to pixel density in graphics).
An axis-aligned gap is closed or open depending on its size and position relative to the grid; see \cref{fig:parallel_axis_aligned_gap}. A gap smaller than half the grid size is always closed. A gap larger than the grid size is always open. Between these, shifting the grid left will cause the mesh to alternate between closed and open.

\begin{figure}[t]
\centering
\subfloat[raw geometry]
{
\includegraphics[trim=36cm 11cm 33cm 10cm,width=0.44\columnwidth,clip]{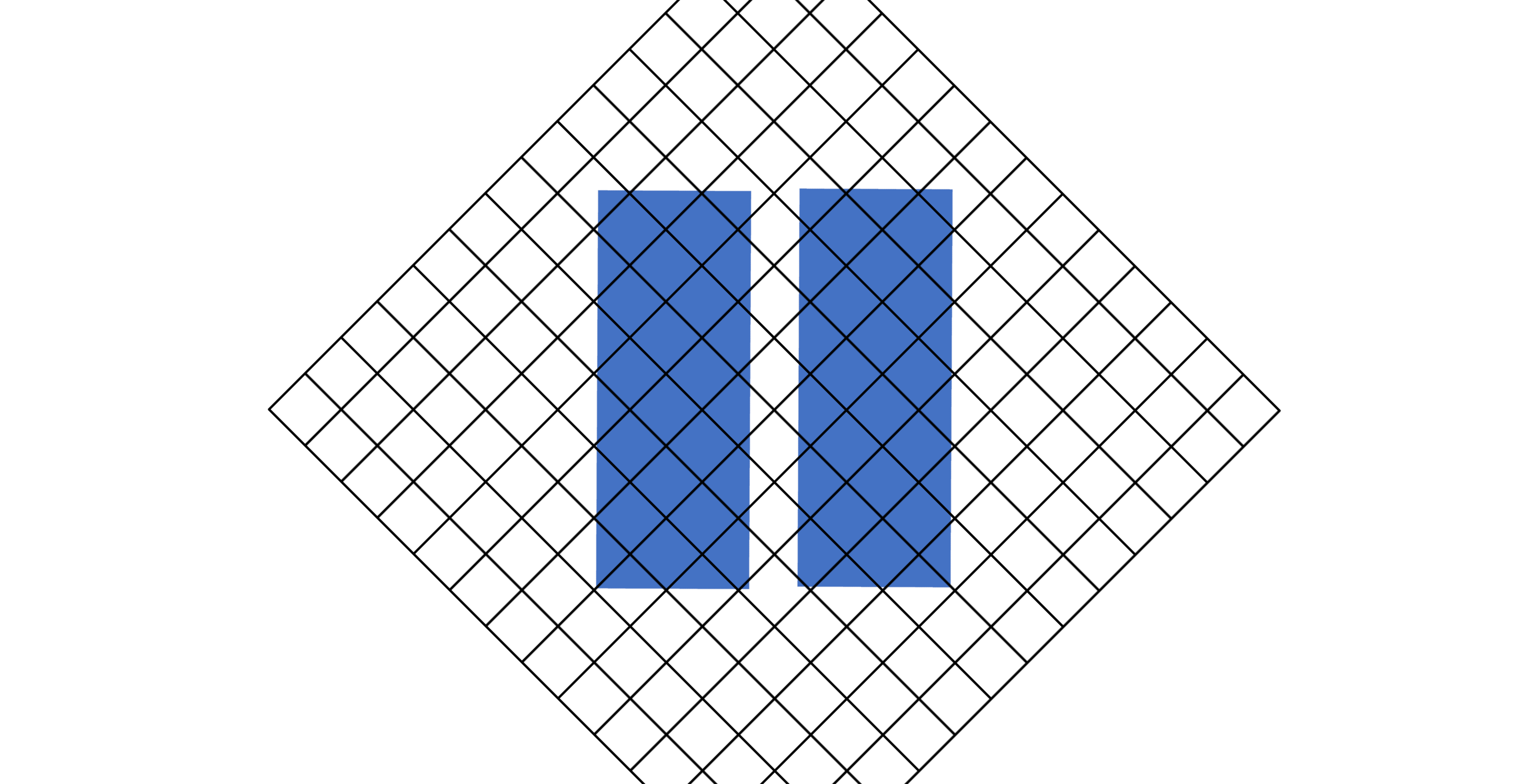}
}
\hspace{8pt}
\subfloat[filled cells]
{
\includegraphics[trim=36cm 11cm 33cm 10cm,width=0.44\columnwidth,clip]{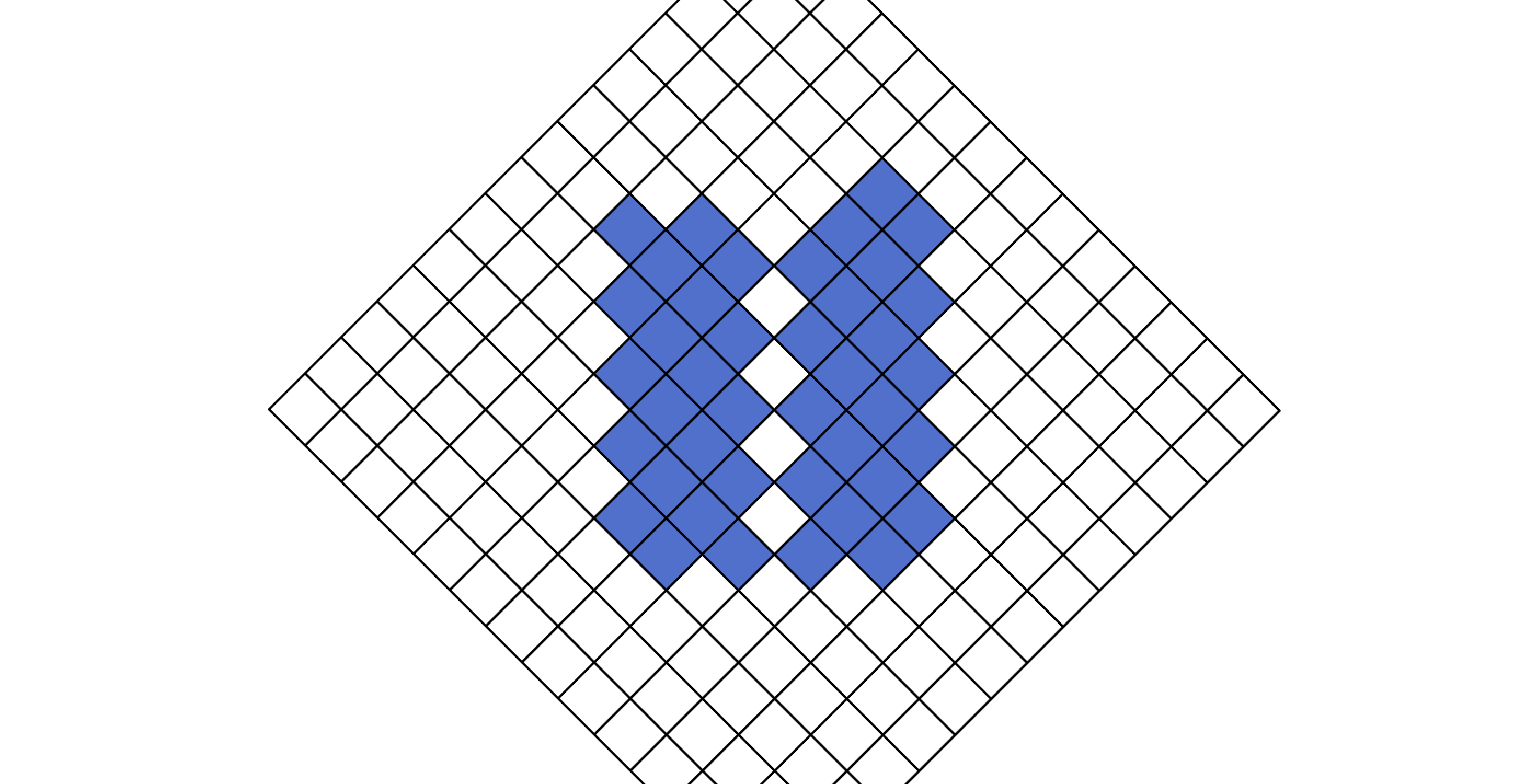}
\label{fig:filled_cells}
}
\caption{This unaligned gap is resolved inconsistently.}
\label{fig:diagonal}
\end{figure}

\begin{figure}[t]
\centering
\includegraphics[trim=13cm 20cm 14cm 10cm,width=0.98\columnwidth,clip]{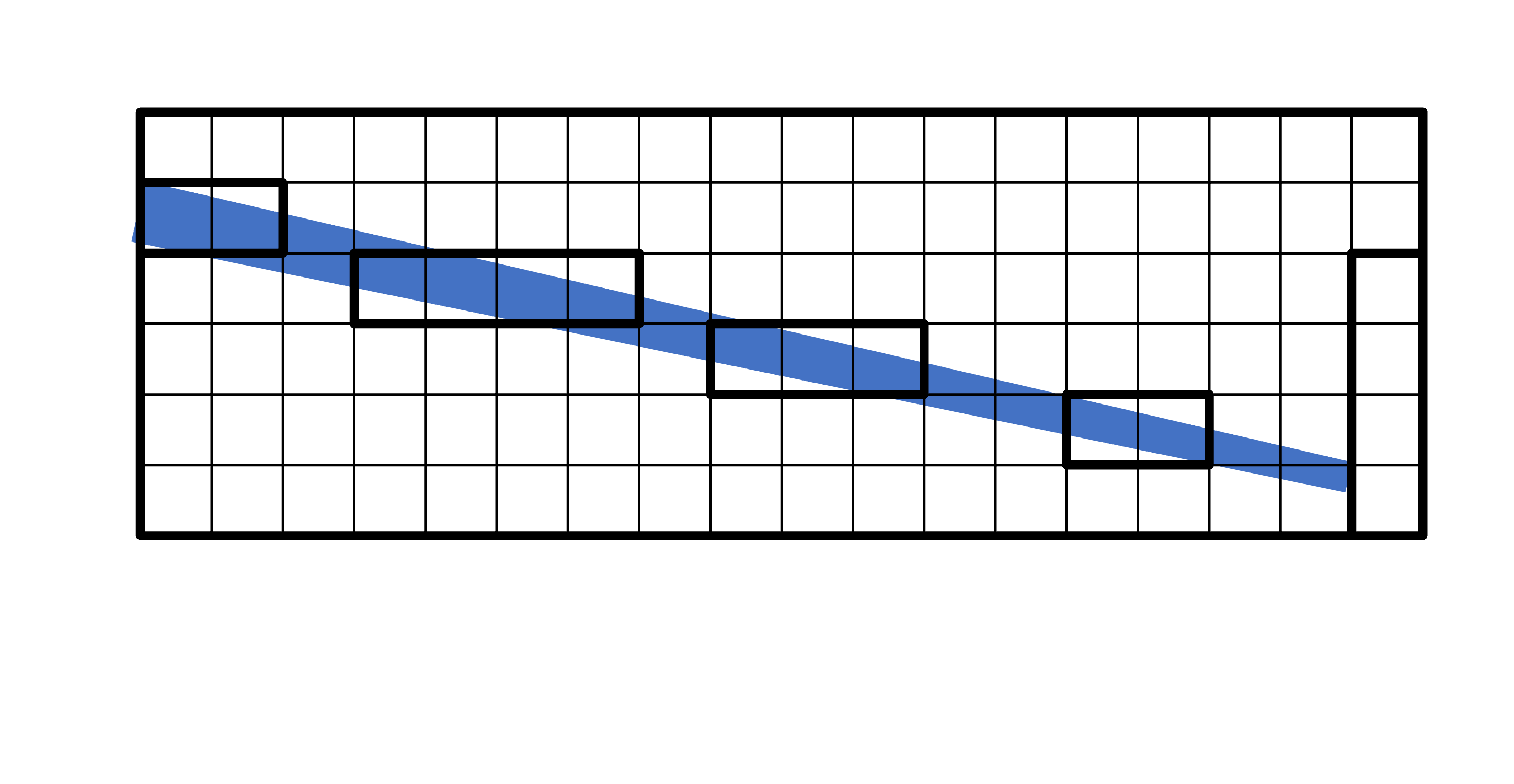}
\caption{Rotational aliasing may cause stair-step patterns, and archipelagos of isolated islands near where two lines meet at a sharp angle. 
Bold-outlined cells are filled, thin are open.}
\label{fig:jaggies}
\end{figure}

In \cref{fig:diagonal,fig:jaggies} we see the aliasing effects of rotations, where the feature is not aligned with the grid.
In \cref{fig:diagonal} the gap is a fixed width, but about the size of the grid cells, leading to the gap being inconsistently resolved as open or closed, with separate sides connected by pinch points.
In \cref{fig:jaggies} we see a similar inconsistency, but exacerbated because the gap width varies.
The boundary lines meet at a sharp angle, so fewer cells are retained as the apex is approached, leading to a chain of small disjoint mesh islands which we call an \emph{archipelago}.

To address these undesirable local topological features, a topological anti-aliasing method is defined and coupled with introducing/removing various templated cells, as described in \cref{sec:methods}.
The first undesirable feature is pinches, where exactly two grid cells meet at a vertex with no shared edge, or exactly two 3D grid cells meet at an edge with no shared faces. 
These must be removed because the mesh is required to be locally connected face-to-face or disjoint.
(The complement is also connected face-to-face or disjoint.)
All other ways in which a mesh can be non-manifold do not occur, because we form the mesh from the union of some cells of a structured grid.
Pinches are either separated into different components by removing small elements, or thickened into meeting face-to-face by adding small elements using data from the anti-aliasing framework.
These small elements come from templates that split background grid cells and perform swaps.
The second undesirable feature is archipelagos.
Our anti-aliasing technique joins some islands with template elements around connecting edges and quads, and removes any small islands that remain.
For comparison, Sculpt resolves pinches by adding or removing entire grid cells,
and resolves small components by removing them~\cite{owen2017hexahedral}.

For simplicity we only discuss domains with a single material and only discuss the retained cells.
In principle our anti-aliasing could be extended and applied to multi-material volume-fraction 
meshing~\cite{ZHANG2010405,owen2017hexahedral}.

\paragraph{Homology.}
The topology of a mesh should contain the significant features of a domain for its intended computational analysis.
Herein, we shall study mesh topology using a cellular complex: nodes are zero-cells, edges are one-cells, faces are two-cells, and volumes are three-cells. 
Specifically, we will make use of simplicial and cubical complexes, in which two-cells are triangles and quadrilaterals, and three-cells are tetrahedra and hexahedra, respectively.
Homology \cite{Hatcher:2001} is a mathematical tool that distinguishes cell complexes using certain algebraic quotient groups. 
The \emph{Betti numbers} $B_{i}$ count the rank of these groups.
Specifically, $B_{0}$ equals the number of connected components, $B_{1}$ is the number of holes, and $B_2$ is the number of cavities or voids.
For planar domains $B_2$ will always be zero.

Persistent homology \cite{Edelsbrunner:2002,Otter:2017} describes homology changes as objects are added and connections are made.
A \emph{filtration} has a ``persistence parameter'' which defines when a cell enters the complex. 
A filtration is monotonic, so no cell may ever leave the complex after entering. 
However, the homology has both additions and removals because adding a cell could, e.g., create a new connected component, or combine two components into one. 
The parameter value at which a group generator is created is called its ``birth,'' while the value it disappears is called its ``death.''
Birth and death coordinates are plotted in a persistence diagram such as \cref{fig:persistence_diagram}.
This not only counts Betti numbers, but tracks individual components and holes.

This work studies the persistent homology of cubical background grids using volume fractions as the persistence parameter.
(In other work the signed distance to a domain boundary was the parameter~\cite{moon2019statistical}.) 
Alternatively, zigzag persistence, which does not require a monotonic filtration 
~\cite{carlsson2010zigzag,dey2024computing}, could be used, but doing so would increase complexity and computational expense.

\begin{figure}[!b]
\centering
\includegraphics[width=0.40\columnwidth]{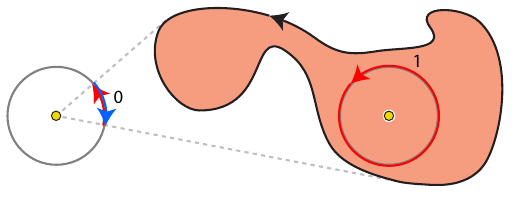}
\includegraphics[width=0.58\columnwidth]{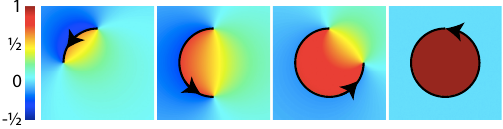}
\caption{Winding number point and field values, courtesy Jacobson et al.~\cite{Jacobson:2013} Figures 4 and 6. Used with permission of the Association for Computing Machinery, conveyed through Copyright Clearance Center, Inc.}\label{fig:jacobson_winding}
\end{figure}

\paragraph{Winding Numbers.}
Volume fractions may be estimated by sampling points and  counting the fraction of them inside the geometry. 
However, for ``ugly'' geometry, what is ``inside'' may be poorly defined.
The generalized winding number \cite{barill2018fast,Jacobson:2013} overcomes this obstacle; see \cref{fig:jacobson_winding}.
It gives answers identical to ray shooting for watertight domains, and gives answers that humans find both reasonable and intuitive for other domains.
In its traditional form, the winding number at a point with respect to a closed curve describes the net number of times the curve encircles the point in the counter-clockwise direction, with negative numbers indicating clockwise encirclement.
The winding number is the integral of the angle of the ray from the point to the curve as it is traversed. 
The generalized winding number extends this definition to sets of open curves.
It yields a continuous value where 0 indicates outside and 1 is inside.
For geometry with gaps, the winding number near a gap is typically between 0 and 1. 
In extreme cases, such as overlapping domain boundaries, invalid geometries may give values beyond $[0,1]$.
In 3D, the winding number integrates the solid angles seen from a point, e.g., for a volume defined by a triangle soup. 
Which normal direction is outward-facing determines the sign of the solid-angle contribution.

\section{Methodology}\label{sec:methods}
%
%

\begin{figure*}[!htb]
\centering
\subfloat[Vol. Frac.: 50\%; $B_0 =  4; B_1 =  2$]{\includegraphics[trim=22.5cm 0cm 25cm 0cm, width=0.6\columnwidth,clip]{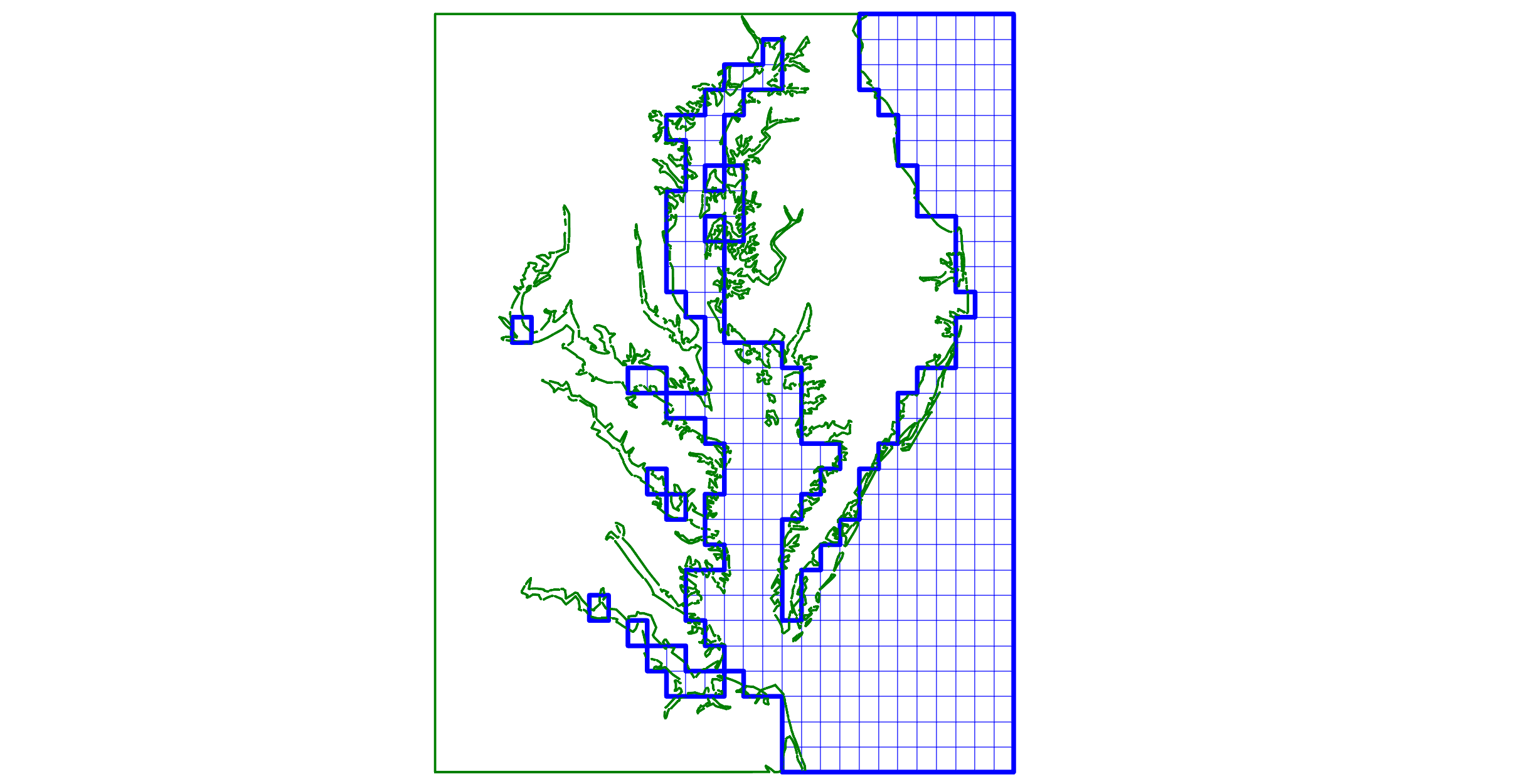}}\qquad
\subfloat[Vol. Frac.: 15\%; $B_0 =  1; B_1 = 7$]{\includegraphics[trim=22.5cm 0cm 25cm 0cm, width=0.6\columnwidth,clip]{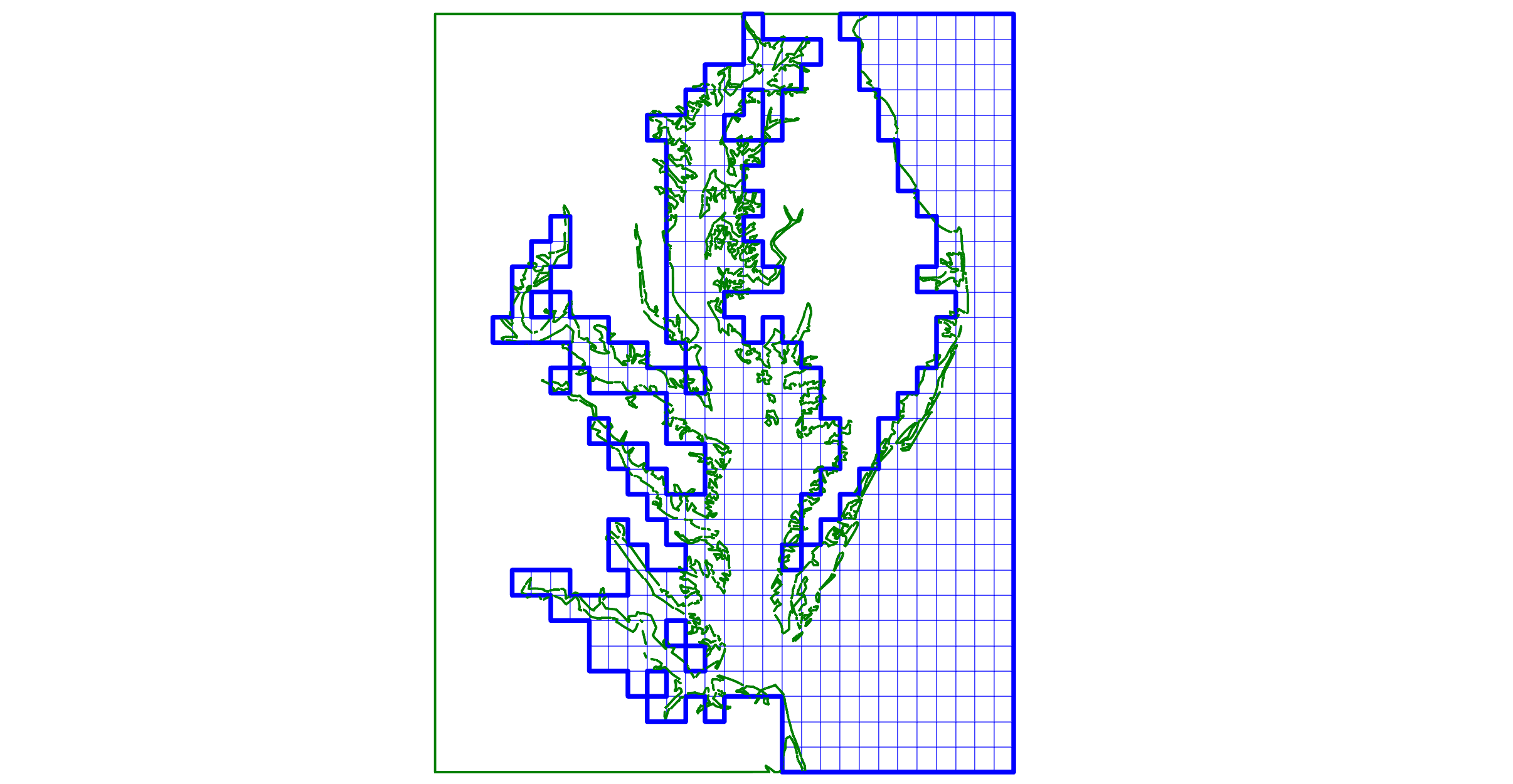}}\qquad
\subfloat[Vol. Frac.: 1\%; $B_0 = 1; B_1 =  12$]{\includegraphics[trim=22.5cm 0cm 25cm 0cm,width=0.6\columnwidth,clip]{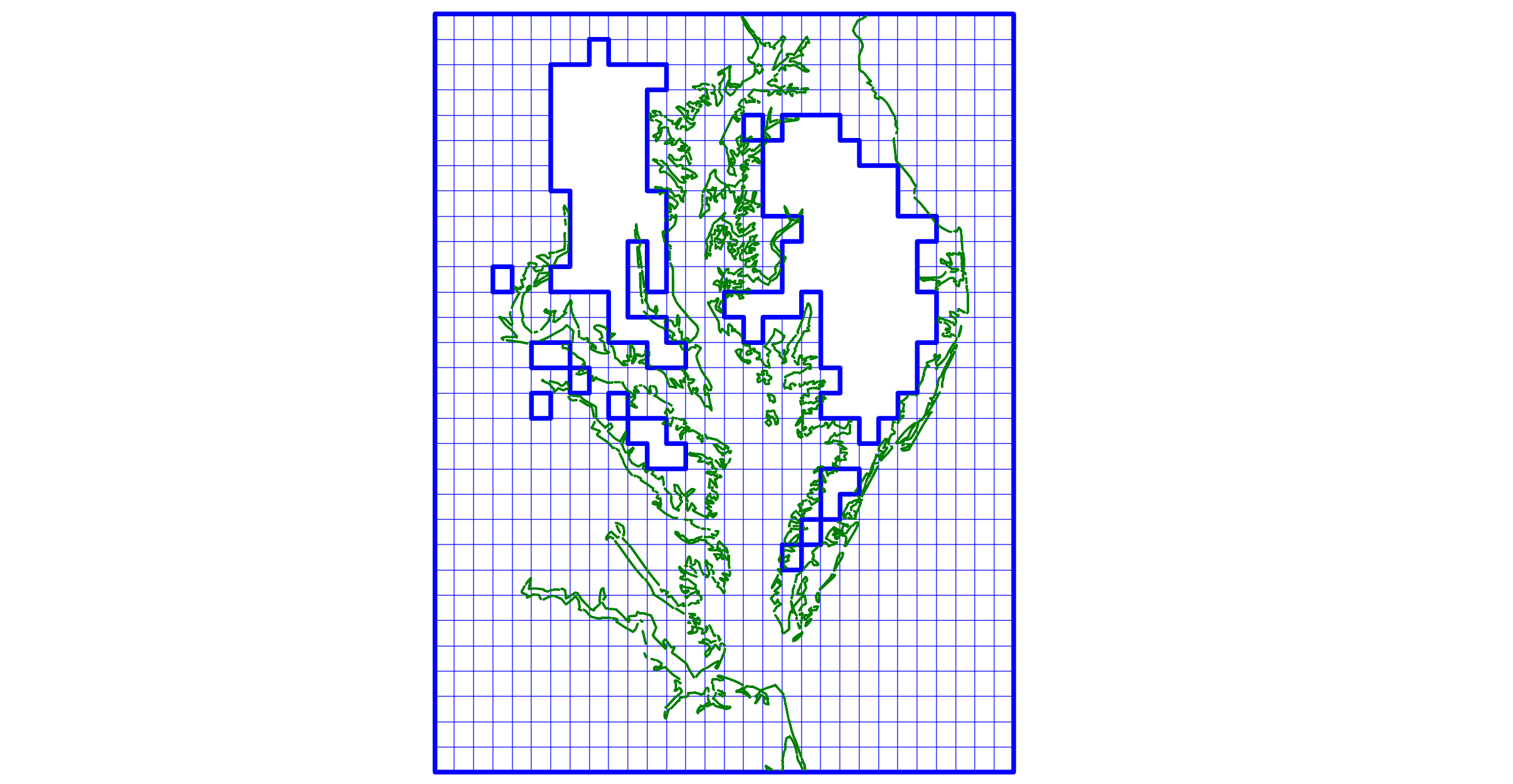}}

\caption{Chesapeake Bay meshes and their Betti numbers change as the volume-fraction threshold is lowered. 
The model contains deliberate errors: gaps, overlaps, and offsets. 
As a result of these errors, the winding numbers of sampled points may be positive in regions that are clearly inland, as in the lower-left region of the rightmost figure.
Also note how rasterization affects the thin bays and peninsulas, where here we have skipped anti-aliasing.}
\label{fig:volfrac_chesapeake}
\end{figure*}

\begin{figure}[!htb]
\includegraphics[width=\columnwidth,trim = 0cm 0cm 0cm 1cm, clip]{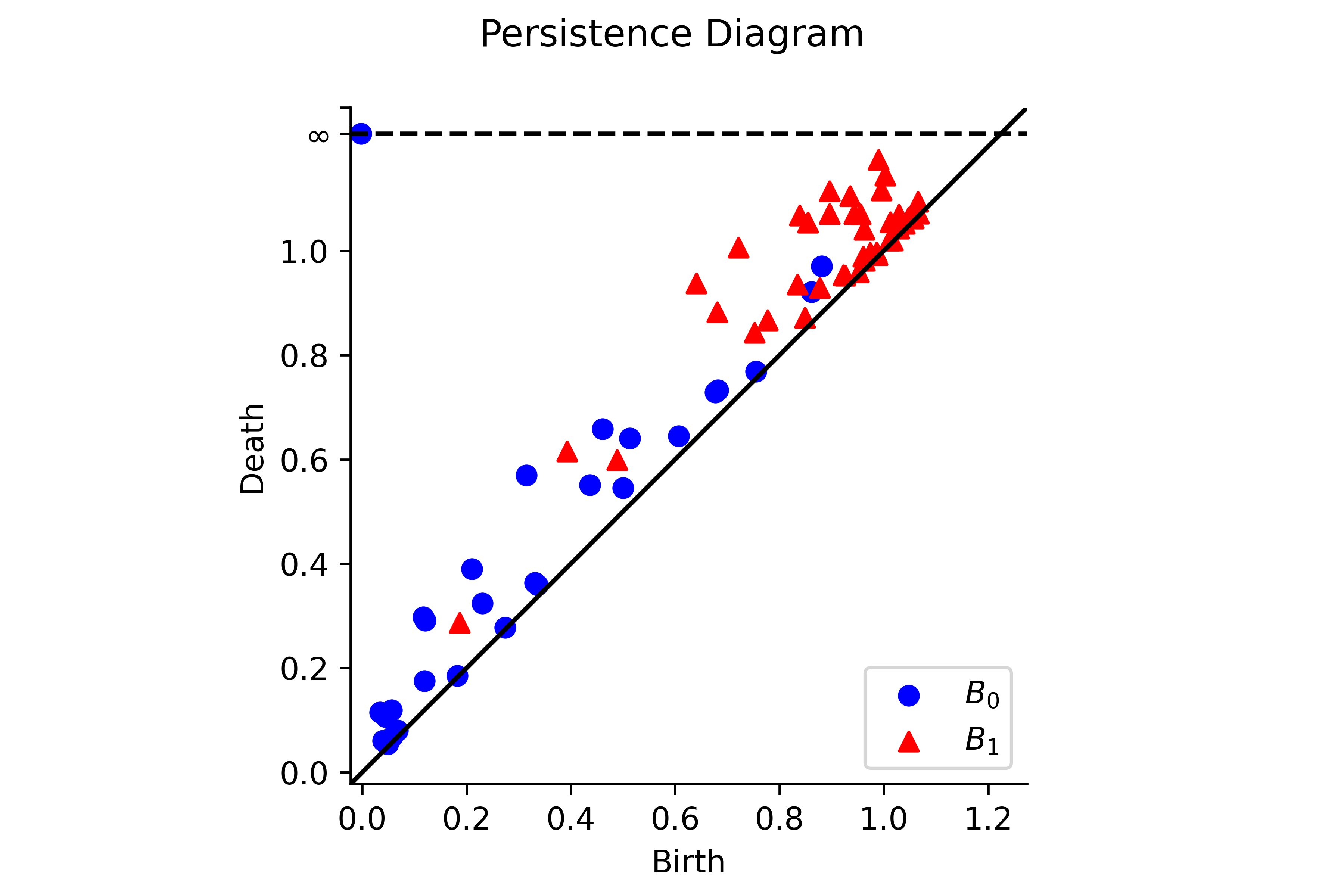}
\caption{The persistence diagram of the Chesapeake Bay grid, with parameter 1 minus the volume fraction.}\label{fig:persistence_diagram}
\end{figure}

\begin{figure}[!htb]
\centering
\subfloat[Vertex, even ]{
	\includegraphics[trim=30cm 2cm 26cm 2cm, width=0.23\columnwidth,,clip]{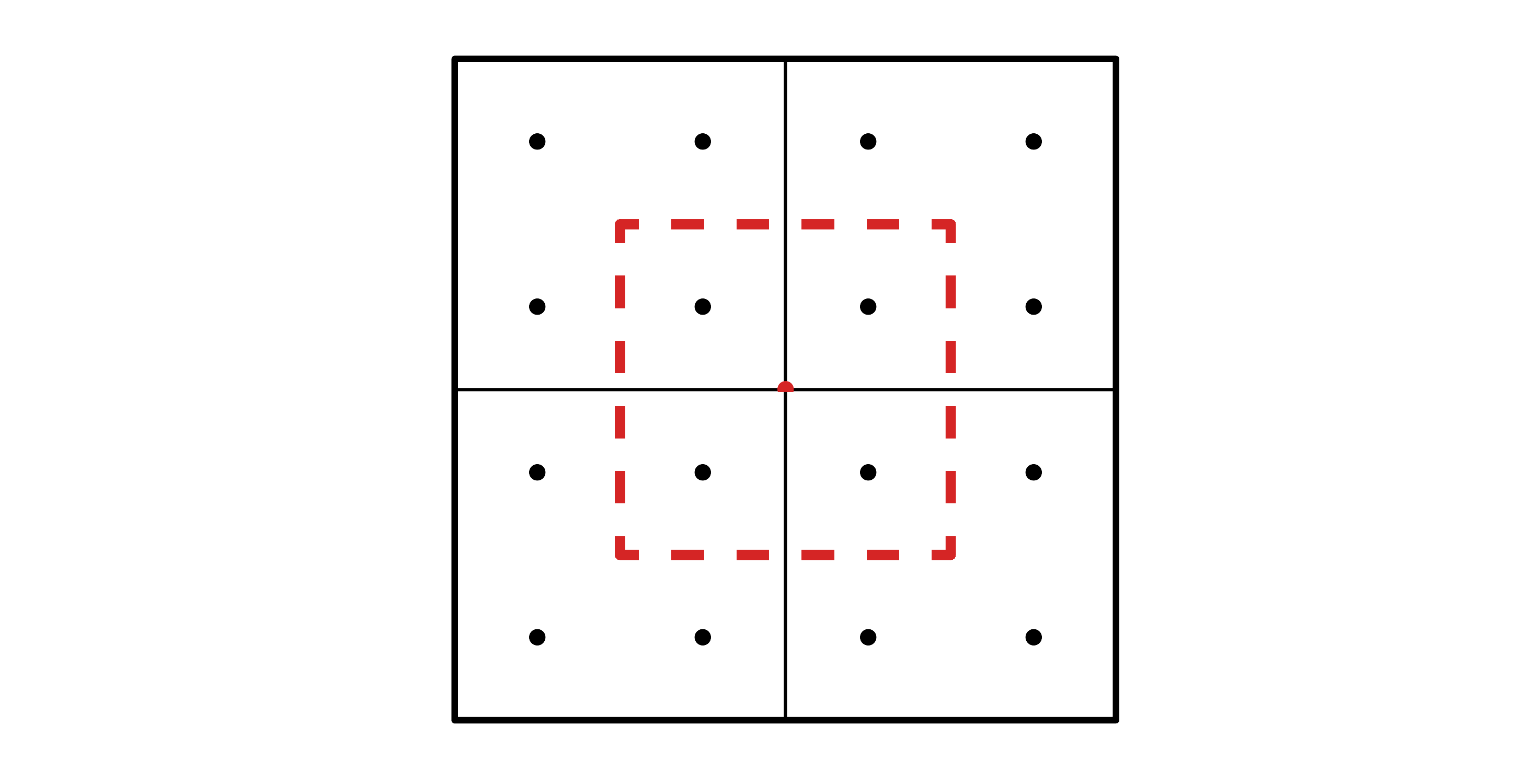}}
\subfloat[Vertex, odd ]{
	\includegraphics[trim=30cm 2cm 26cm 2cm, width=0.23\columnwidth,,clip]{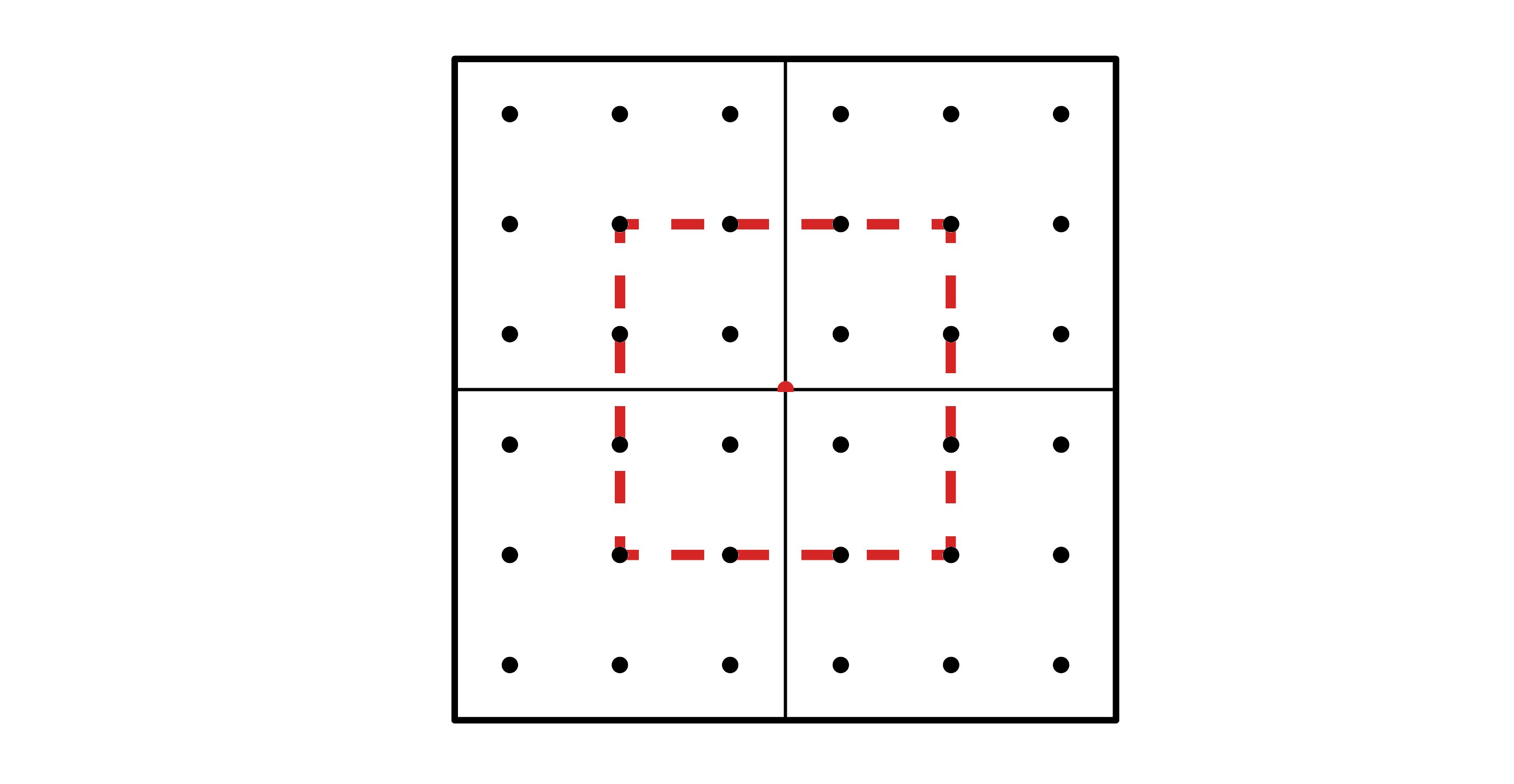}
	\label{fig:odd_vert}}
\subfloat[Edge, even ]{
	\includegraphics[trim=30cm 2cm 26cm 2cm, width=0.23\columnwidth,,clip]{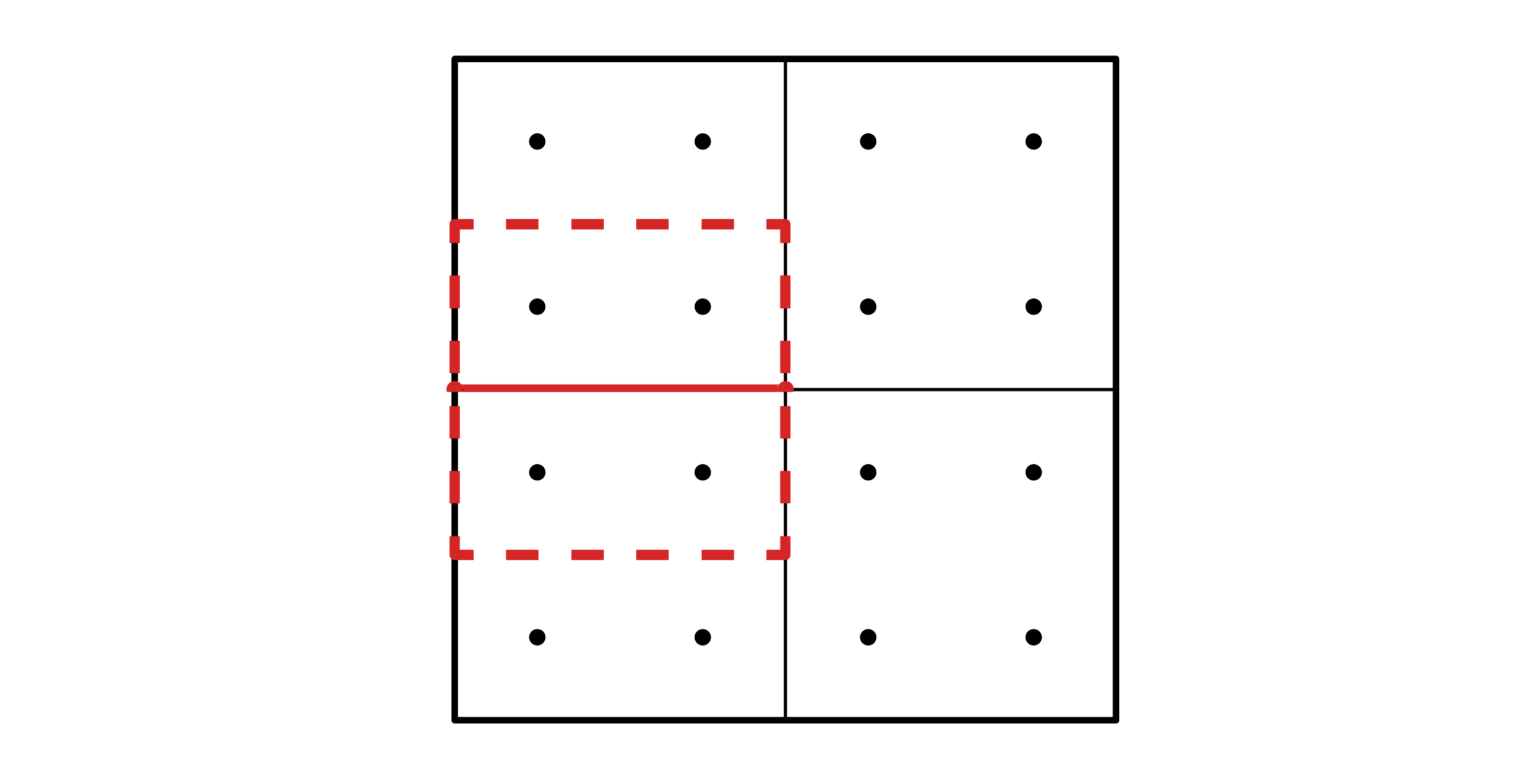}}
\subfloat[Edge, odd ]{
	\includegraphics[trim=30cm 2cm 26cm 2cm, width=0.23\columnwidth,,clip]{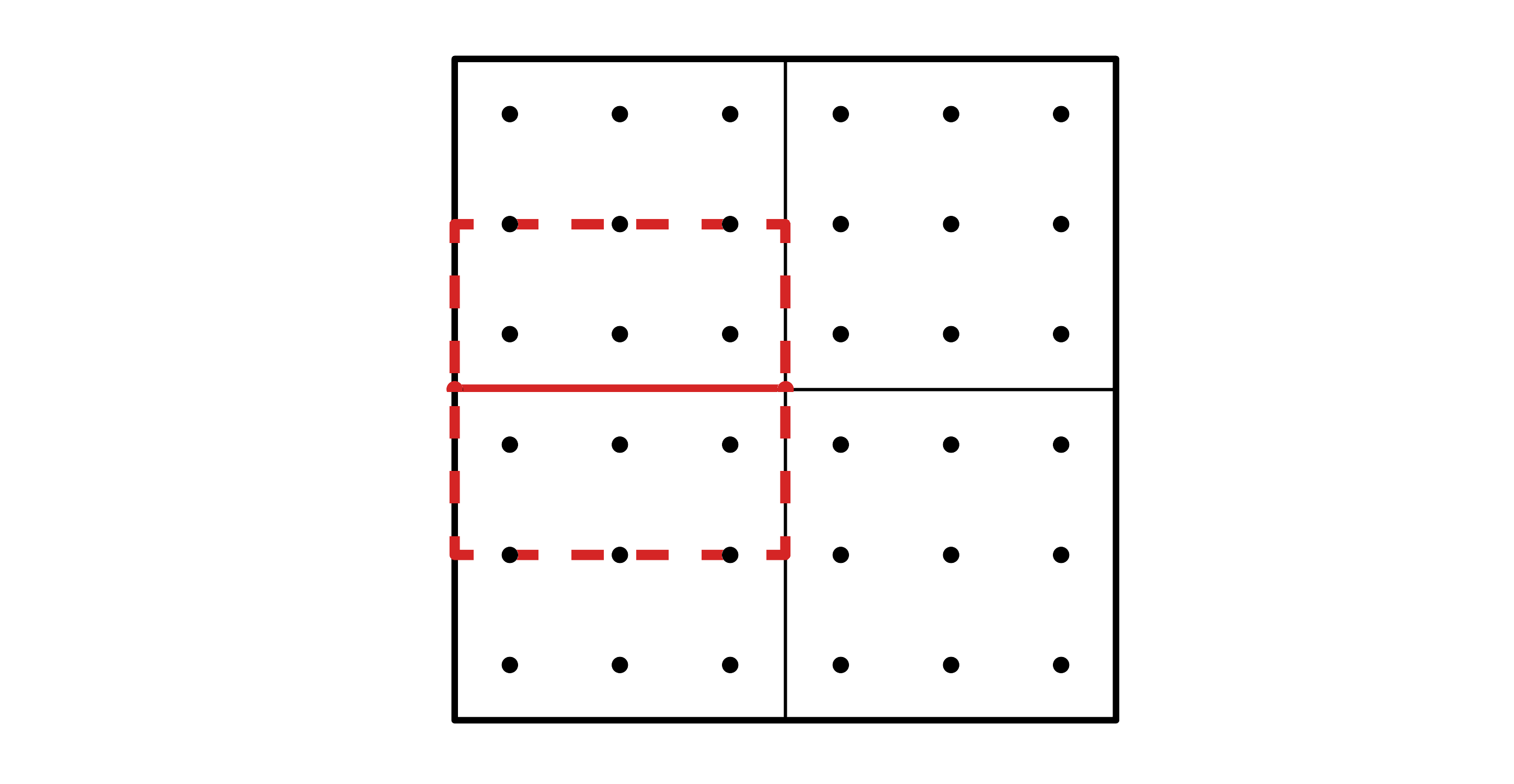}
	\label{fig:odd_edge}}
\caption{Sample $s \times s$ arrays are shown. Samples contained by the red dashed lines define vertex and edge volume fractions in 2D. We use only even sample arrays.}
\label{fig:subcell_pts}
\end{figure}

\paragraph{Volume Fractions.}
Herein we study both 2D and 3D domains.
We define a regular background grid, e.g. by subdividing an axis-aligned bounding box.
This grid is a cubical quadrilateral or hexahedral complex, depending on the domain dimension.
The volume fraction of each maximal-dimension cell is computed as the average of the winding numbers of its sample points.
Sample points lie in an $s^d$ array, as shown in \cref{fig:subcell_pts}.  
(Recall we calculate persistent homology based on volume fractions, with the goal that the user may select the volume fraction that achieves their desired mesh topology.)

\paragraph{Volume-Fraction Persistence Parameter.}
The persistence parameter used herein is the volume-fraction threshold, ordered from 1 down to 0 (i.e. by decreasing value).
By defining volume fractions only for cells of maximal dimension, a mesh is defined by including all cells with volume fraction greater than or equal to the chosen threshold, and removing all others.
The order that cells are added to the mesh as a function of the persistence parameter is demonstrated on a topologically invalid representation of the Chesapeake Bay in \cref{fig:volfrac_chesapeake}.
The persistent homology diagram is displayed in \cref{fig:persistence_diagram}.
These images illustrate significant topological aliasing, which limits the utility of these meshes.
In what follows, we aim to mitigate these rasterization effects.

\paragraph{Sub-cell Volume Fractions.}
To assist in topological anti-aliasing, we also define volume fractions for all lower-dimensional grid cells.
For any such $n$-cell, its sample points are those in a (fictitious) grid cell centered at that $n$-cell; see \cref{fig:subcell_pts}.
We use only even numbers $s$ of sample points because these samples do not lie on cell boundaries.
Volume fractions for these lower-dimensional cells are computed as averages of winding numbers associated with sample points contained in the fictitious grid cell, in a manner analogous to cells of maximal dimension.
As before, only cells with volume fraction greater than or equal to a prescribed threshold will remain in the cell complex. All others are omitted.
Omitted cells are called ``exterior'' cells, while remaining cells are called ``interior'' cells.
We call this sampling process ``\emph{subgrid sampling}.''


\paragraph{Anti-aliasing.}
To address the undesirable rasterization effects introduced by our background grid (including pinch and archipelago removal), we employ subgrid sampling as an anti-aliasing method.


\begin{figure}[!t]
	\centering
	\subfloat[2D vertex]{
		\includegraphics[trim=0cm 5cm 0cm 5cm, height=1in,clip]{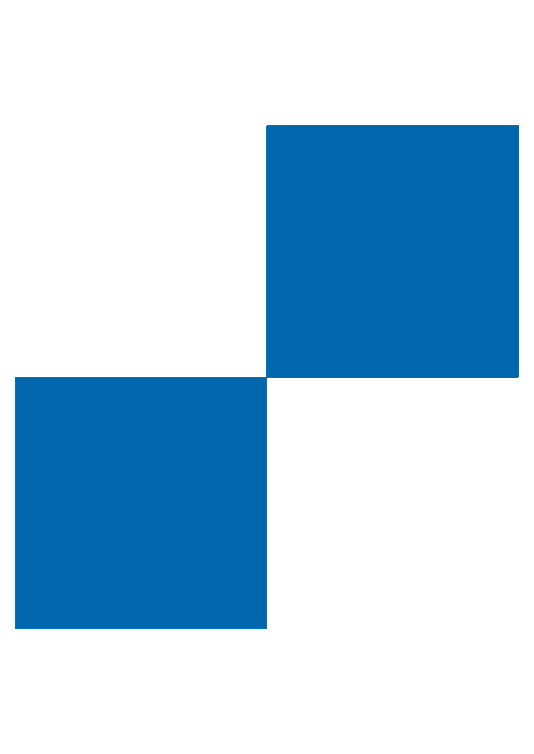}
		\label{fig:two_d_nonmanifold}
	}
	\hfill
	\subfloat[1 vertex]{
		\includegraphics[trim=0cm 0.2cm 0cm 0.7cm, height=1in,clip]{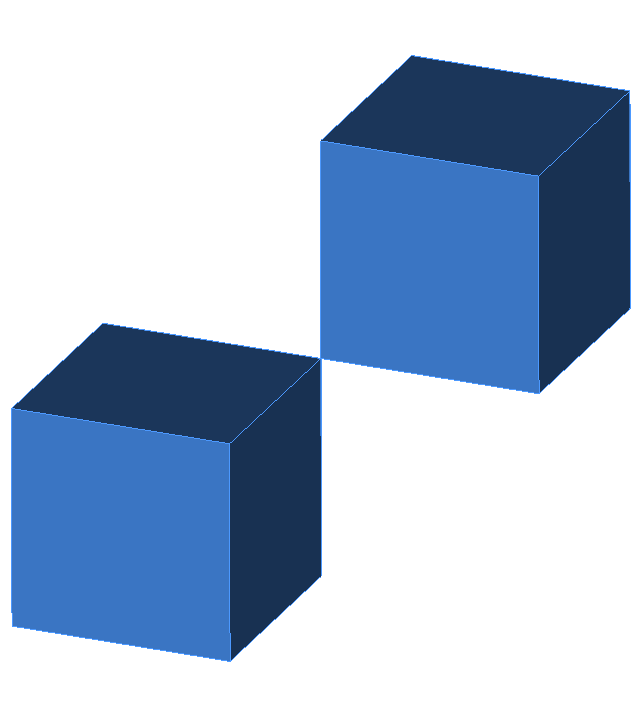}
	\label{fig:case1}
	}
	\hfill
	\subfloat[1 edge]{
		\includegraphics[trim=0cm 0.2cm 0cm 0.7cm, height=1in,clip]{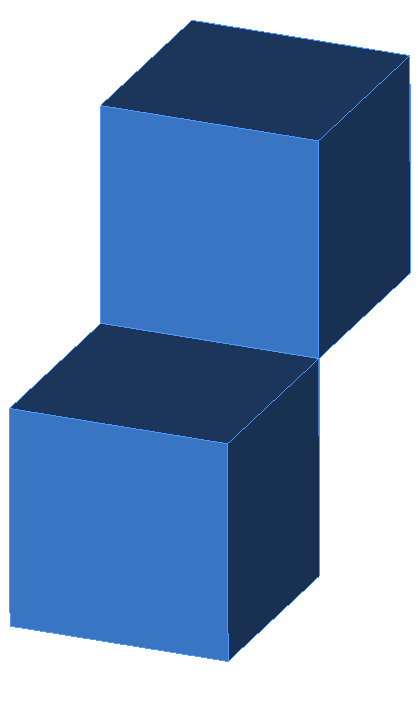}
	\label{fig:case2}
	}
	\hfill
	\subfloat[1 edge]{
		\includegraphics[trim=0cm 0.2cm 0cm 0.7cm, height=1in,clip]{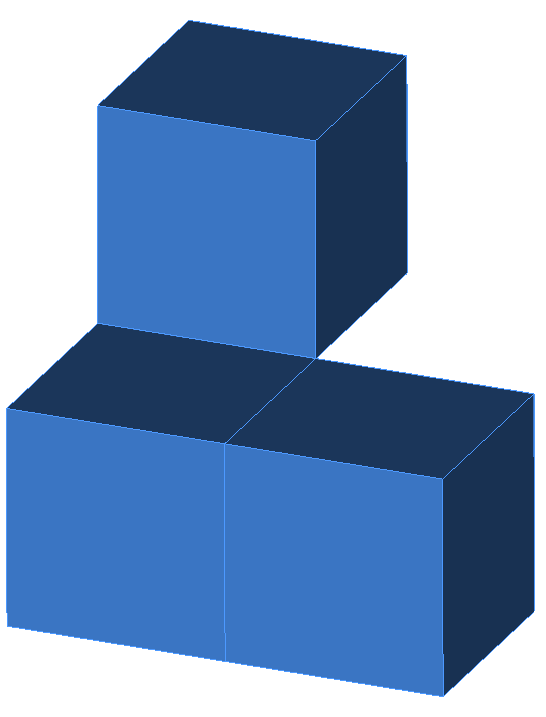}
	\label{fig:case3}
	}
	\hfill
	\subfloat[3 edges]{
		\includegraphics[trim=0cm 0.2cm 0cm 0.7cm, height=1in,clip]{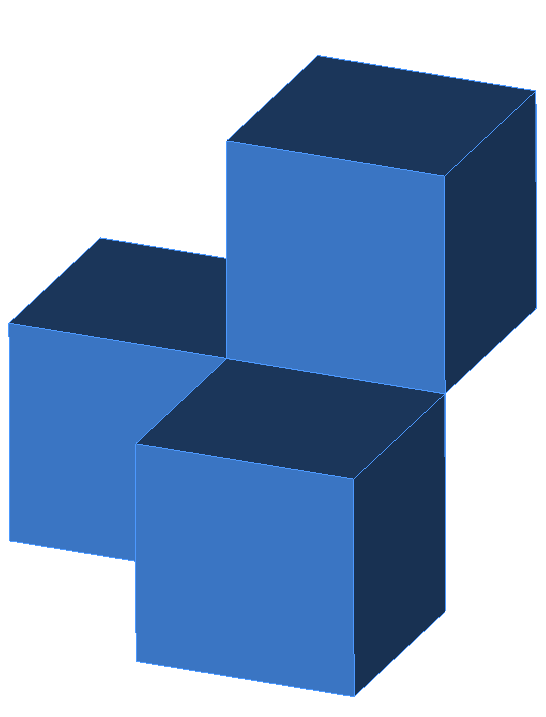}
	\label{fig:case4}
	}
	\hfill
	\subfloat[2 edges]{
		\includegraphics[trim=0cm 0.2cm 0cm 0.7cm, height=1in,clip]{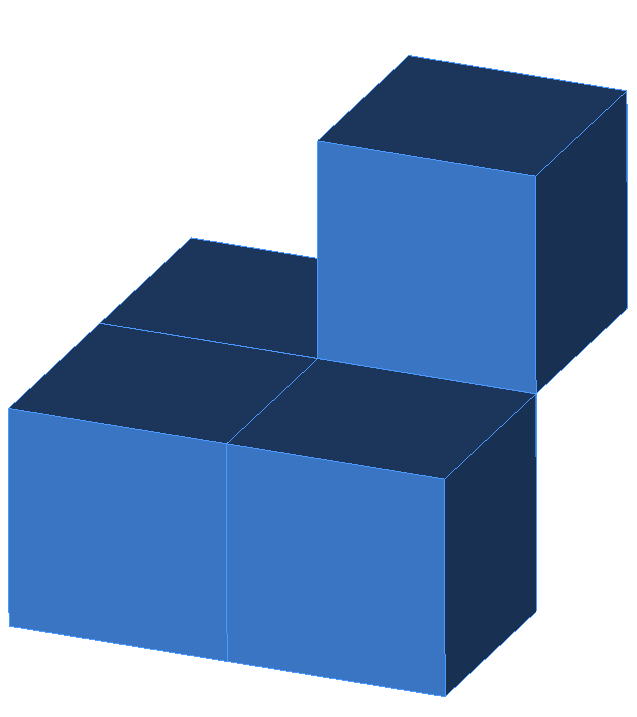}
	\label{fig:case5}
	}
	\hfill
	\subfloat[2 edges]{
		\includegraphics[trim=0cm 0.2cm 0cm 0.7cm, height=1in,clip]{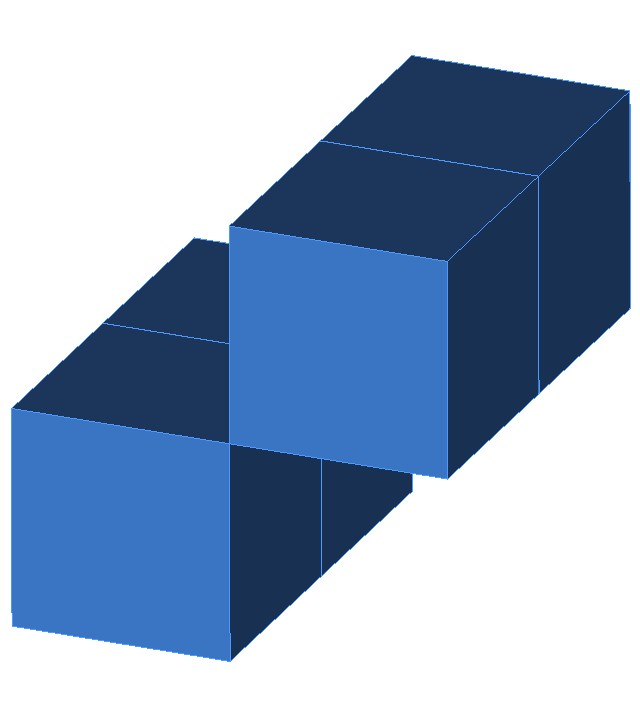}
	\label{fig:case6}
	}
	\hfill
	\subfloat[\hspace{-2pt}2\hspace{-1pt} edges,\hspace{-2pt} 1\hspace{-2pt} vertex]{
		\includegraphics[trim=0cm 0.2cm 0cm 0.7cm, height=1in,clip]{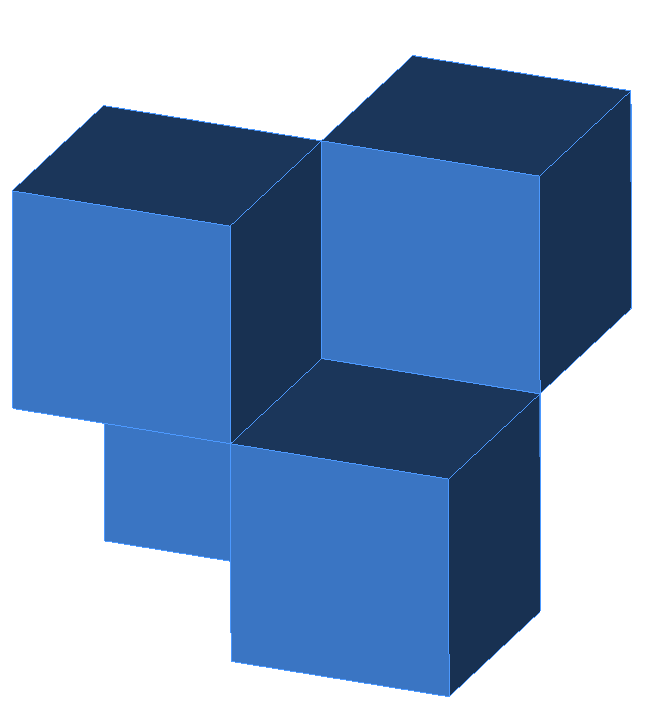}
	\label{fig:case7}
	}
	\hfill
	\subfloat[1 edge]{
		\includegraphics[trim=0cm 0.2cm 0cm 0.7cm, height=1in,clip]{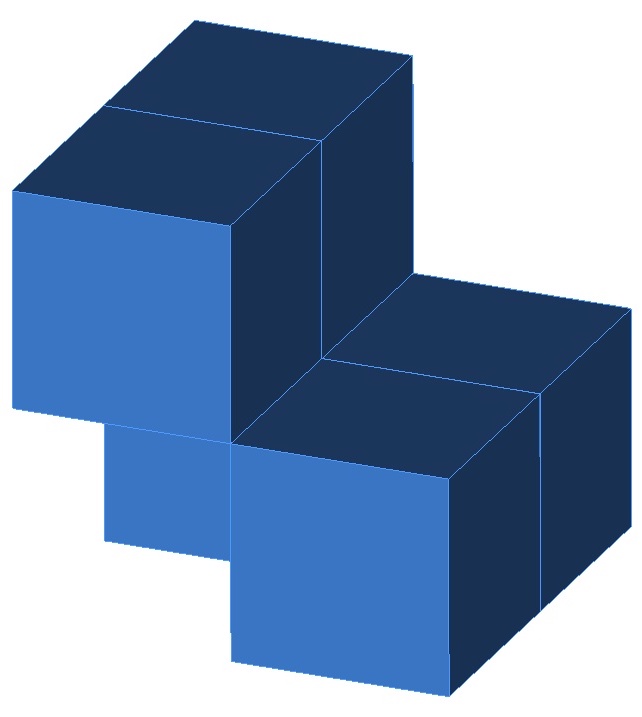}
	\label{fig:case8}
	}
	\hfill
	\subfloat[3 edges]{
		\includegraphics[trim=0cm 0.2cm 0cm 0.7cm, height=1in,clip]{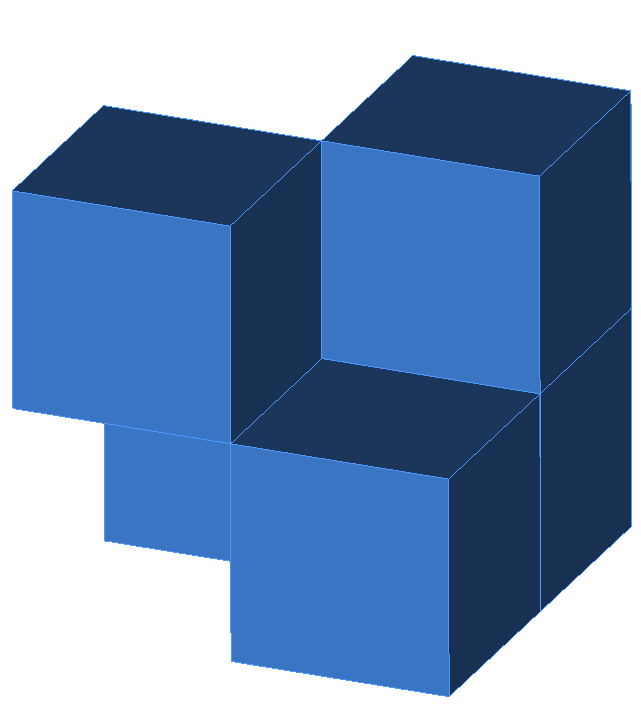}
	\label{fig:case9}
	}
	\hfill
	\subfloat[1 edge]{
		\includegraphics[trim=0cm 0.2cm 0cm 0.7cm, height=1in,clip]{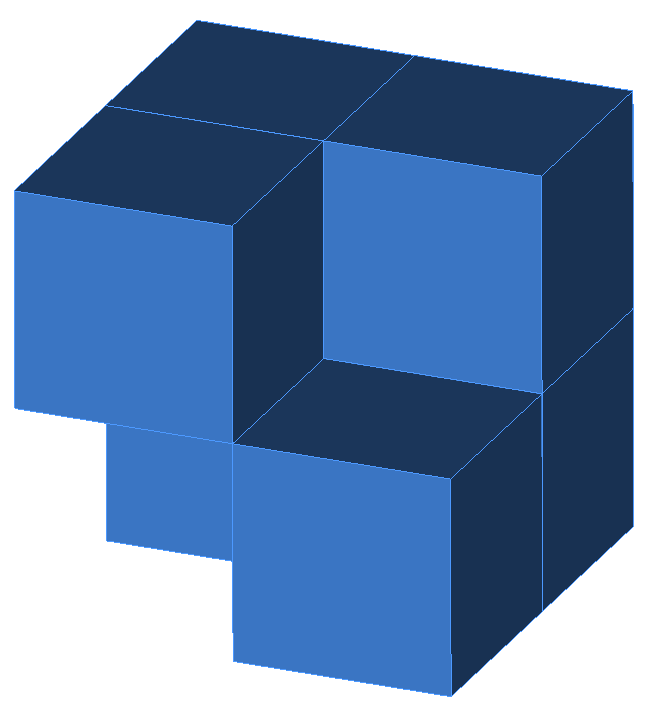}
	\label{fig:case10}
	}
	\hfill
	\subfloat[1 vertex]{
		\includegraphics[trim=0cm 0.2cm 0cm 0.7cm, height=1in,clip]{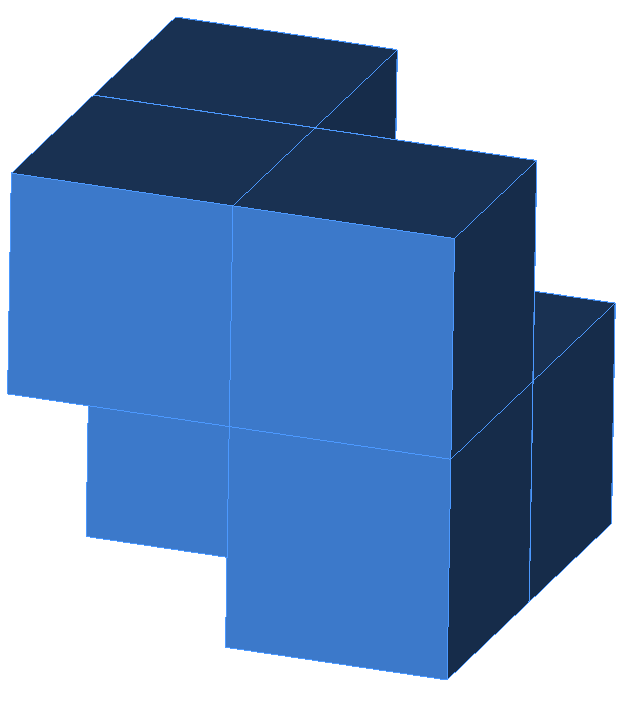}
	\label{fig:case11}
	}
	\caption{All possible pinches in 2D and 3D are shown.}
	\label{fig:3d_nonmanifold}
\end{figure}

\begin{figure}[!htbp]
	\centering
	\subfloat[2D 1--5]{
		\includegraphics[height=0.58in]{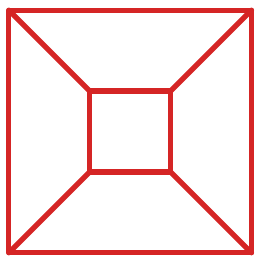}
	\label{fig:2d_split}
	}
	\hfill
	\subfloat[3D 1--7]{
		\includegraphics[height=0.58in]{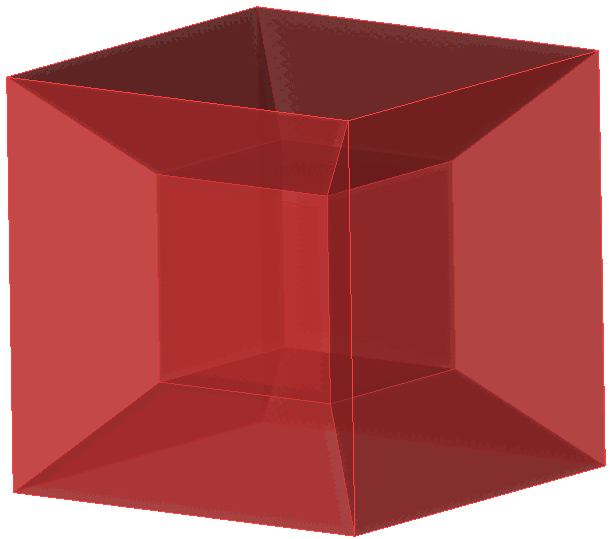}
	\label{fig:3d_split_1_7}
	}
	\hfill
	\subfloat[3D 2--6]{
		\includegraphics[height=0.63in]{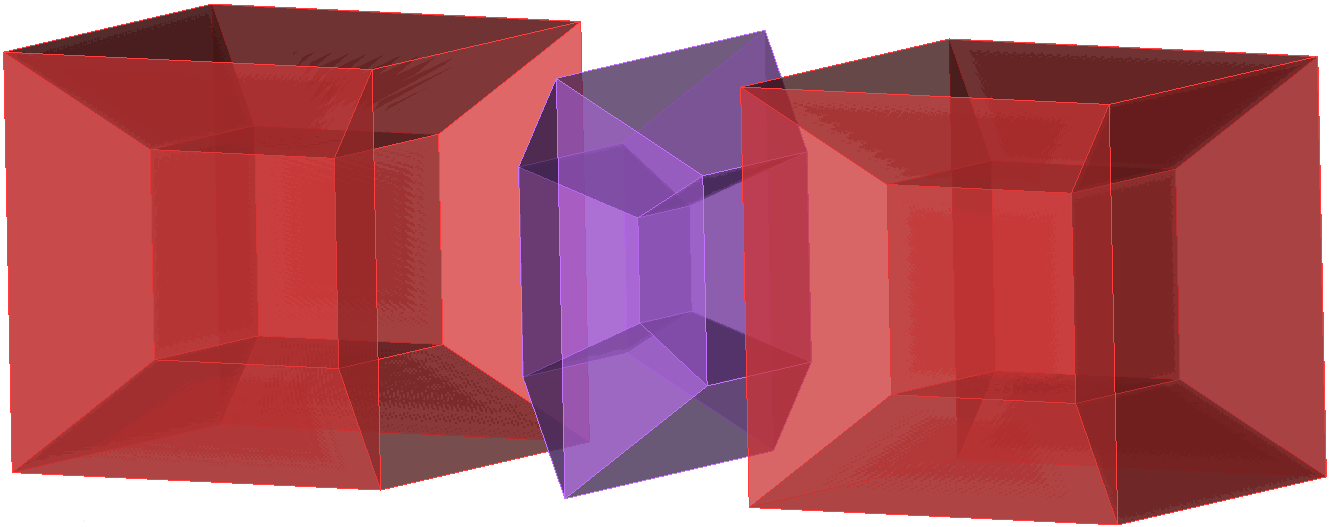}
	\label{fig:3d_split_2_6}
	}
	\caption{Templates for fixing pinches are depicted.
	\label{fig:templates}}
\end{figure}

\begin{figure}[!t]
	\centering
	\subfloat[2D vertex]{
		\includegraphics[trim=0cm 2cm 0cm 2cm, height=1in,clip]{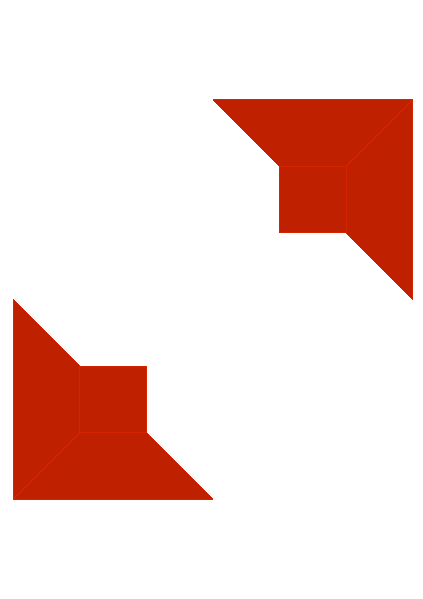}
		\label{fig:two_d_separated}
	}
	\hfill
	\subfloat[1 vertex]{
		\includegraphics[trim=0cm 1.1cm 0cm 1.25cm, height=1in,clip]{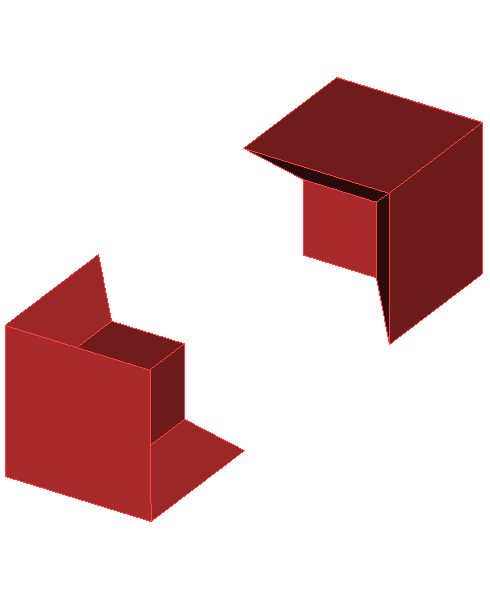}
	\label{fig:case1_separated}
	}
	\hfill
	\subfloat[1 edge]{
		\includegraphics[trim=0cm 1.1cm 0cm 1.25cm, height=1in,clip]{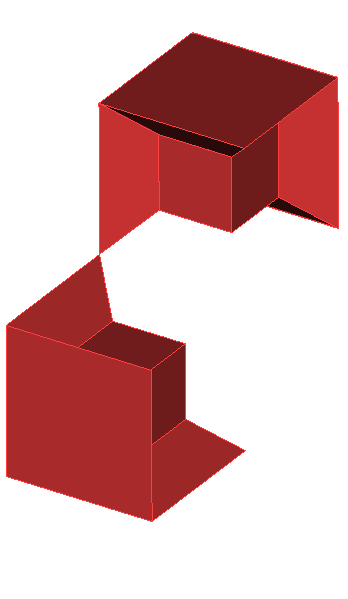}
	\label{fig:case2_separated}
	}
	\hfill
	\subfloat[1 edge]{
		\includegraphics[trim=0cm 1.1cm 0cm 1.25cm, height=1in,clip]{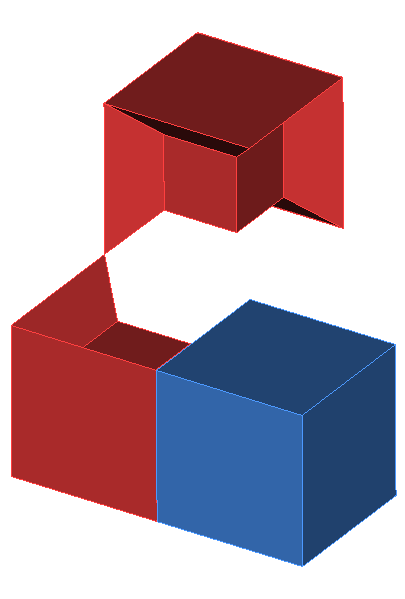}
	\label{fig:case3_separated}
	}
	\hfill
	\subfloat[3 edges]{
		\includegraphics[trim=0cm 1.1cm 0cm 1.25cm, height=1in,clip]{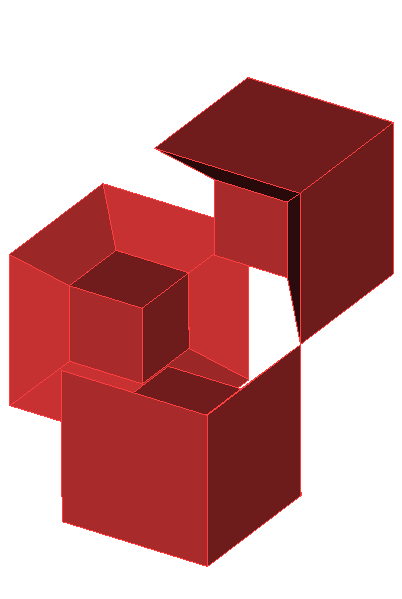}
	\label{fig:case4_seaprated}
	}
	\hfill
	\subfloat[2 edges]{
		\includegraphics[trim=0cm 1.1cm 0cm 1.25cm, height=1in,clip]{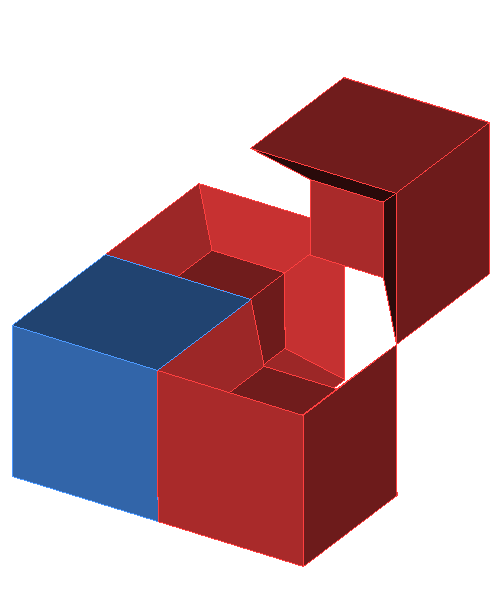}
	\label{fig:case5_separated}
	}
	\hfill
	\subfloat[2 edges]{
		\includegraphics[trim=0cm 1.1cm 0cm 1.25cm, height=1in,clip]{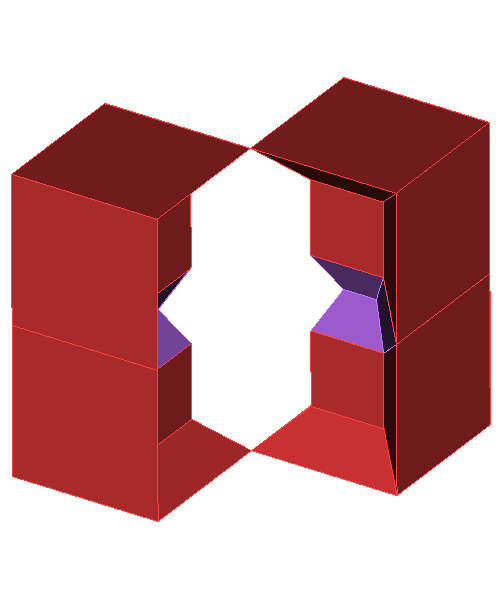}
	\label{fig:case6_separated}
	}
	\hfill
	\subfloat[\hspace{-2pt}2\hspace{-1pt} edges,\hspace{-2pt} 1\hspace{-2pt} vertex]{
		\includegraphics[trim=0cm 1.1cm 0cm 1.25cm, height=1in,clip]{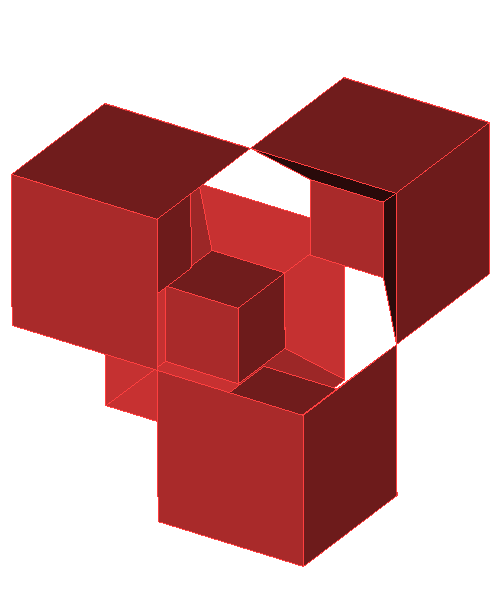}
	\label{fig:case7_separated}
	}
	\hfill
	\subfloat[1 edge]{
		\includegraphics[trim=0cm 1.1cm 0cm 1.25cm, height=1in,clip]{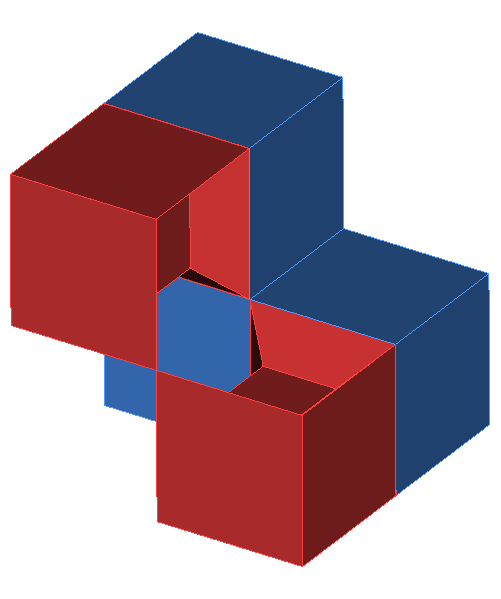}
	\label{fig:case8_separated}
	}
	\hfill
	\subfloat[3 edges]{
		\includegraphics[trim=0cm 1.1cm 0cm 1.25cm, height=1in,clip]{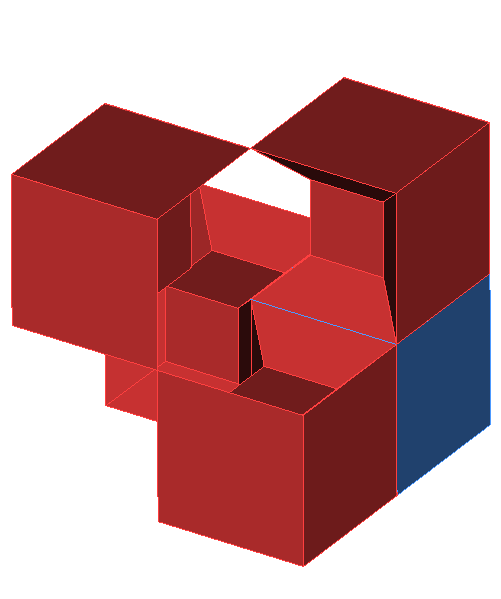}
	\label{fig:case9_separated}
	}
	\hfill
	\subfloat[1 edge]{
		\includegraphics[trim=0cm 1.1cm 0cm 1.25cm, height=1in,clip]{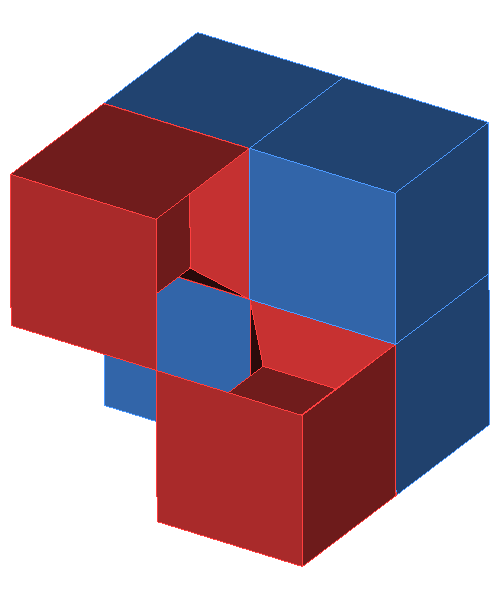}
	\label{fig:case10_separated}
	}
	\hfill
	\subfloat[1 vertex]{
		\includegraphics[trim=0cm 1.1cm 0cm 1.25cm, height=1in,clip]{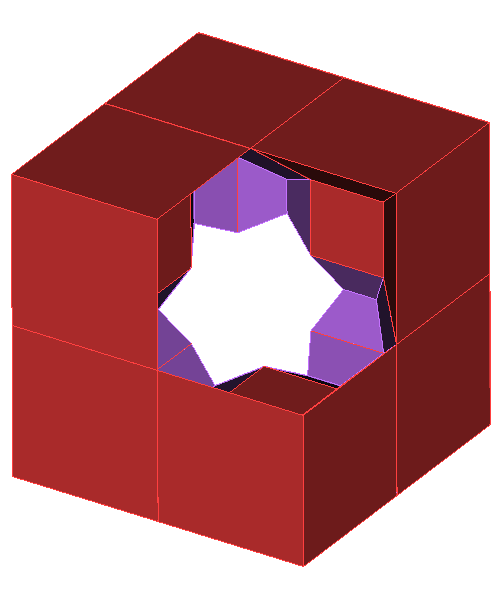}
	\label{fig:case11_separated}
	}
  \caption{Pinch shrinking templates are shown. Pinches on the boundary of the neighborhoods (non-centered vertices) are resolved by neighborhoods centered on them.}\label{fig:cases_separated}
\end{figure}

\begin{figure}[!t]
\centering
	\subfloat[2D vertex]{
		\includegraphics[trim=0cm 2cm 0cm 2cm, height=1in,clip]{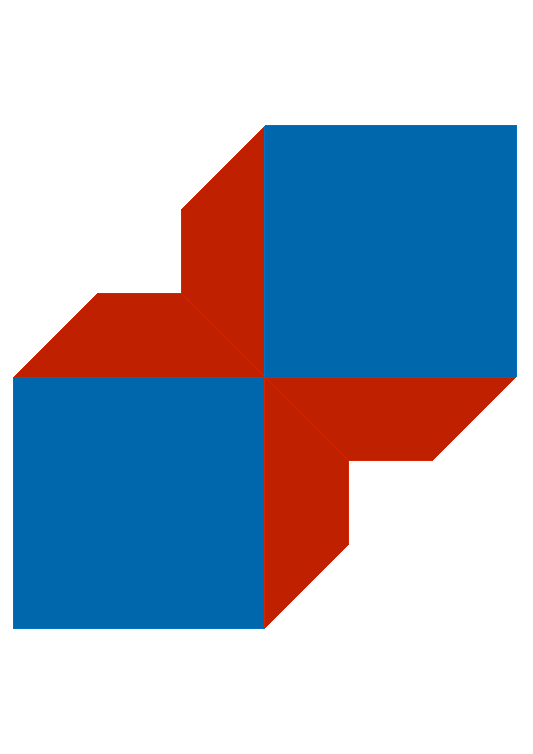}
		\label{fig:two_d_connected}
	}
	\hfill
	\subfloat[1 vertex]{
		\includegraphics[trim=0cm 1.1cm 0cm 1.25cm, height=1in,clip]{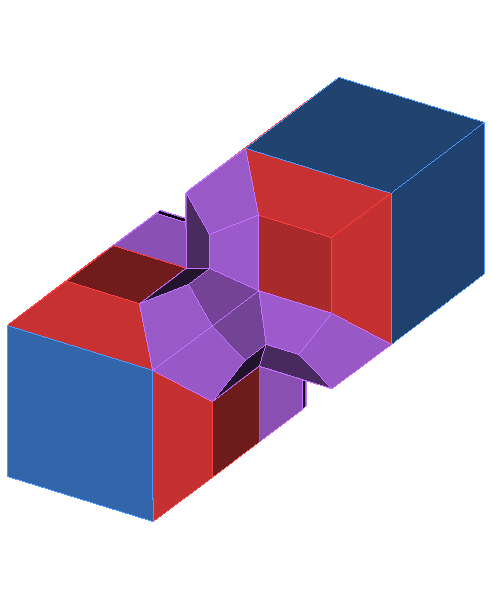}
	\label{fig:case1_connected}
	}
	\hfill
	\subfloat[1 edge]{
		\includegraphics[trim=0cm 1.1cm 0cm 1.25cm, height=1in,clip]{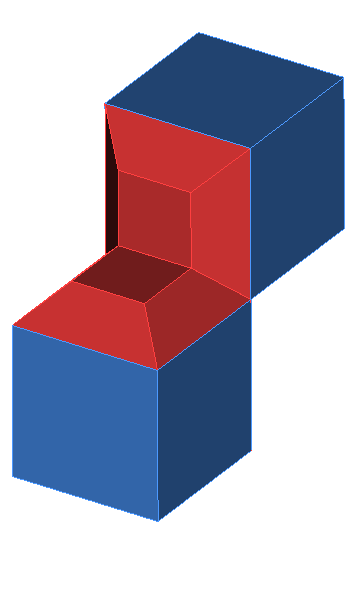}
	\label{fig:case2_connected}
	}
	\hfill
	\subfloat[1 edge]{
		\includegraphics[trim=0cm 1.1cm 0cm 1.25cm, height=1in,clip]{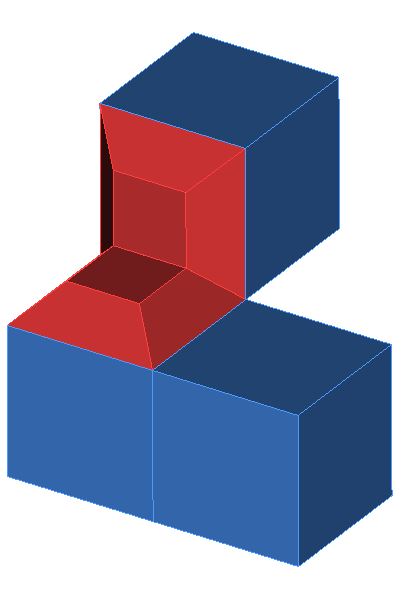}
	\label{fig:case3_connected}
	}
	\hfill
	\subfloat[3 edges]{
		\includegraphics[trim=0cm 1.1cm 0cm 1.25cm, height=1in,clip]{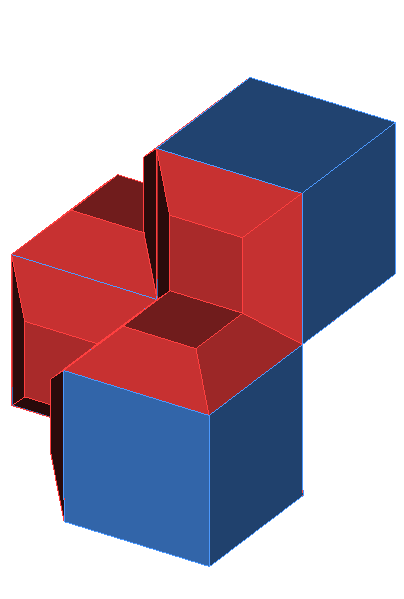}
	\label{fig:case4_connected}
	}
	\hfill
	\subfloat[2 edges]{
		\includegraphics[trim=0cm 1.1cm 0cm 1.25cm, height=1in,clip]{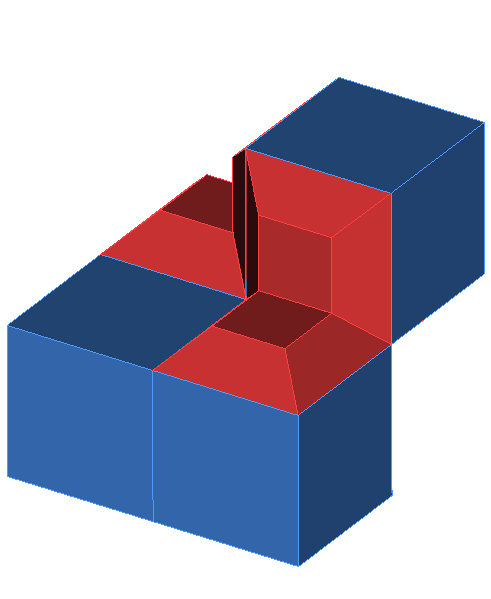}
	\label{fig:case5_connected}
	}
	\hfill
	\subfloat[2 edges]{
		\includegraphics[trim=0cm 1.1cm 0cm 1.25cm, height=1in,clip]{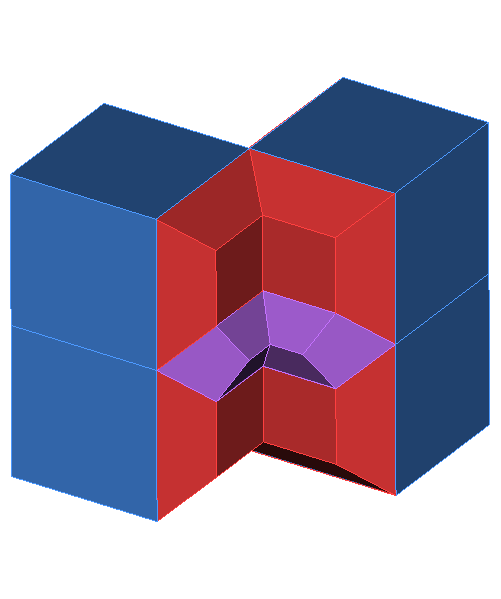}
	\label{fig:case6_connected}
	}
	\hfill
	\subfloat[\hspace{-2pt}2\hspace{-1pt} edges,\hspace{-2pt} 1\hspace{-2pt} vertex]{
		\includegraphics[trim=0cm 1.1cm 0cm 1.25cm, height=1in,clip]{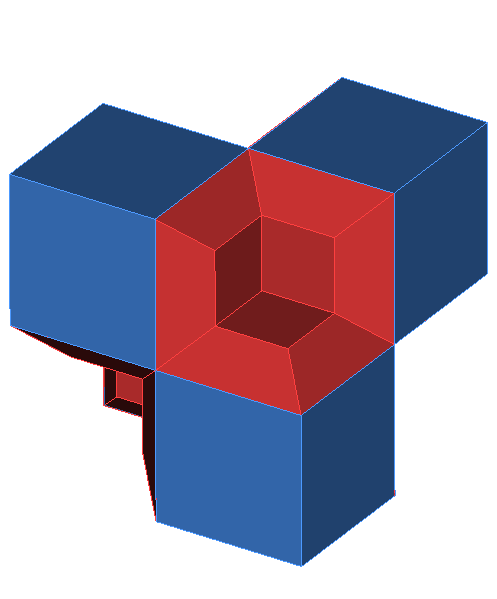}
	\label{fig:case7_connected}
	}
	\hfill
	\subfloat[1 edge]{
		\includegraphics[trim=0cm 1.1cm 0cm 1.25cm, height=1in,clip]{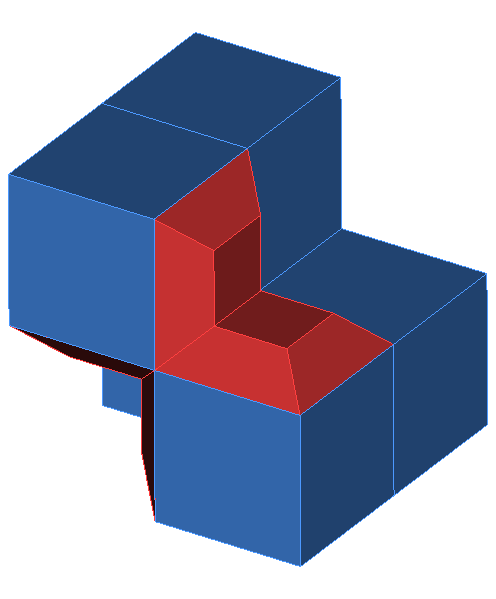}
	\label{fig:case8_connected}
	}
	\hfill
	\subfloat[3 edges]{
		\includegraphics[trim=0cm 1.1cm 0cm 1.25cm, height=1in,clip]{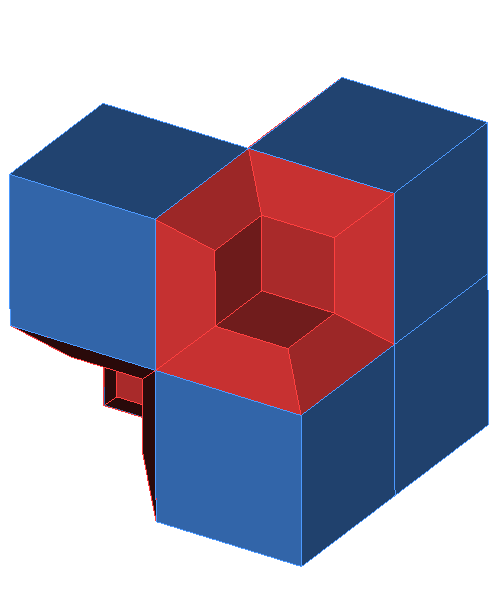}
	\label{fig:case9_connected}
	}
	\hfill
	\subfloat[1 edge]{
		\includegraphics[trim=0cm 1.1cm 0cm 1.25cm, height=1in,clip]{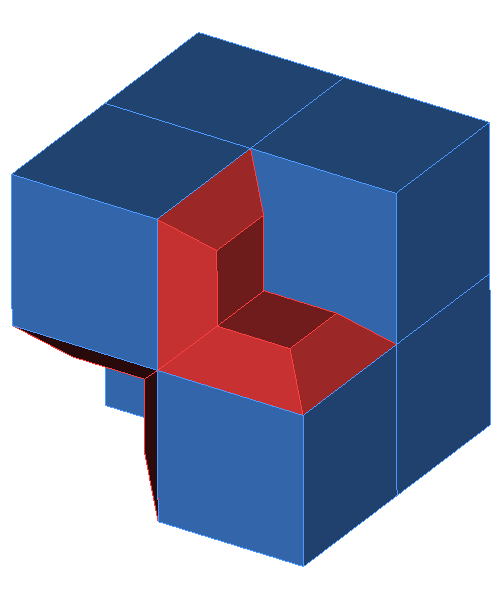}
	\label{fig:case10_connected}
	}
	\hfill
	\subfloat[1 vertex]{
		\includegraphics[trim=0cm 1.1cm 0cm 1.25cm, height=1in,clip]{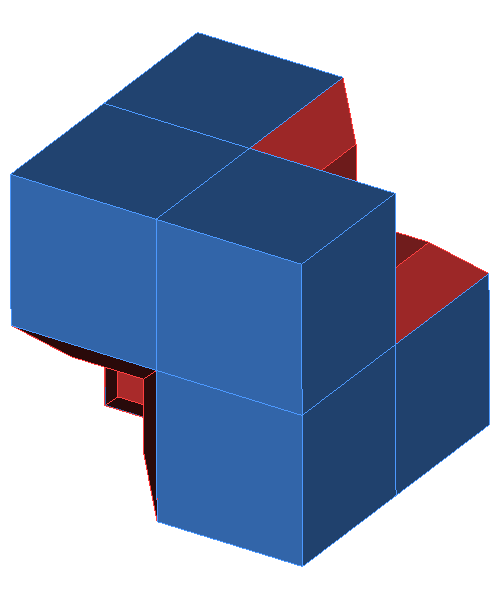}
	\label{fig:case11_connected}
	}
  \caption{Pinch growing templates are shown.}\label{fig:cases_connected}
\end{figure} 

In 2D, the only possible pinch is two quads meeting at a pinch vertex, whereas in 3D there are 11 possible configurations of pinched edges and vertices.
These are shown in \cref{fig:3d_nonmanifold}.
To find pinches, we consider each vertex and the neighborhood of cells containing it, 
i.e., $2 \times  2$ quads in 2D and $2 \times 2 \times 2$ hexes in 3D.
If the neighborhood corresponds to a pinch case, the pinch vertices and edges are queued.
Each pinch in the queue is processed in a way that is compatible with processing nearby pinches.
The  pinches are connected if the subcells (vertices or edges) are interior, and disconnected if they are exterior. 
Each pinch is repaired by splitting cells (either mesh cells or their complement) using predefined splits, and discarding or adding some of the split cells.
The splits are shown in \cref{fig:templates}.
In 2D, the template is a one-to-five split.
In 3D, pinch edges are repaired before vertices.
The edge-repair template is a one-to-seven split. 
For pinch vertices, we follow with a two-to-six split of any pairs of hexes from two different one-to-seven splits that share a face; see \cref{fig:3d_split_2_6}.

\begin{figure}[!htb]
\centering
\subfloat[Mesh]{\includegraphics[trim=17cm 4cm 8cm 4cm,width=0.45\columnwidth,clip]{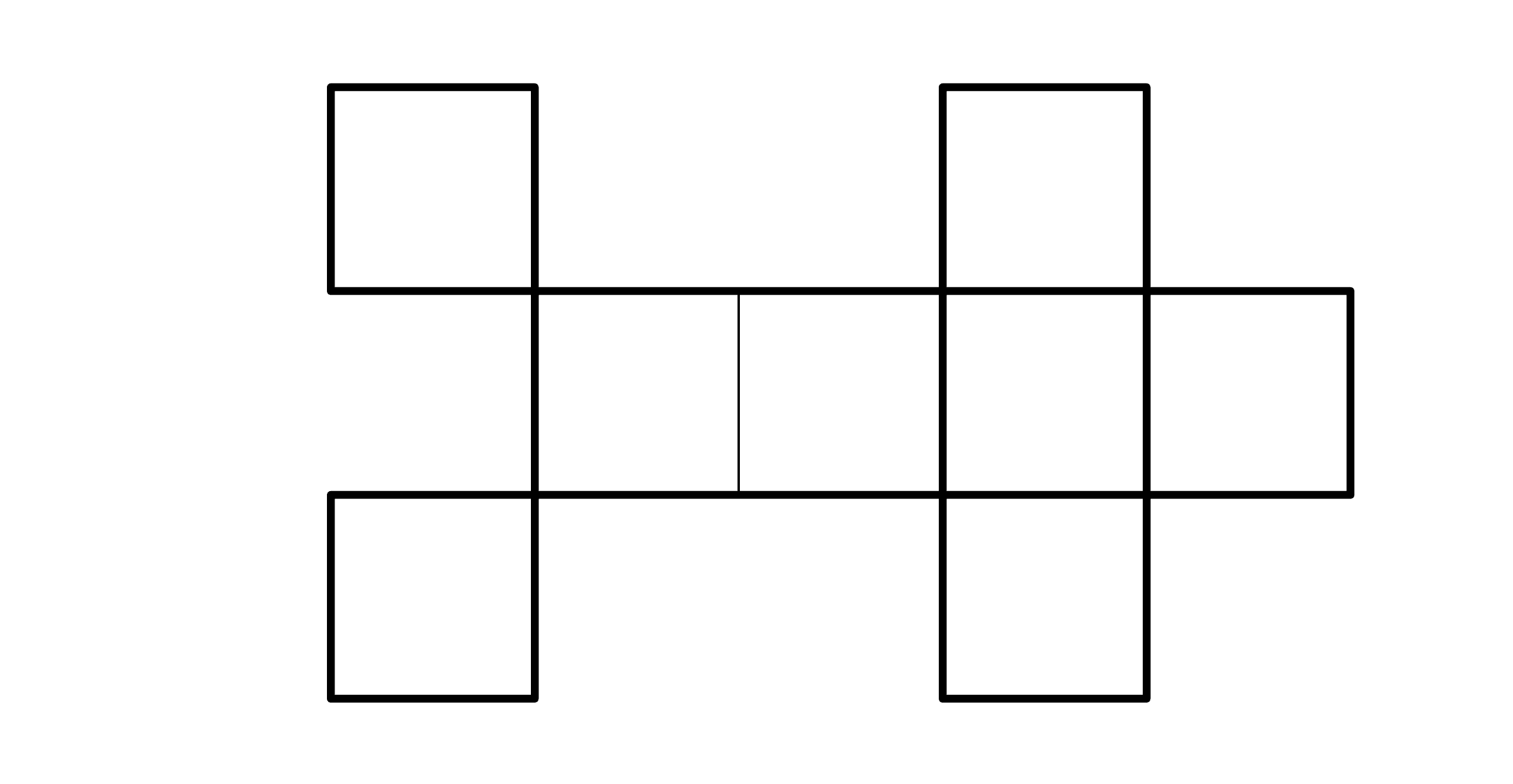}}
\hfill
\subfloat[Separated]{\includegraphics[trim=17cm 4cm 8cm 4cm,width=0.45\columnwidth,clip]{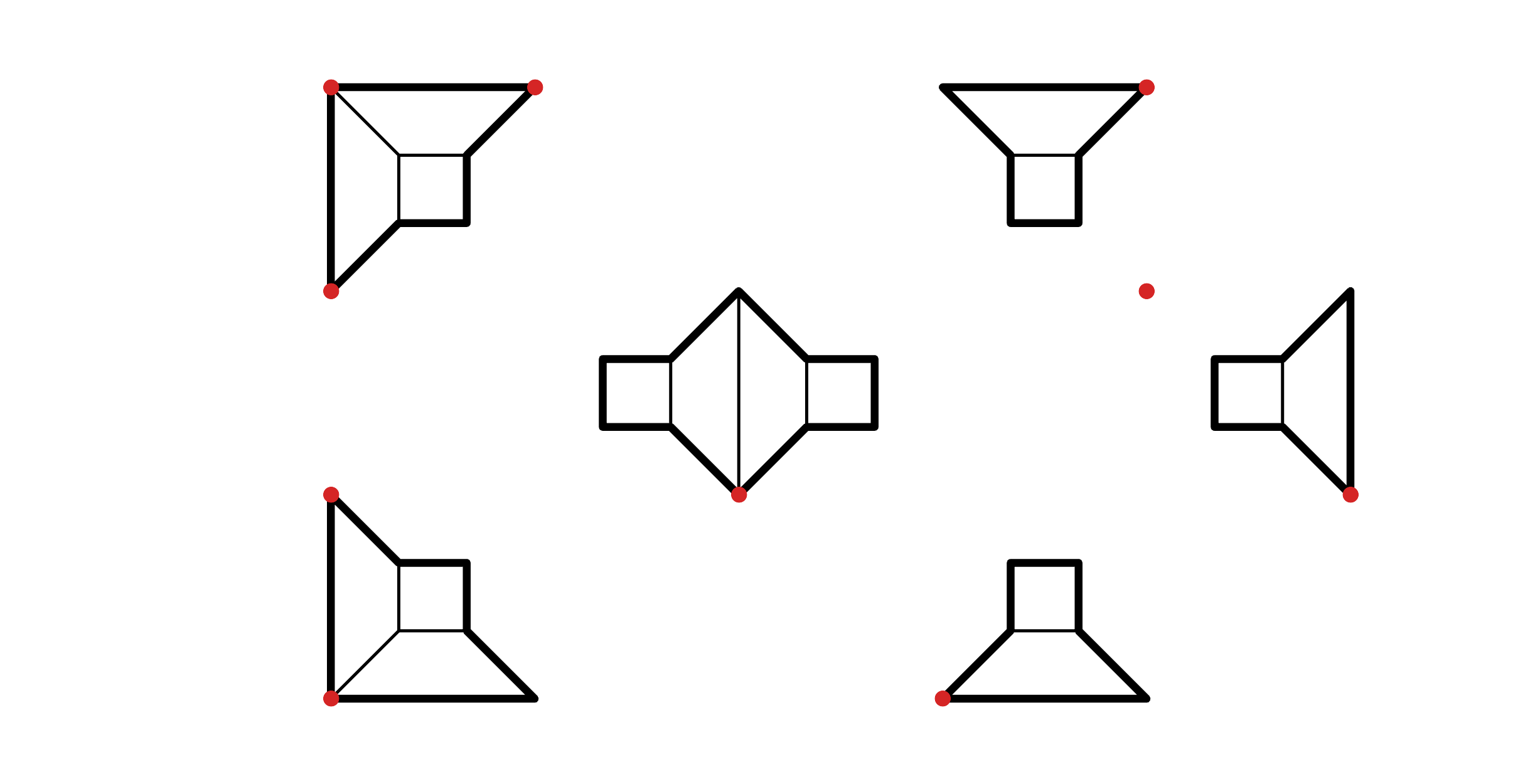}}
\\
\subfloat[Connected]{\includegraphics[trim=17cm 4cm 8cm 4cm, width=0.45\columnwidth,clip]{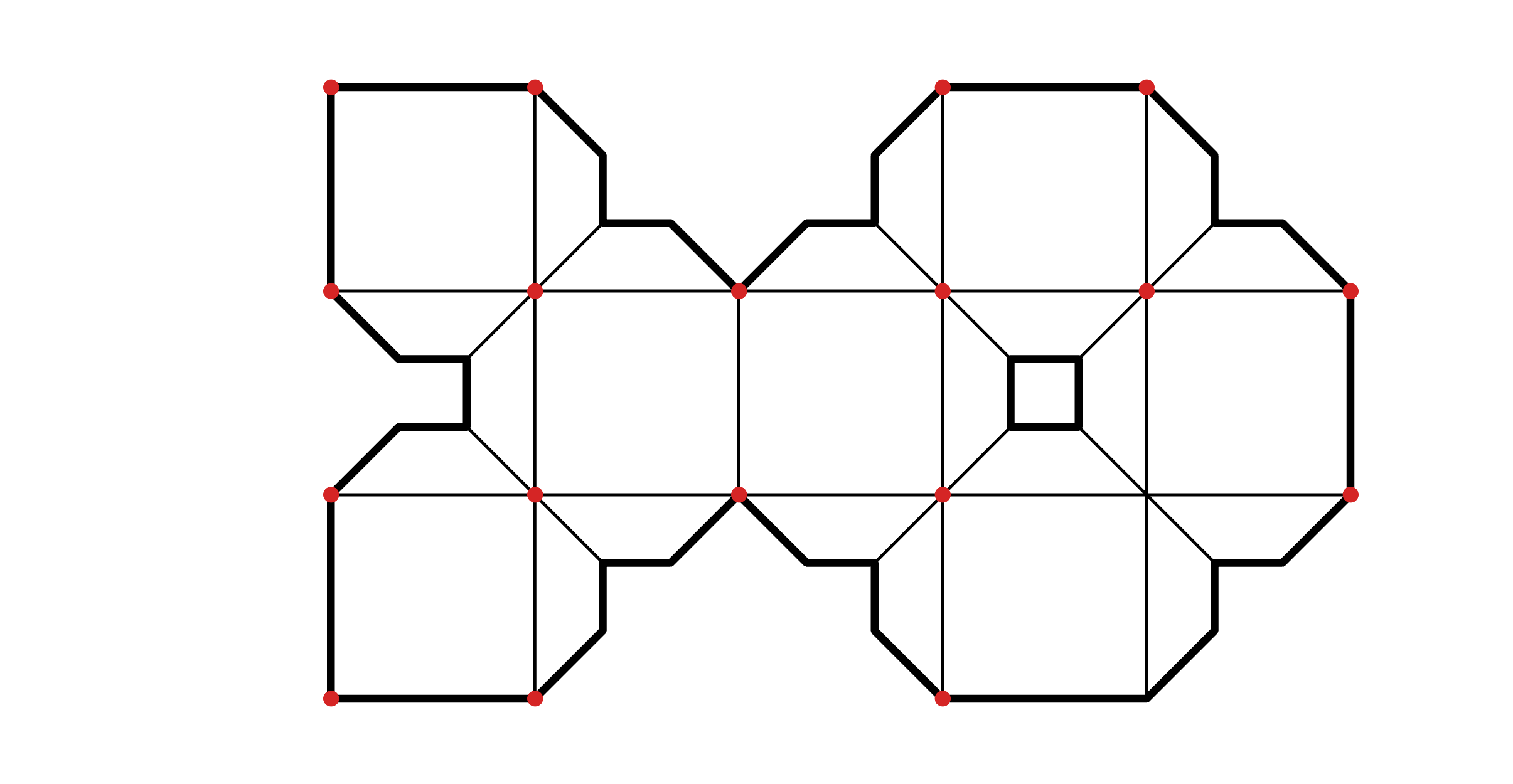}}
\hfill
\subfloat[Both]{\includegraphics[trim=17cm 4cm 8cm 4cm, width=0.45\columnwidth,clip]{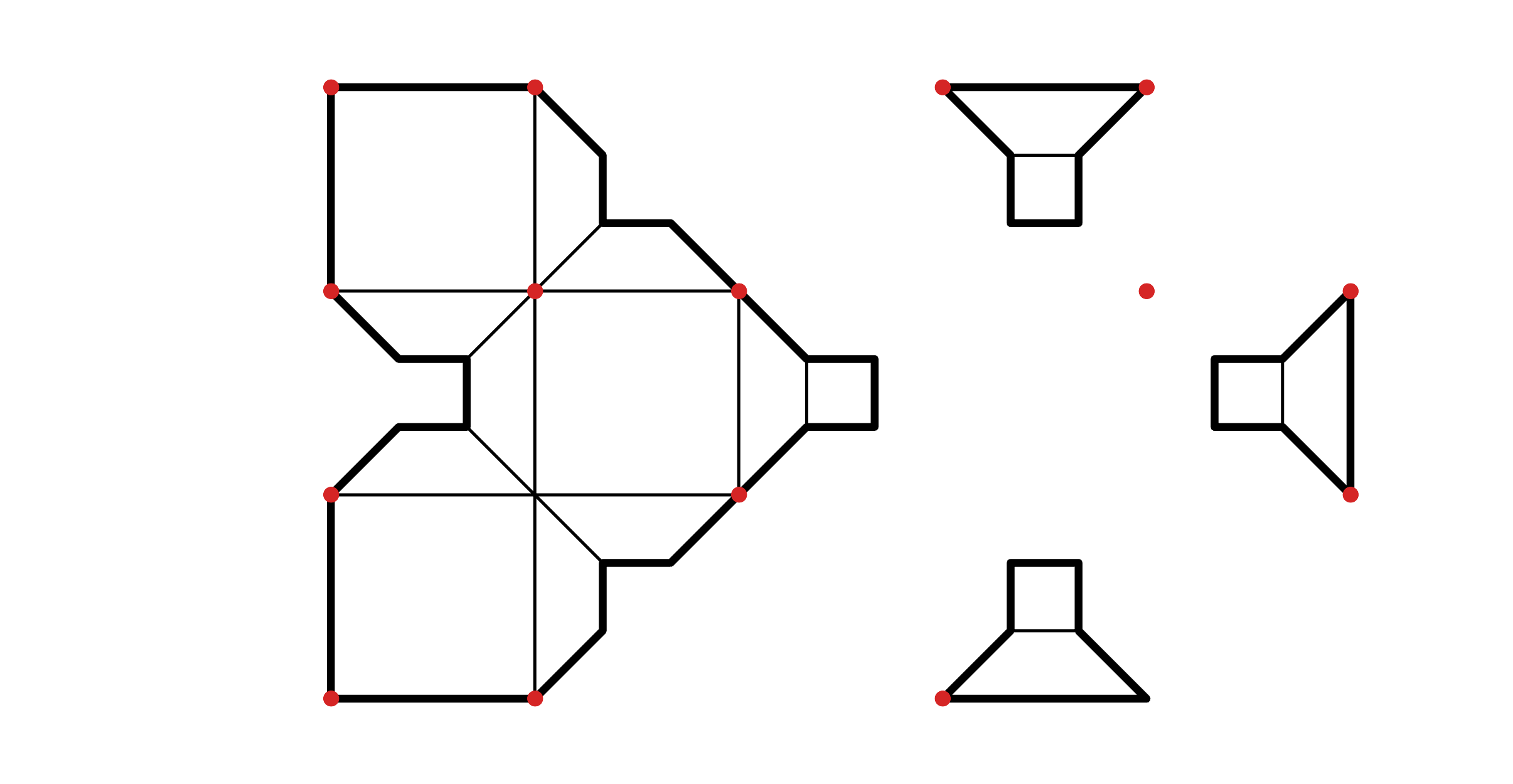}}
\caption{Adjacent pinches must be resolved the same way. Here the resolved mesh is shown for different configurations of ``inside'' vertices (red).}
\label{fig:pinches_in_series}
\end{figure}

\begin{figure}[!htb]
\centering
\subfloat[Vol. Frac.: 1.0]{\includegraphics[trim=15cm 2cm 11cm 2cm,width=0.4\columnwidth,clip]{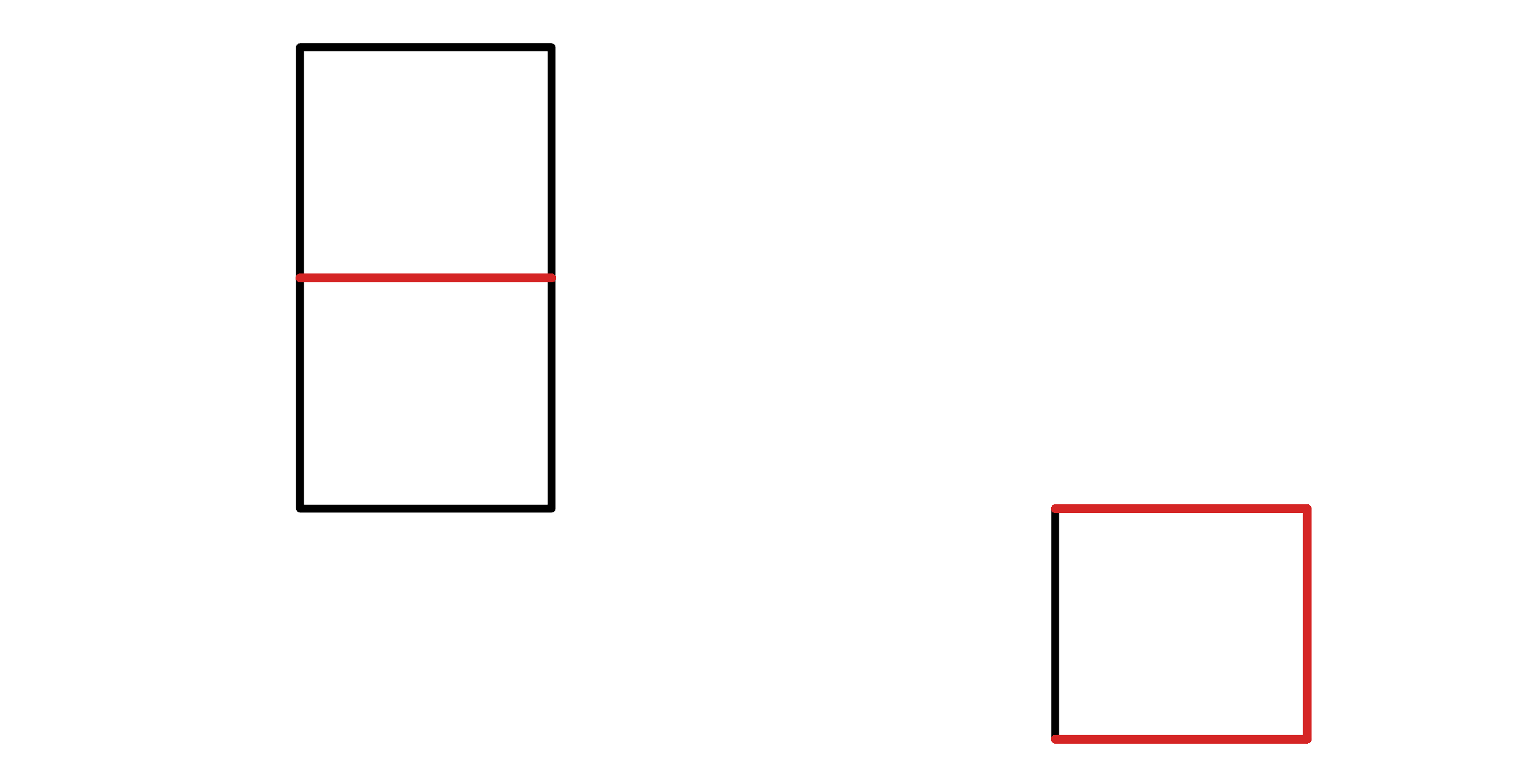}}
\hfill
\subfloat[Vol. Frac.: 0.75]{\includegraphics[trim=15cm 2cm 11cm 2cm,width=0.4\columnwidth,clip]{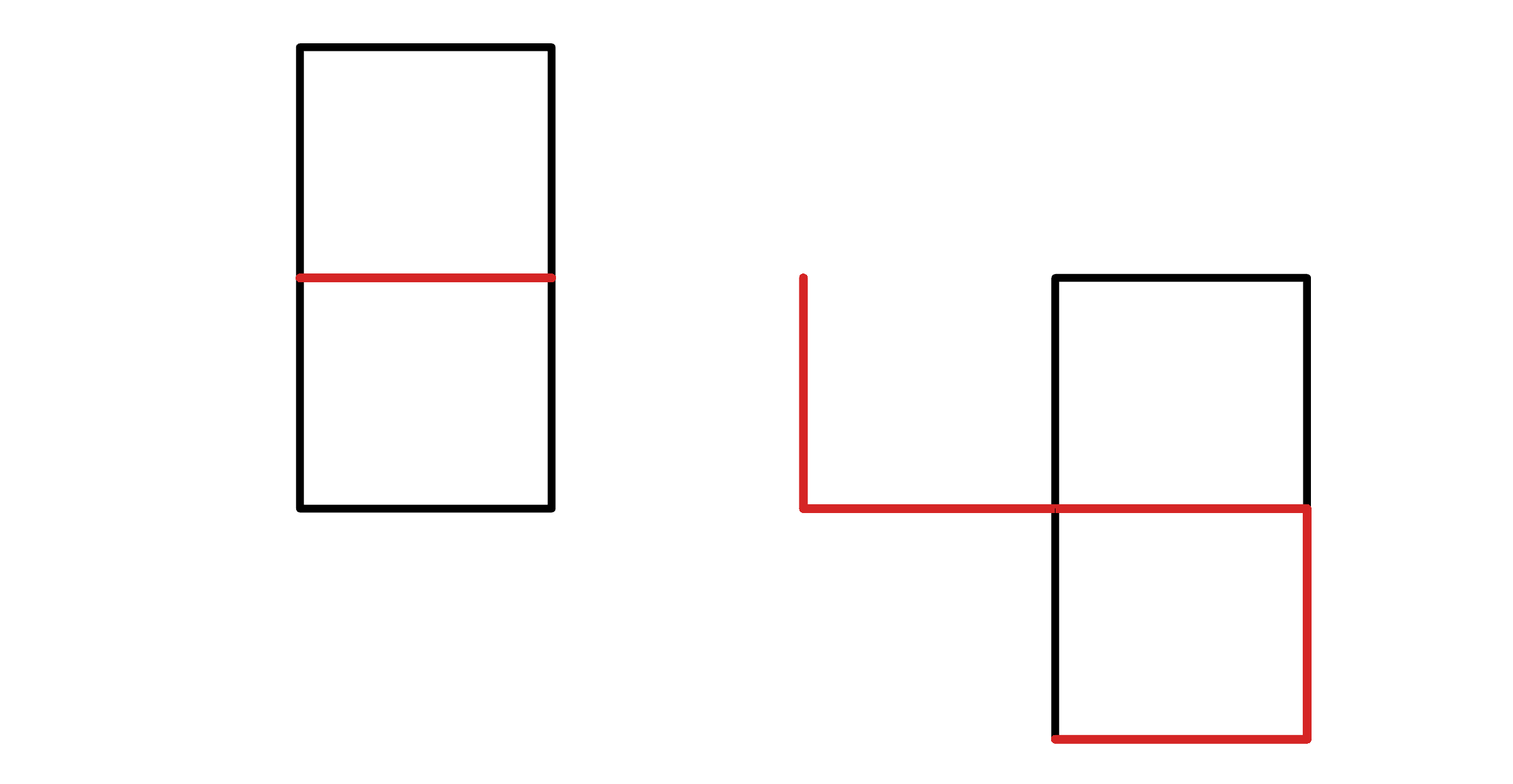}}
\\
\subfloat[Vol. Frac.: 0.5]{\includegraphics[trim=15cm 2cm 11cm 2cm,width=0.4\columnwidth,clip]{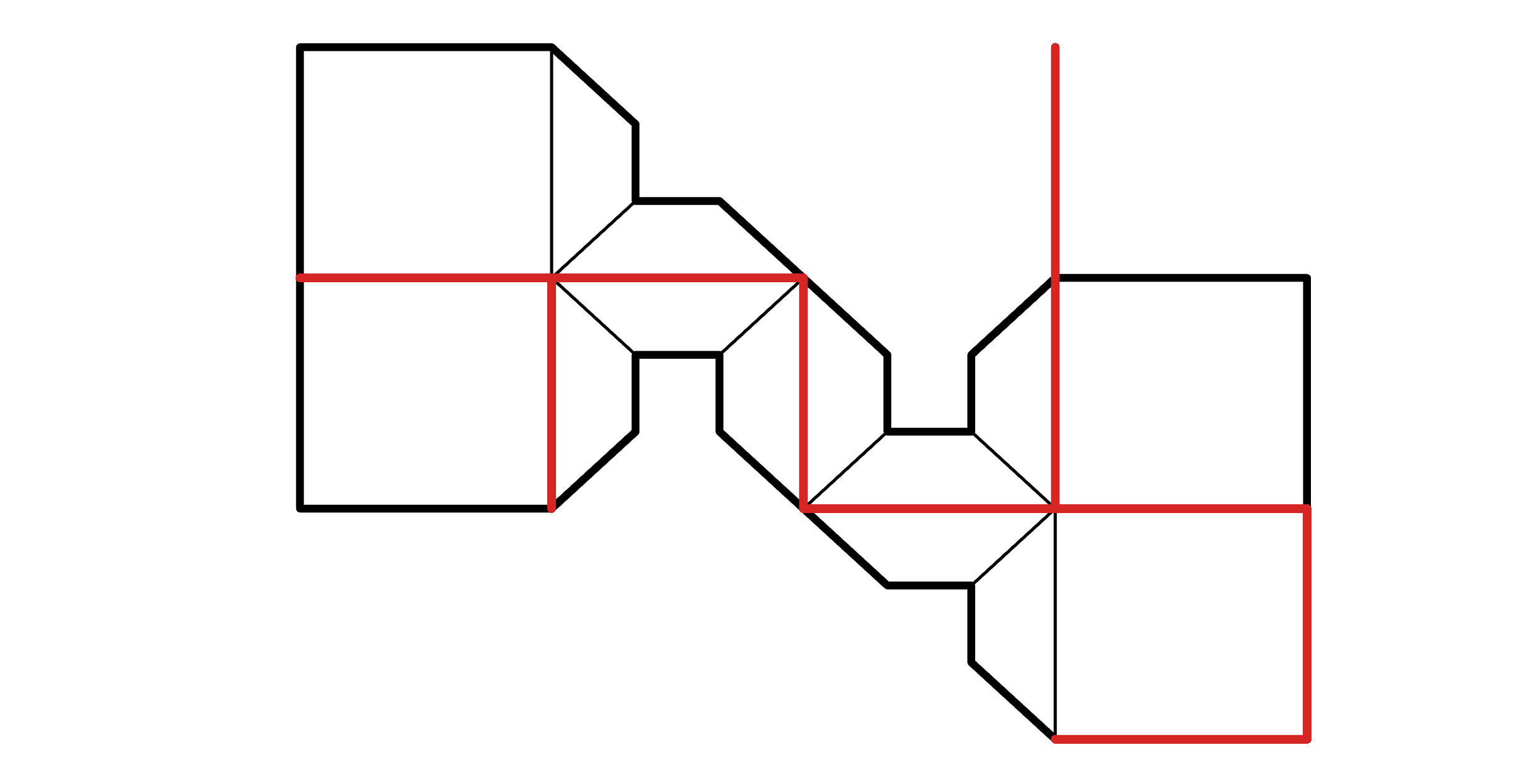}}
\hfill
\subfloat[Vol. Frac.: 0.25]{\includegraphics[trim=15cm 2cm 11cm 2cm,width=0.4\columnwidth,clip]{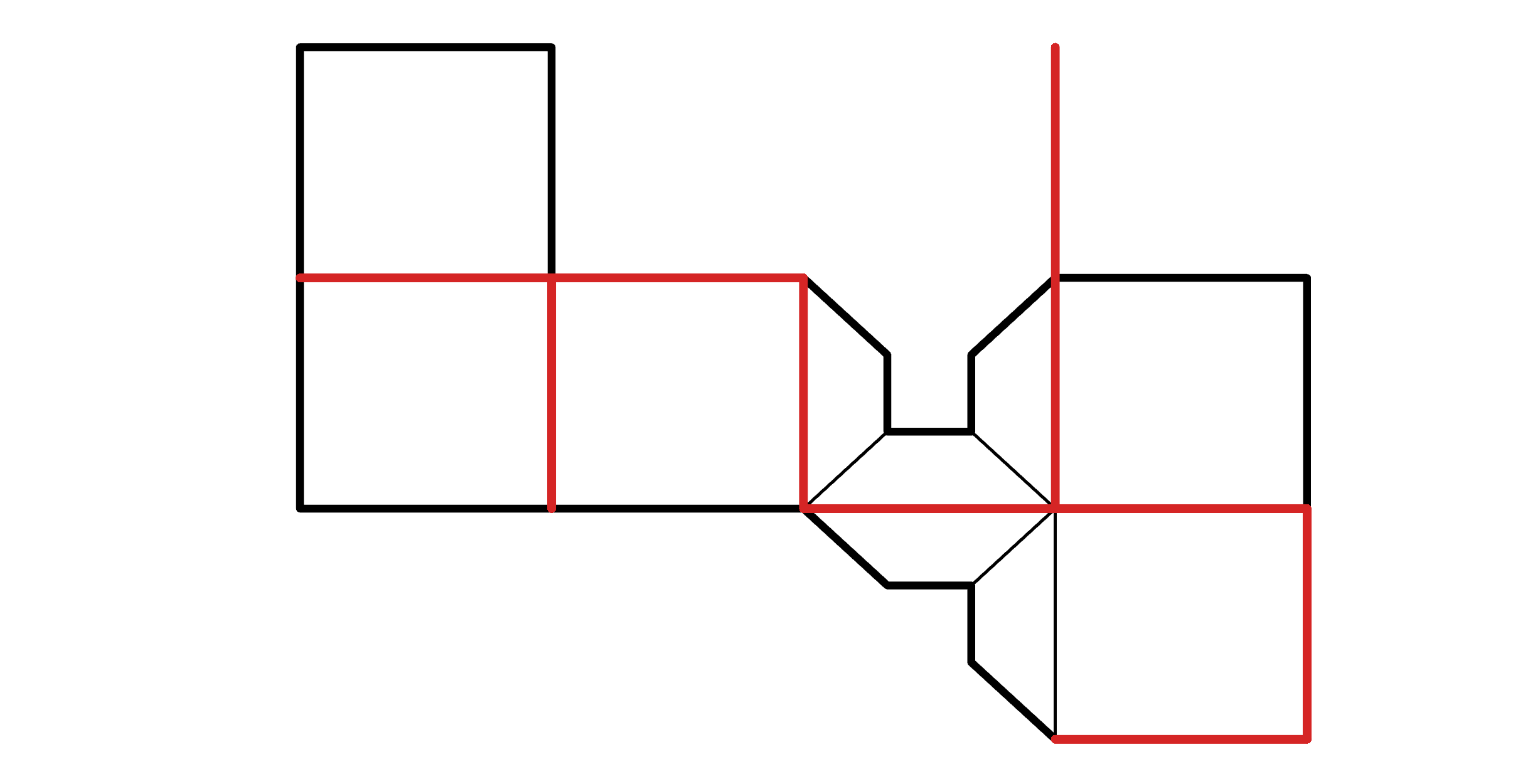}}
\caption{Faces are connected using templates when ``inside'' edges (red) create a continuous bridge between components.}
\label{fig:connecting_edges}
\end{figure}

To separate cells, splits are performed on the cells of the mesh itself, and child cells that contain pinch vertices or edges are removed, as shown in \cref{fig:cases_separated}.
To connect cells, splits are performed on the complement of the mesh, and child cells that contain pinches are added to the mesh, as illustrated in \cref{fig:cases_connected}.
A single mesh can use both separations and connections in different regions.
However, we require that all adjacent pinches must be resolved in the same way to ensure validity of the resulting mesh.
Two sets of pinches separated by cells without pinches can be resolved in opposite ways.
Our rules occasionally indicate that adjacent pinches should be resolved in opposite ways.
We pre-select whether we connect or separate these cases, see \cref{fig:pinches_in_series}.

\begin{figure}[!htb]
\centering
\subfloat[Subcells]{
	\includegraphics[trim=10.833cm 16.667cm 11.662cm 8.333cm,width=0.98\columnwidth,clip]{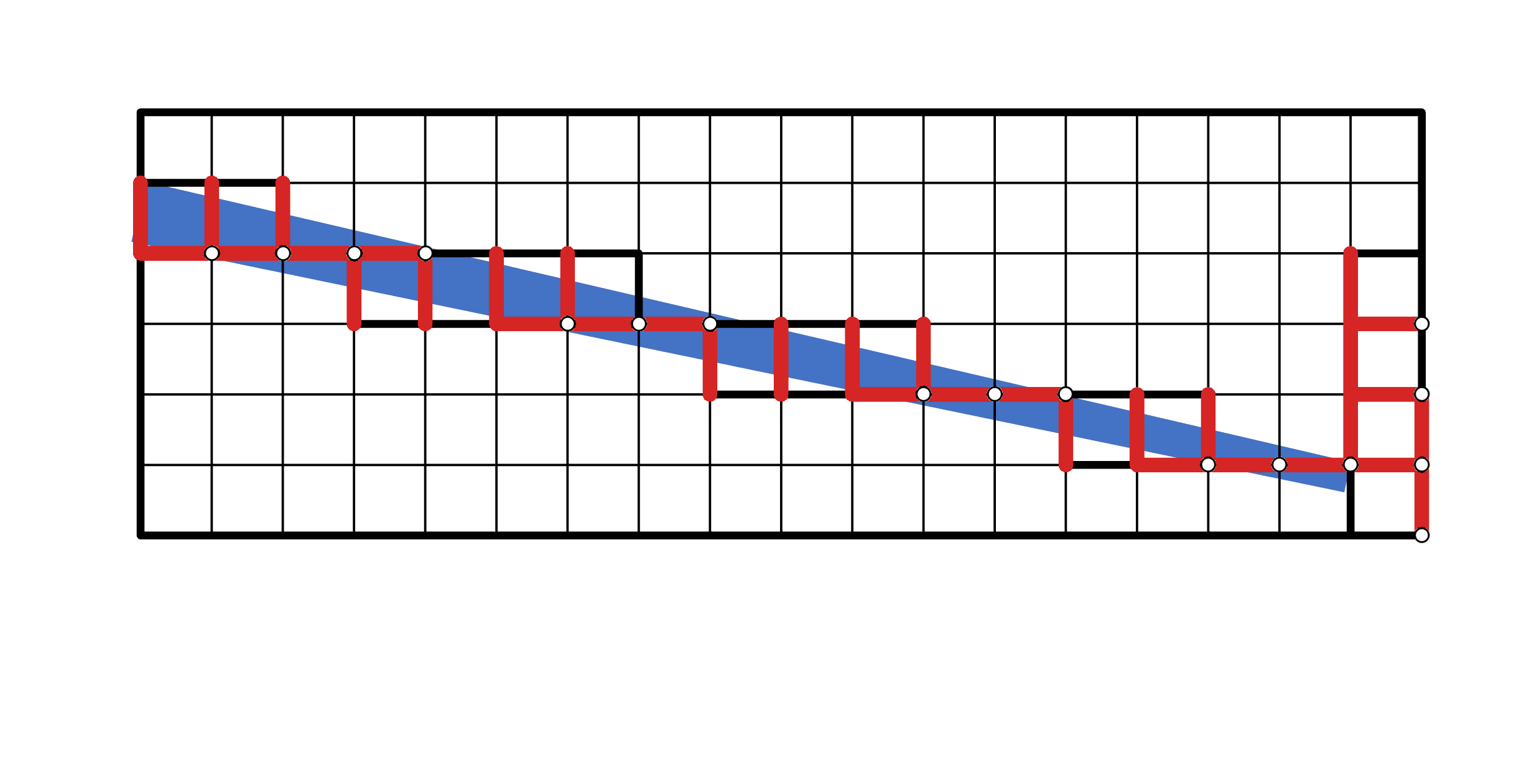}}
\\
\subfloat[Anti-aliased]{
	\includegraphics[trim=13cm 20cm 14cm 10cm,width=0.98\columnwidth,clip]{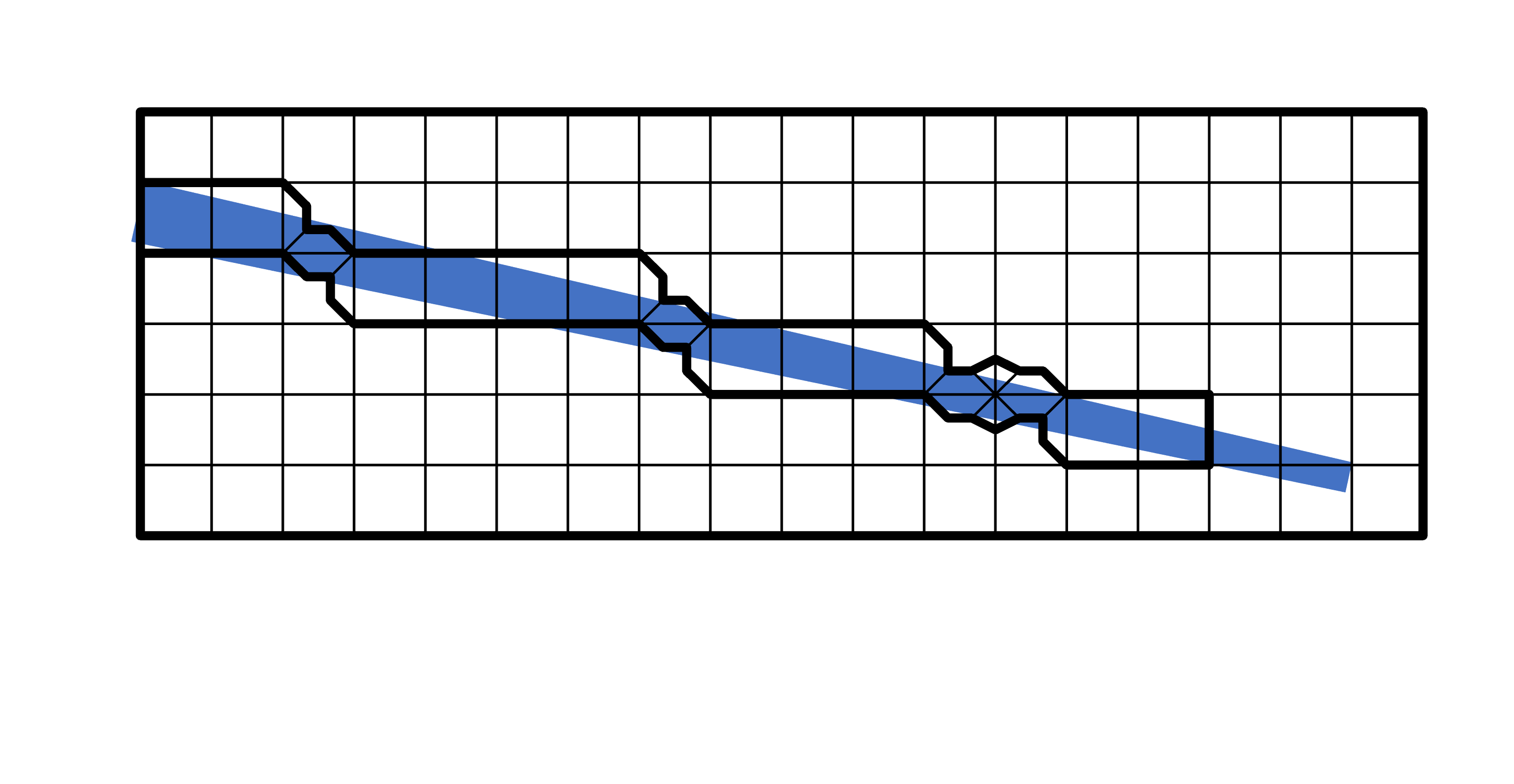}}
\caption{Subgrid sampling and anti-aliasing performed on \cref{fig:jaggies}.}
\label{fig:jaggies_subcells}
\end{figure}

\begin{figure}[!htb]
\centering
\subfloat[Subcells]{
	\includegraphics[trim=30cm 9.167cm 27.5cm 8.333cm,width=0.46\columnwidth,clip]{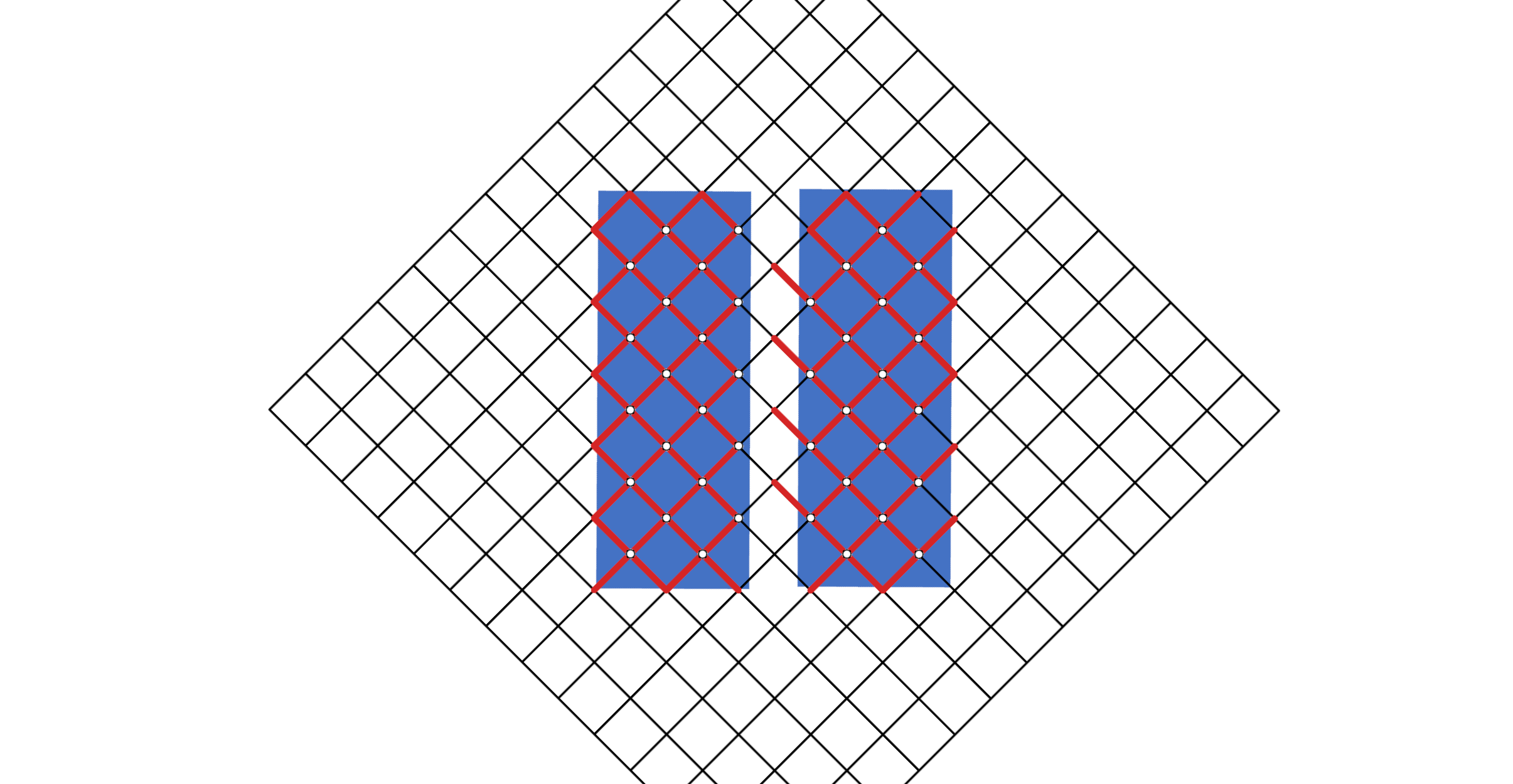}}
\hspace{8pt}
\subfloat[Anti-aliased]{
	\includegraphics[trim=36cm 11cm 33cm 10cm, width=0.46\columnwidth,clip]{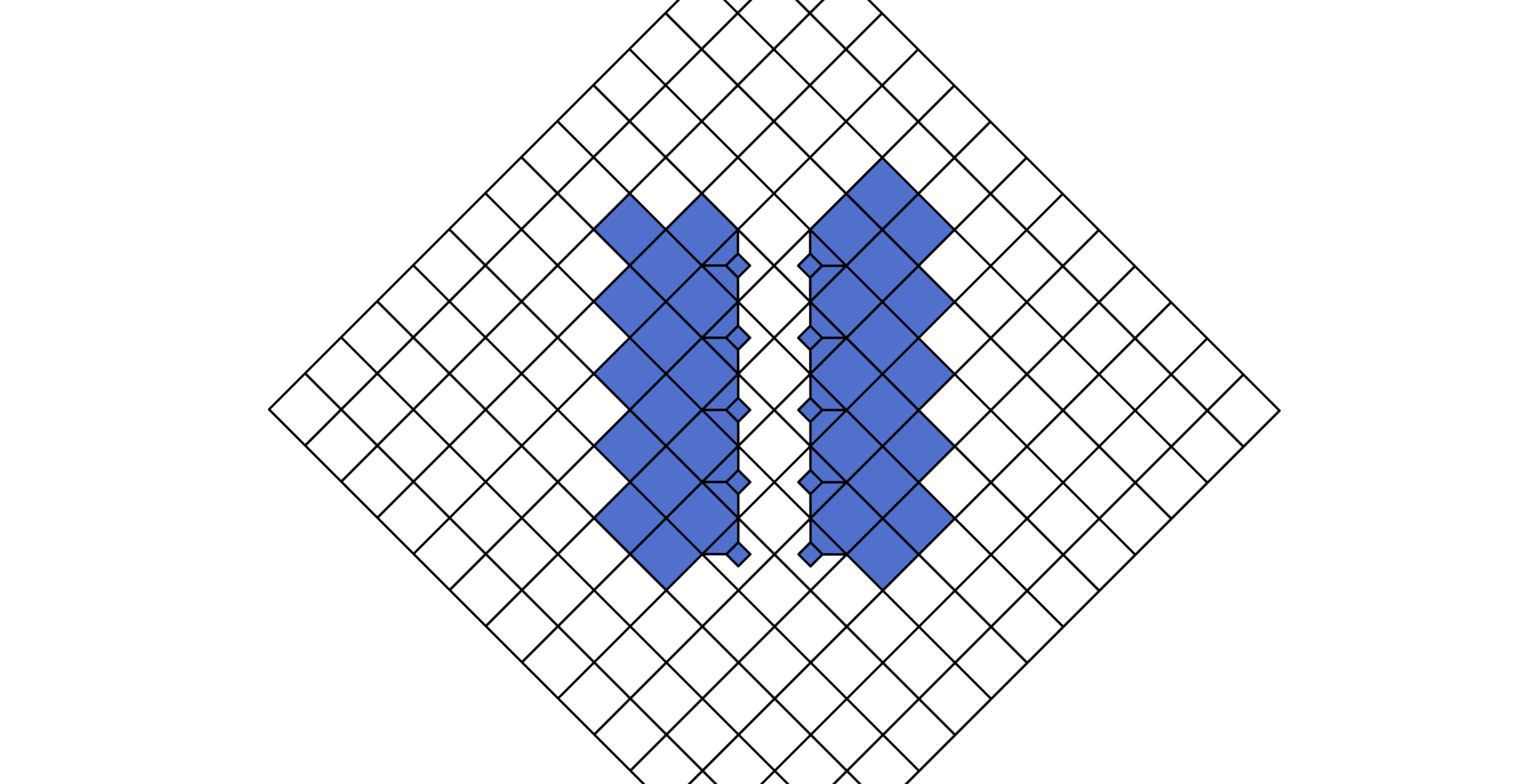}}
\caption{Subgrid sampling and anti-aliasing are performed on \cref{fig:diagonal}.}
\label{fig:diagonal_subcells}
\end{figure}

When separating pinches, the configuration in \cref{fig:case3} is the only exception to the rule of removing all the child cells that share the pinch edge. 
The one-to-seven split is performed on the hexahedra sharing the pinch edge, and all the child hexahedra that contain that edge are removed. 
This turns the central vertex into a pinch. To prevent that, one additional child hexahedron is removed. 
In the orientation of  \cref{fig:case3}, the removed hex is the rightmost child of the top hex, which contains the central vertex; see \cref{fig:case3_separated}.


For resolving archipelagos, all the connected components of the mesh are identified.
For any pair of connected components, if the edges that connect them are interior to the geometry, the components are joined using templates along those edges.
The remaining connected components that contain fewer than a user-defined number of highest-dimensional cells are removed.
This process is demonstrated in \cref{fig:connecting_edges}. 
In \cref{fig:diagonal_subcells,fig:jaggies_subcells} the utility of the anti-aliasing algorithm is demonstrated on the unaligned gap and sharp angle of \cref{fig:diagonal,fig:jaggies} respectively.


\paragraph{Transferring Persistent Parameters to Simplices.}
Having a framework by which topological anti-aliasing can be performed on the mesh by use of subgrid sampling and templates, we now turn our attention to ensuring that the topological anti-aliasing defined above is accurately represented in persistent homology calculations.
Because most open-source persistent homology software employs simplicial complexes, we first 
transform the cubical filtration into a topologically-equivalent simplicial filtration.
In what follows, we prioritize consistent pinch resolution, over archipelago resolution.
As a result, we make the assumption in 2D that an edge shared by two interior quadrilaterals will also be interior, and that an edge shared by two interior vertices must also be interior.
Similarly, in 3D, a face shared by two interior hexahedra will be interior, as will a face bounded by four interior edges and vertices.
We first focus on the 2D framework, followed by 3D.

In 2D, a primal vertex induces a dual 2-cell, and a primal edge induces a dual edge, and a primal quad induces a dual vertex.
Each dual vertex is assigned the filtration value of its corresponding primal face's volume fraction.
The dual mesh is then further subdivided into a simplicial mesh.
To create the simplicial mesh, an additional simplicial vertex is introduced at the centroid of each dual 2-cell. 
Simplices are then formed as the join of each dual face's edge with the simplicial centroid vertex.
This simplicial vertex corresponds to a vertex on the primal mesh, and takes the filtration value of the corresponding primal vertex's volume fraction.
However, to preclude the introduction of spurious topological artifacts (and consistent with previous computations), we also require that this volume fraction be between the maximal and minimal volume fraction of the surrounding four vertices.
The filtration value of this simplicial vertex then informs whether two primal faces connected with a pinch should be topologically separated or connected.
Having thus defined filtration values for all vertices of the induced simplicial complex, we then use a Vietoris–Rips complex to calculate persistent homology, meaning that 
at persistence value $k \in \mathbb{R}$, all vertices with persistence parameter value greater than or equal to $k$ are added to the filtration, then all edges between already-added vertices, then all triangles formed by already-added edges.

\begin{figure}[!tb]
	\centering
	\subfloat[Primal volume fractions]{
		\includegraphics[trim=0.6cm 2.8cm 22.8cm 1.4cm, width=0.4\columnwidth,clip]{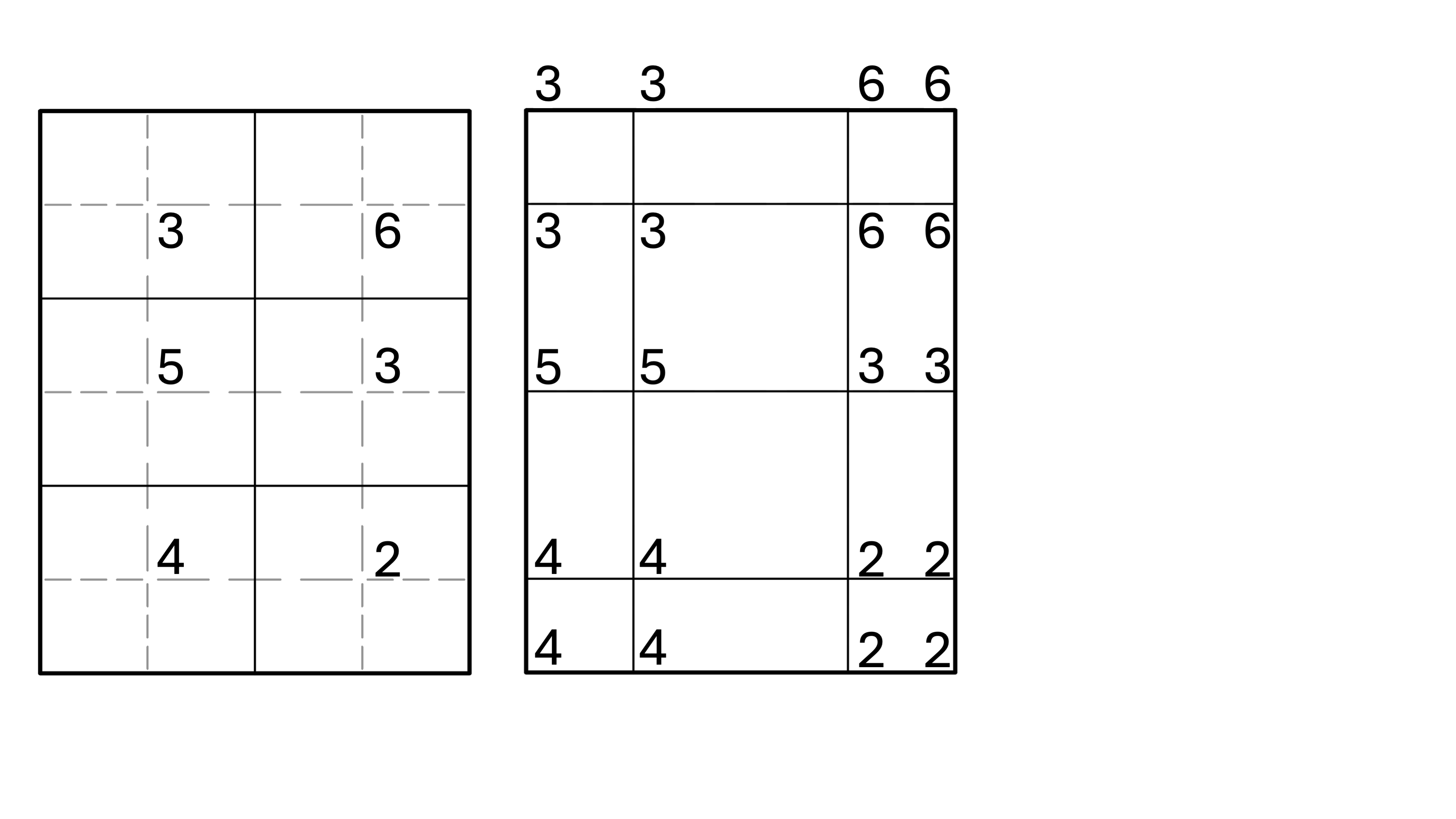}\label{fig:primalVF}
	}
	\;\;
	\subfloat[Dual volume fractions]{
		\includegraphics[trim=12cm 2.8cm 11.4cm 1.4cm, width=0.4\columnwidth,clip]{subdivision_to_triangulation_setup.png}\label{fig:dualVF}
	}
	\;\;
	\subfloat[Connected simplices]{
		\includegraphics[trim=0.6cm 2.2cm 22.8cm 1.4cm, width=0.4\columnwidth,clip]{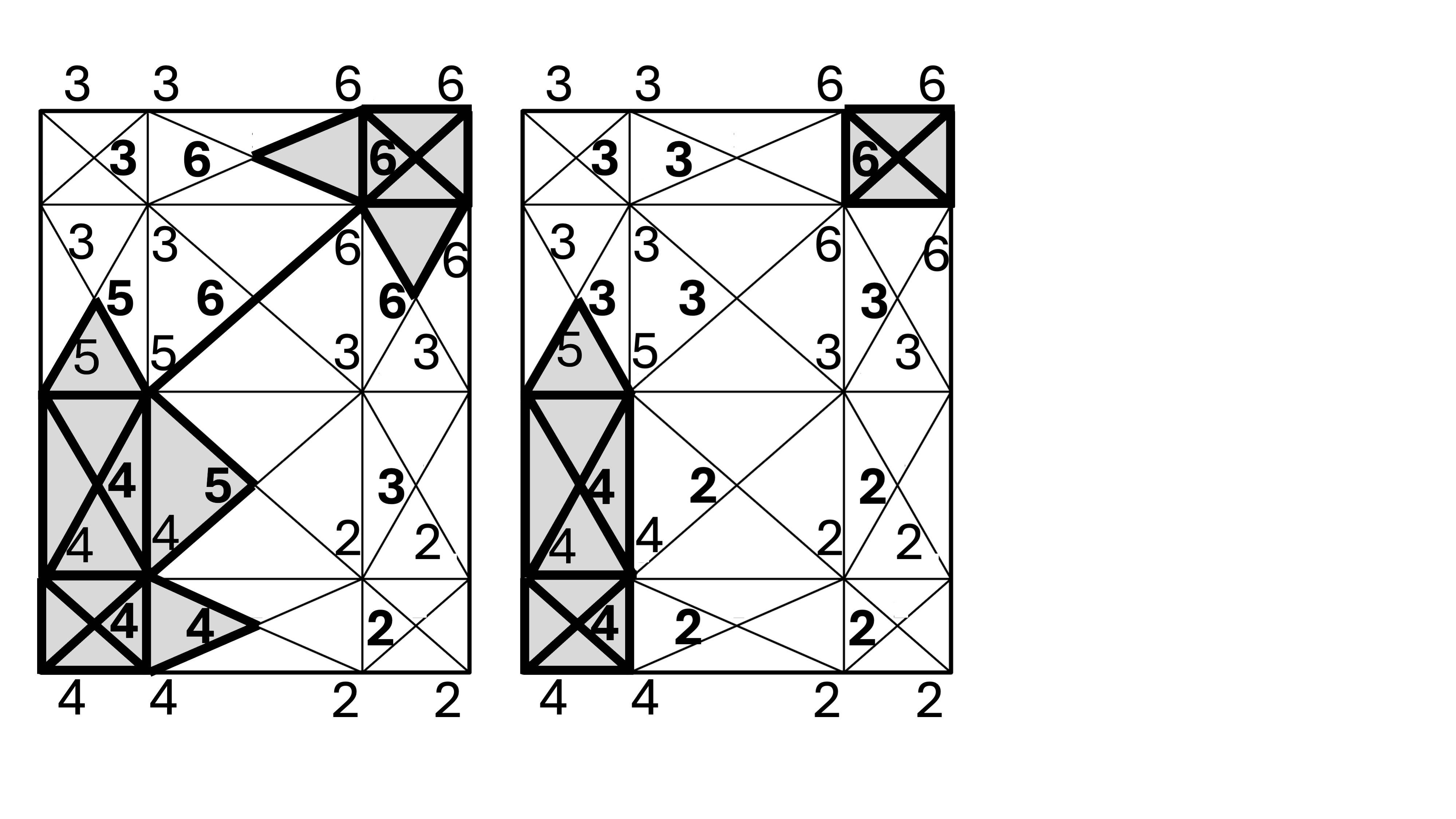}
		\label{fig:connected_simplicies}
	}
	\;\;
	\subfloat[Disconnected simplices]{
		\includegraphics[trim=12cm 2.2cm 11.4cm 1.4cm, width=0.4\columnwidth,clip]{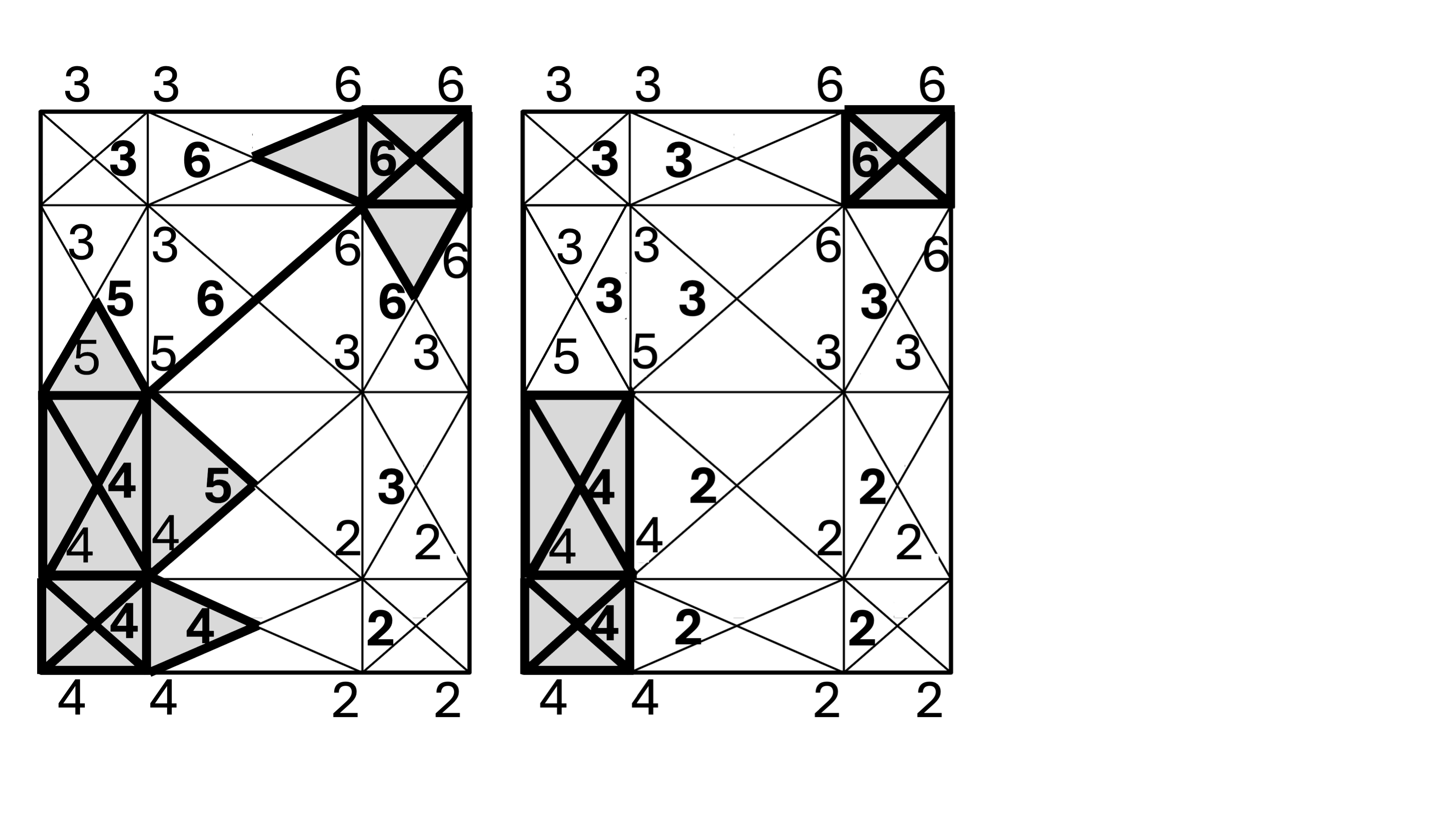}
		\label{fig:separated_simplicies}
	}
\caption{An example of converting 2D cell volume fractions into a filtration of a simplicial complex. 
	Numbers are proportional to volume fraction.
	\protect{\Cref{fig:primalVF}} shows the ordering of grid cells from high to low. 
	\protect{\Cref{fig:dualVF}} shows the dual grid with vertex values transferred from grid cells.
	In \cref{fig:connected_simplicies,fig:separated_simplicies} the dual complex is subdivided into a simplicial complex
	and the threshold is set so that all cells with value 4 or more are added.
	The choice of volume fraction for the introduced vertices (in bold) determines the connectivity near pinches. 
	}\label{fig:subdivision}
\end{figure}

\begin{figure}[!h]
	\centering
		\includegraphics[trim=5cm 0cm 8cm 0cm, width=0.3\columnwidth,clip]{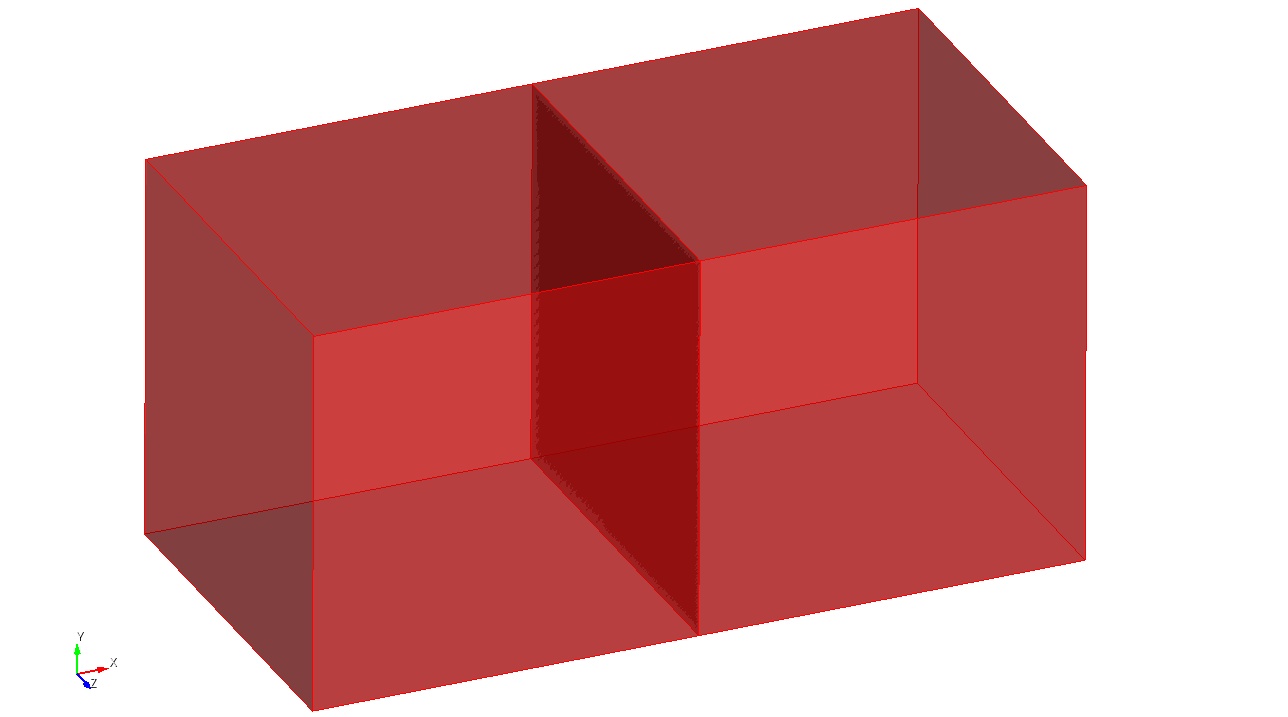}
		\hfill
		\includegraphics[trim=5cm 0cm 8cm 0cm, width=0.3\columnwidth,clip]{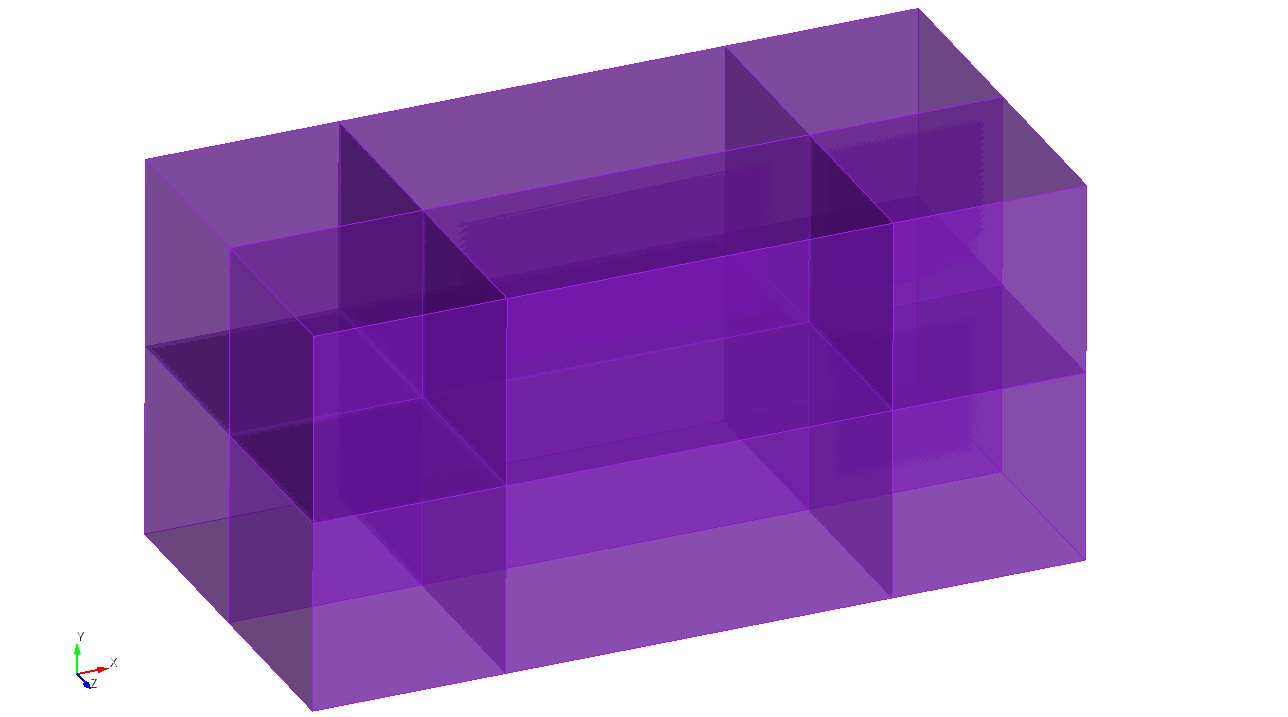}
		\hfill
		\includegraphics[trim=5cm 0cm 8cm 0cm, width=0.3\columnwidth,clip]{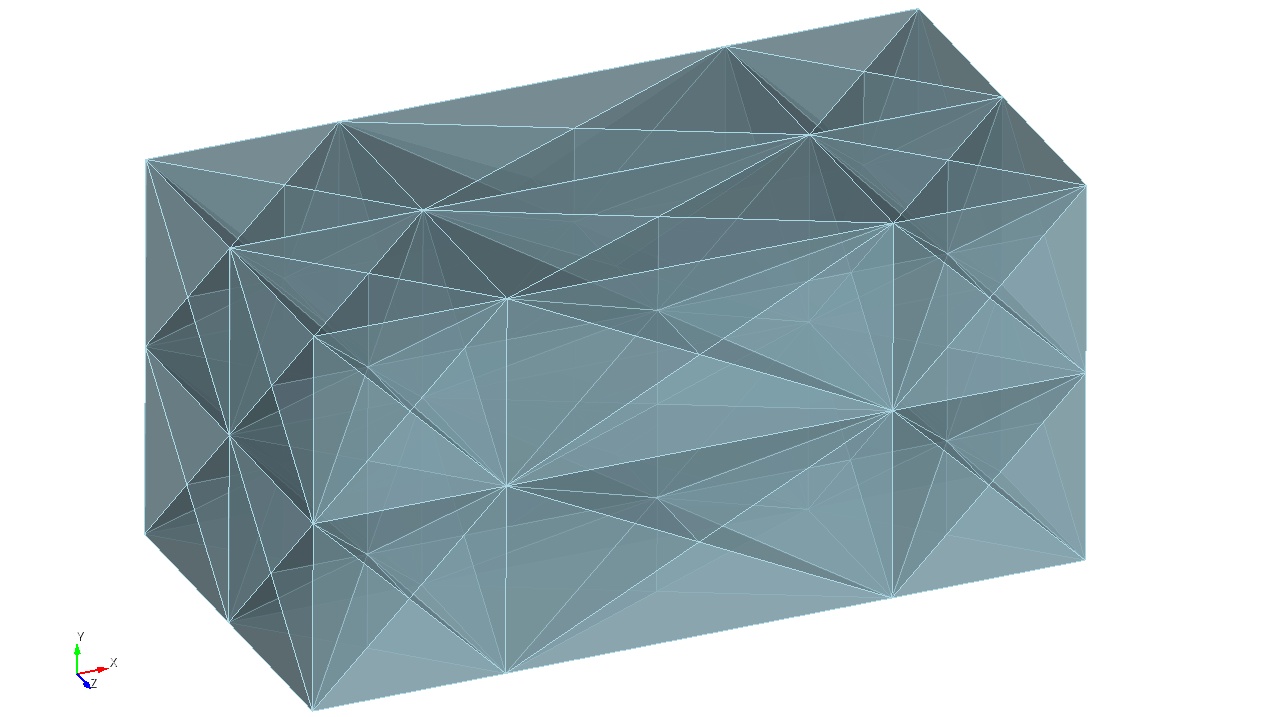}
	\caption{The conversion process taking two adjacent 3D hexes into a tetrahedral simplicial complex with an equivalent filtration is shown.}
\label{fig:3d_subdivision}
\end{figure}

In 3D, a primal vertex induces a dual volume cell, and a primal edge induces a dual face, a primal face induces a dual edge, and a primal volume induces a dual vertex.
As in 2D, each dual vertex is assigned the filtration value of its primal volume.
To generate a simplicial mesh from the dual mesh, each dual face is subdivided into four triangles with a new vertex introduced, as in 2D.
Here, the additional vertex corresponds to a primal edge and will take the filtration value of the volume fraction of this primal edge, subject to constraints keeping the filtration value between the maximal and minimal values of the surrounding four simplices of the dual face.
Each dual volume is then subdivided into 24 tetrahedra by introducing a single simplicial vertex at the centroid of the dual volume and taking the join of this vertex with each of the 24 triangles defined on the (subdivided) faces of the dual volume.
Again, this new simplicial vertex will correspond to a vertex on the primal mesh, and consequently takes a filtration value of the volume fraction of this primal mesh vertex (again subject to the constraint that the filtration value must be between the maximal and minimal values of the 26 surrounding vertices).
The filtration values of the simplicial vertices corresponding to primal vertices and primal edges is then locally consistent with the procedures performed resolve pinch points, and will have identical persistent homology.
This conversion process, from a cubical primal cell complex into a simplicial one with an identical filtration is illustrated in \cref{fig:subdivision,fig:3d_subdivision}.

For the sake of completeness, we also note that similar primal-to-dual-to-simplicial operations could be performed on unstructured background meshes.
In the 2D case, each dual cell of maximal dimension and with $k$ sides would be subdivided into $k$ triangles.
In 3D, each dual cell of maximal dimension and with $\ell$ faces,  with the $i$th face having $k_i$ sides, would subdivide into $\sum_{i=1}^\ell k_i$ tetrahedra.

Finally, we note that for a truly general framework, a vertex in the above-defined simplicial complexes would need to be defined for each cell in the primal complex.
Particularly, in 2D we currently introduce vertices corresponding to primal faces and vertices, but not for primal edges.
This would require splitting every dual face into 8 simplices, rather than 4.
In 3D, we introduce vertices for all cells except for primal faces, and would require splitting every dual volume into 48 tetrahedra, rather than 24.
Given the challenges of navigating this topological space in a meaningful way (as well as the additional computational expense incurred by such a navigation), we leave this for future work.

%

\begin{figure}[!tb]
\centering
\subfloat[\hspace{-2pt}Bearings]{\includegraphics[trim= 14cm 5cm 14cm 5cm, width=0.3\columnwidth,clip]{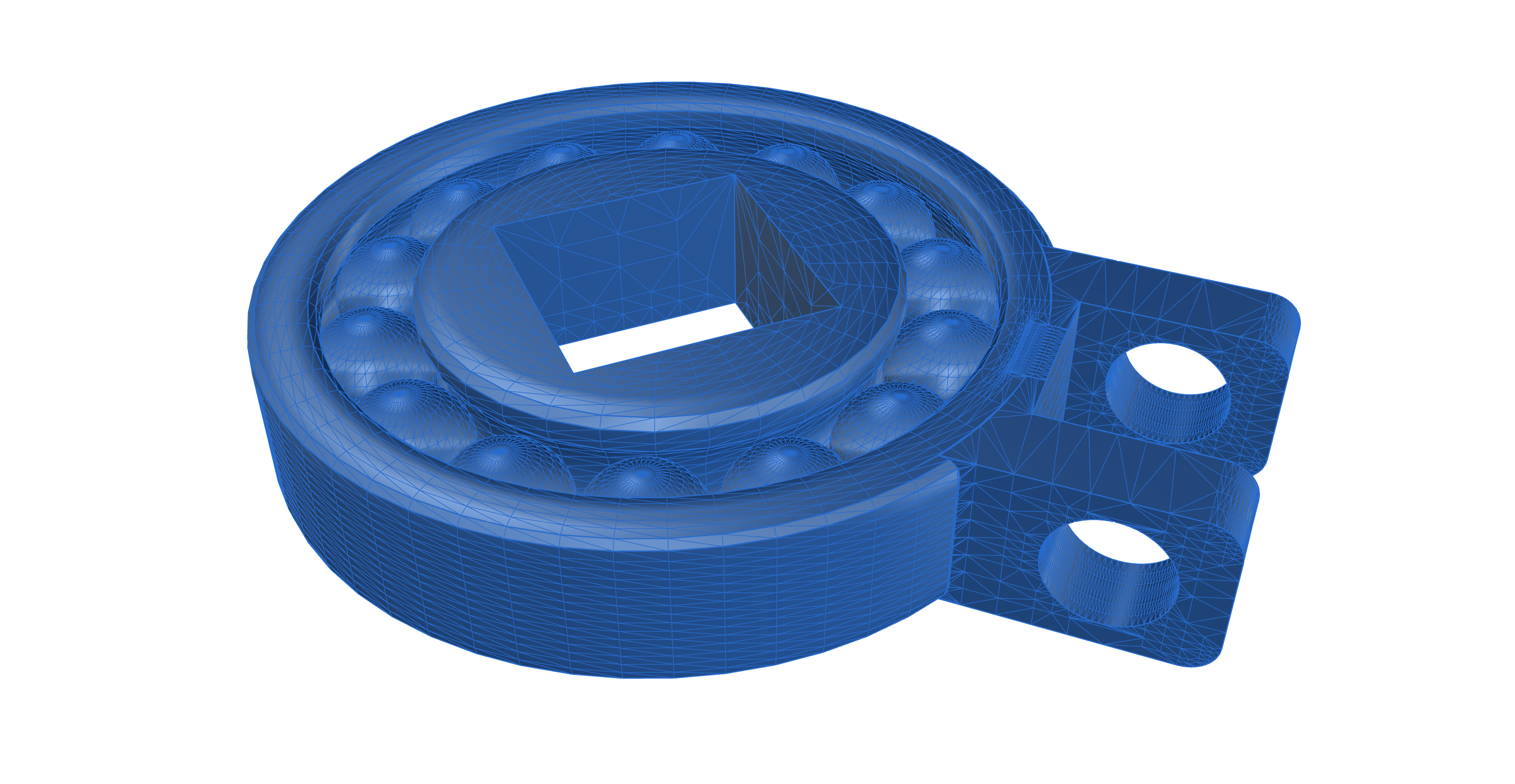}}\hfill
\subfloat[\hspace{-2pt}Vol.\hspace{-2pt} Frac.:\hspace{-2pt} 12.5\%]{\includegraphics[trim= 14cm 5cm 14cm 5cm, width=0.3\columnwidth,clip]{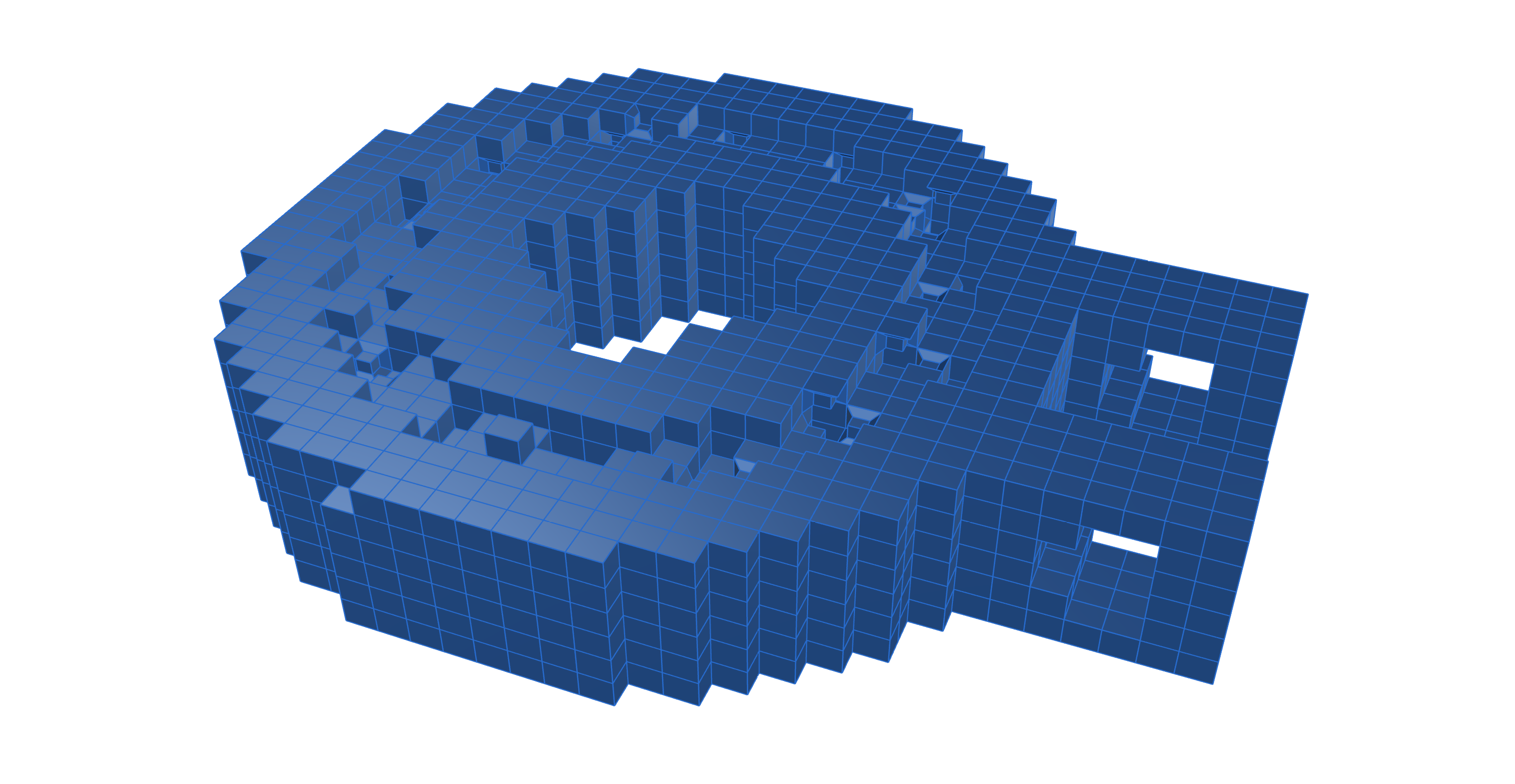}}\hfill
\subfloat[\hspace{-2pt}Vol.\hspace{-2pt} Frac.:\hspace{-2pt} 25\%]{\includegraphics[trim= 14cm 5cm 14cm 5cm, width=0.3\columnwidth,clip]{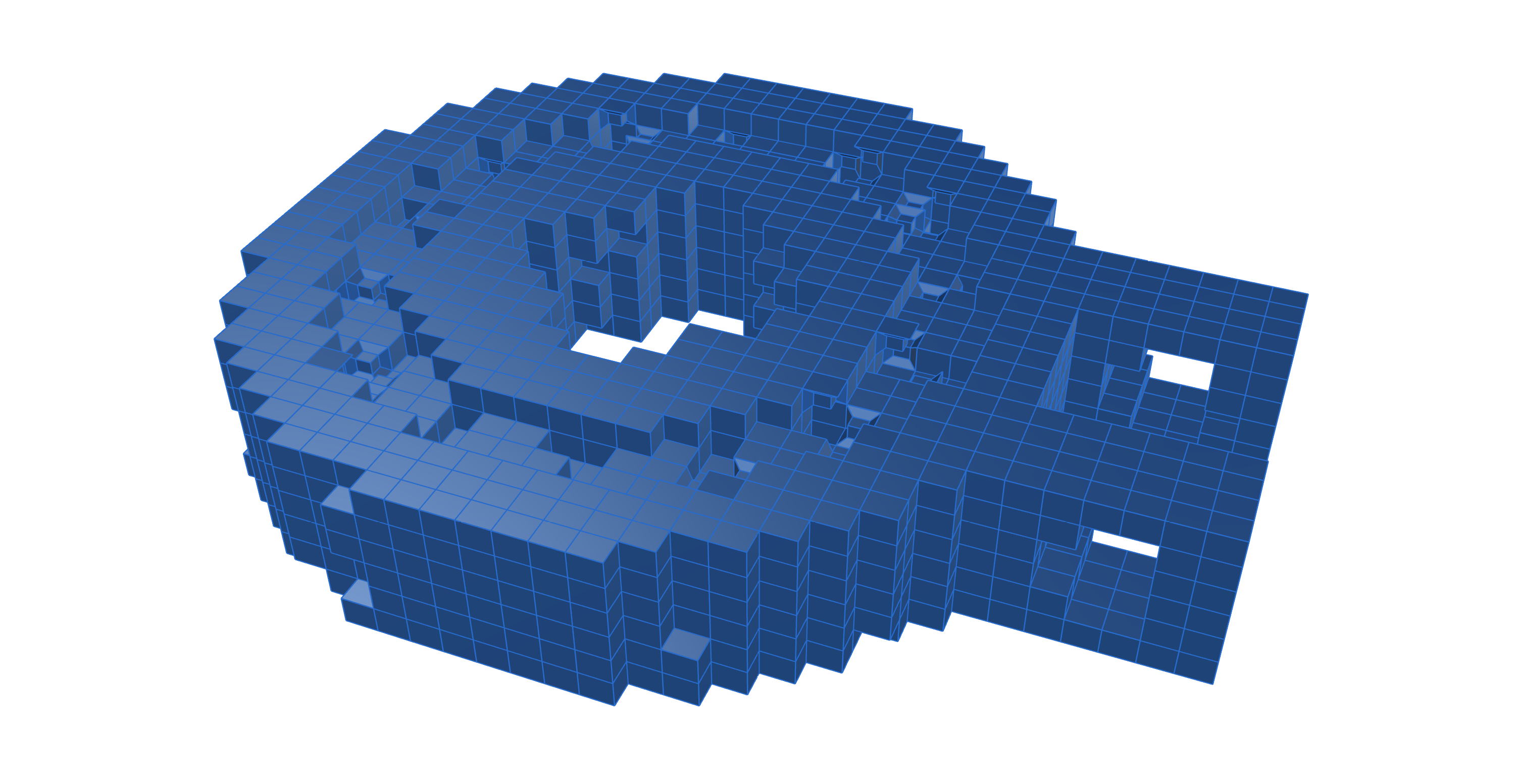}}
\\
\subfloat[\hspace{-2pt}Vol.\hspace{-2pt} Frac.:\hspace{-2pt} 37.5\%]{\includegraphics[trim= 14cm 5cm 14cm 5cm, width=0.3\columnwidth,clip]{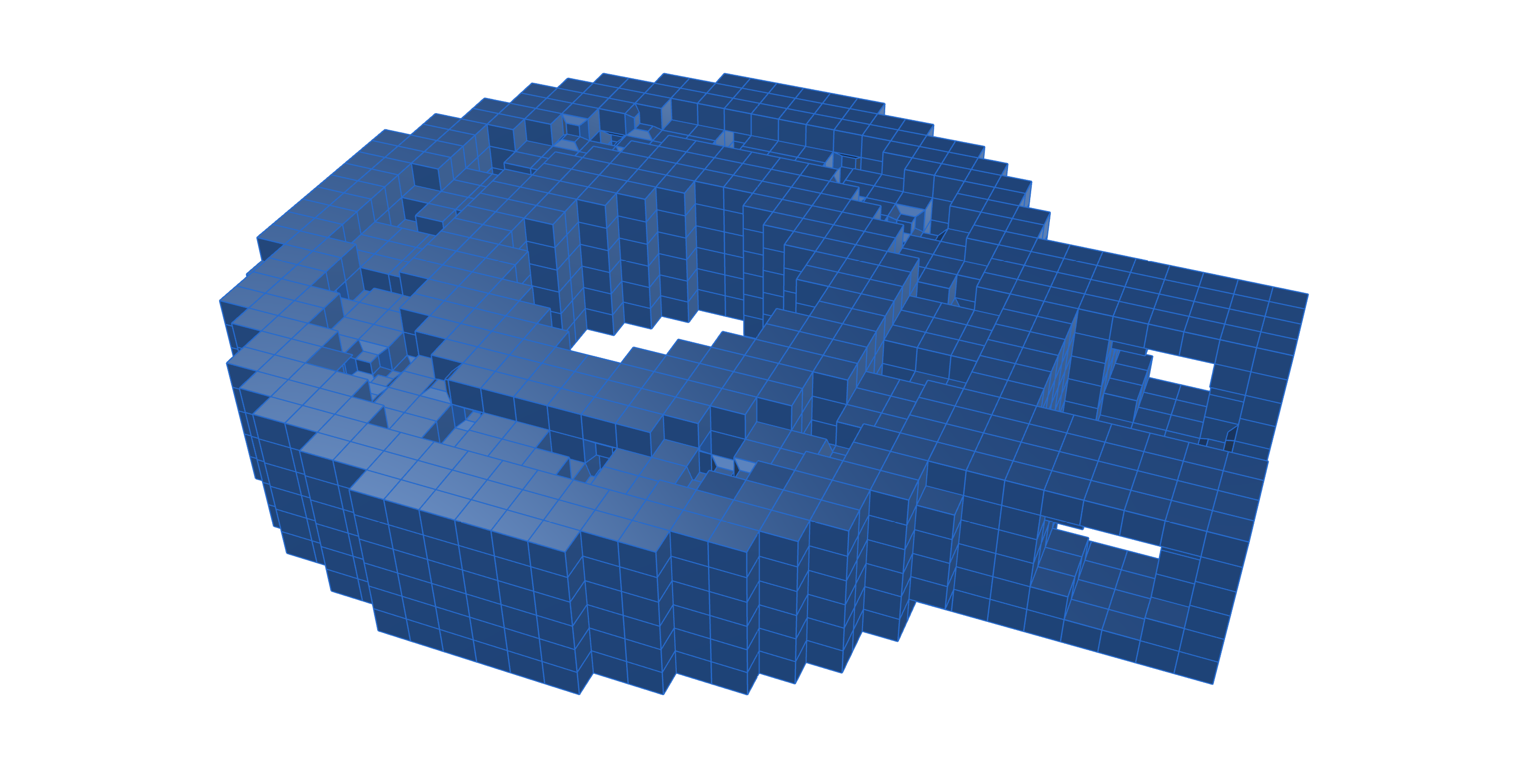}}\hfill
\subfloat[\hspace{-2pt}Vol.\hspace{-2pt} Frac.:\hspace{-2pt} 50\%]{\includegraphics[trim= 14cm 5cm 14cm 5cm, width=0.3\columnwidth,clip]{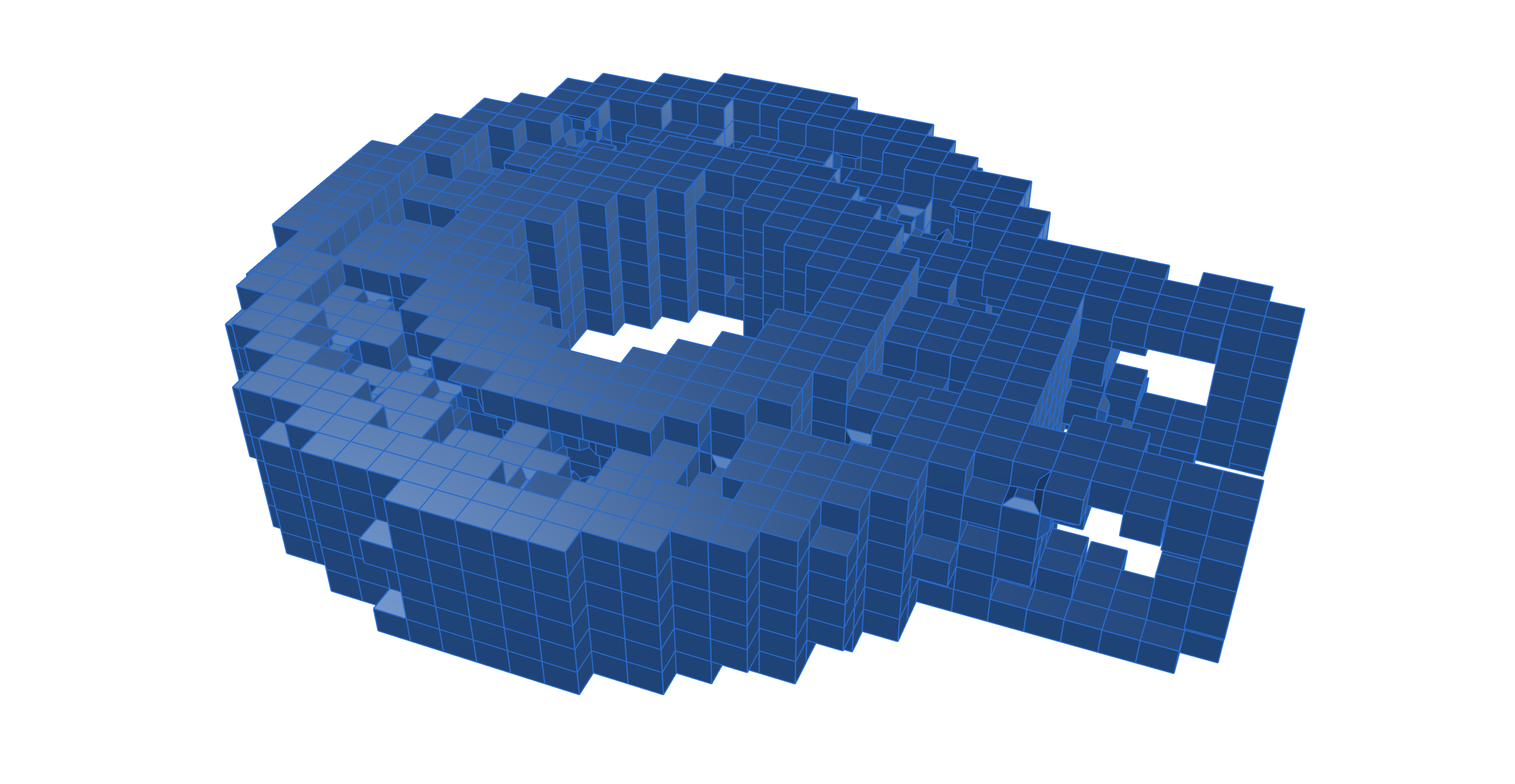}}\hfill
\subfloat[\hspace{-2pt}Vol.\hspace{-2pt} Frac.:\hspace{-2pt} 62.5\%]{\includegraphics[trim= 14cm 5cm 14cm 5cm, width=0.3\columnwidth,clip]{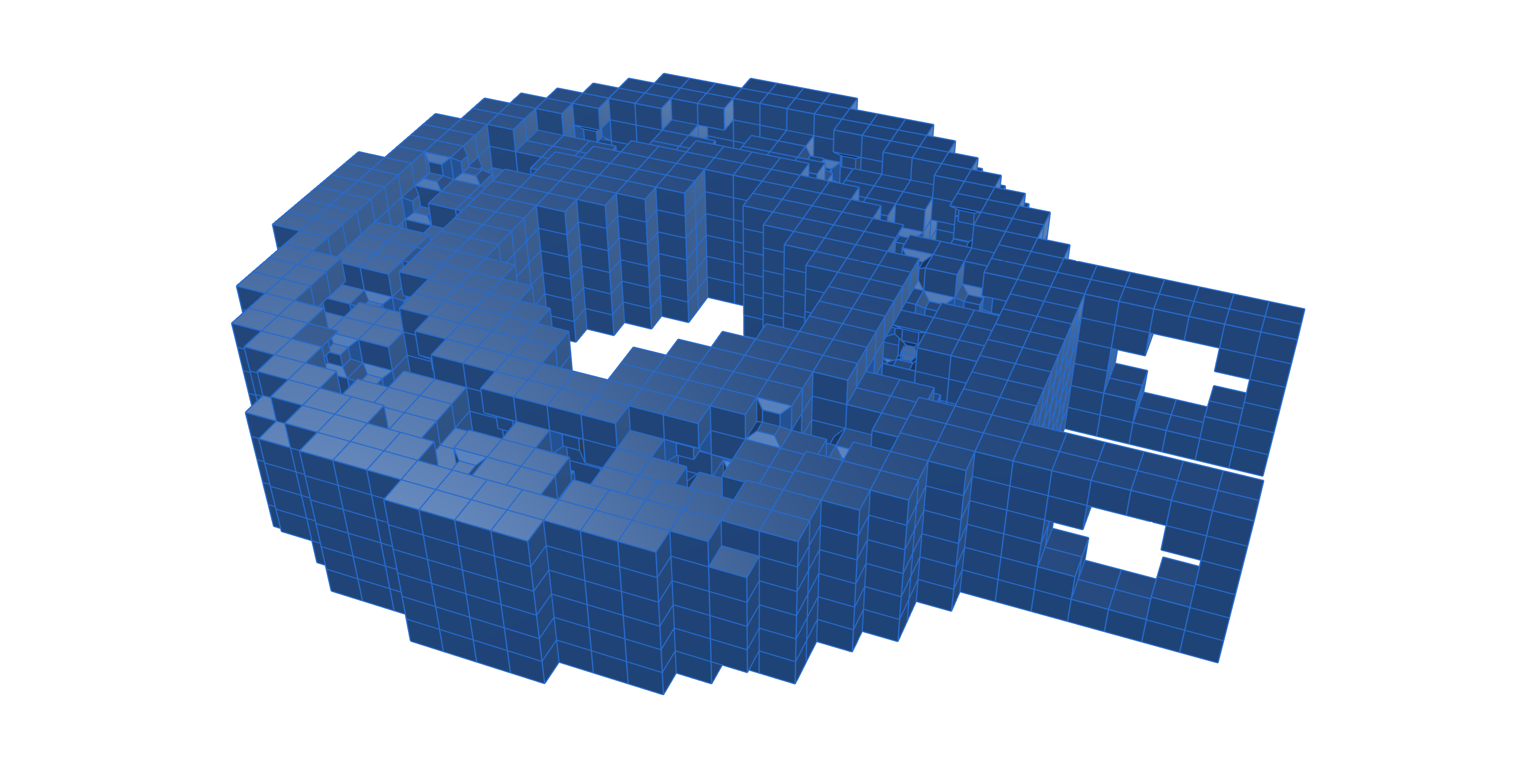}}
\\
\subfloat[\hspace{-2pt}Vol.\hspace{-2pt} Frac.:\hspace{-2pt} 75\%]{\includegraphics[trim= 14cm 5cm 14cm 5cm, width=0.3\columnwidth,clip]{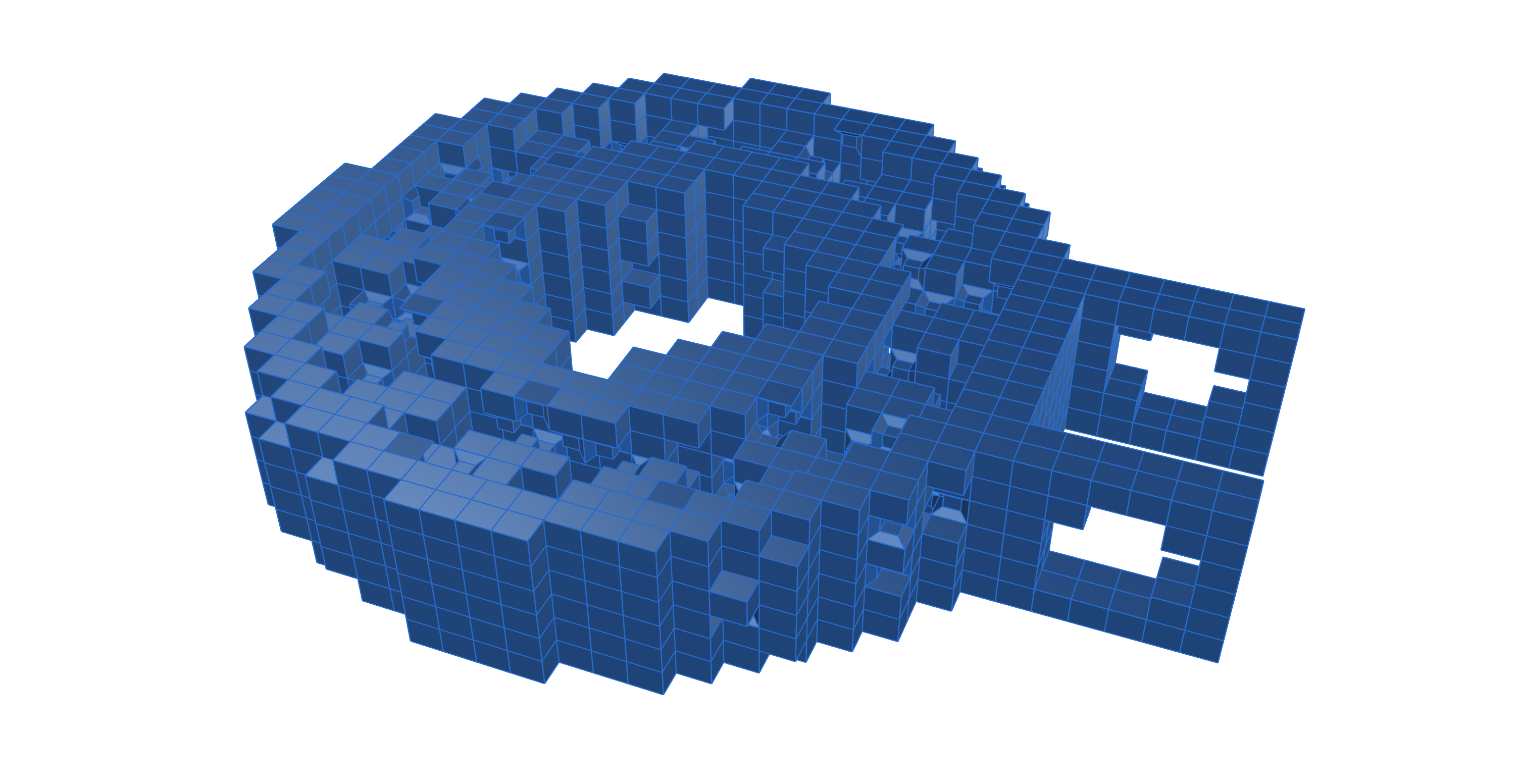}}\hfill
\subfloat[\hspace{-2pt}Vol.\hspace{-2pt} Frac.:\hspace{-2pt} 87.5\%]{\includegraphics[trim= 14cm 5cm 14cm 5cm, width=0.3\columnwidth,clip]{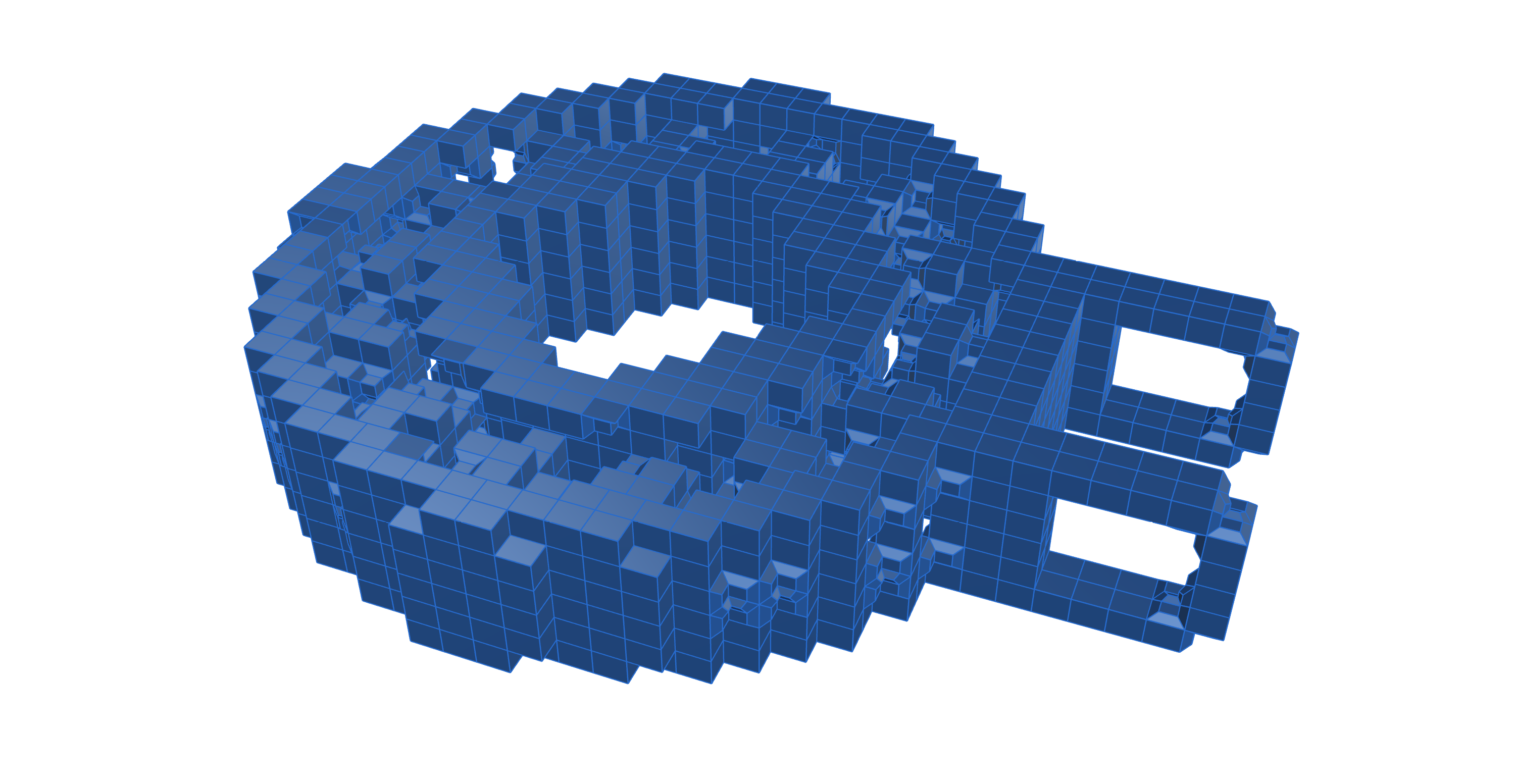}}\hfill
\subfloat[\hspace{-2pt}Vol.\hspace{-2pt} Frac.:\hspace{-2pt} 100\%]{\includegraphics[trim= 14cm 5cm 14cm 5cm, width=0.3\columnwidth,clip]{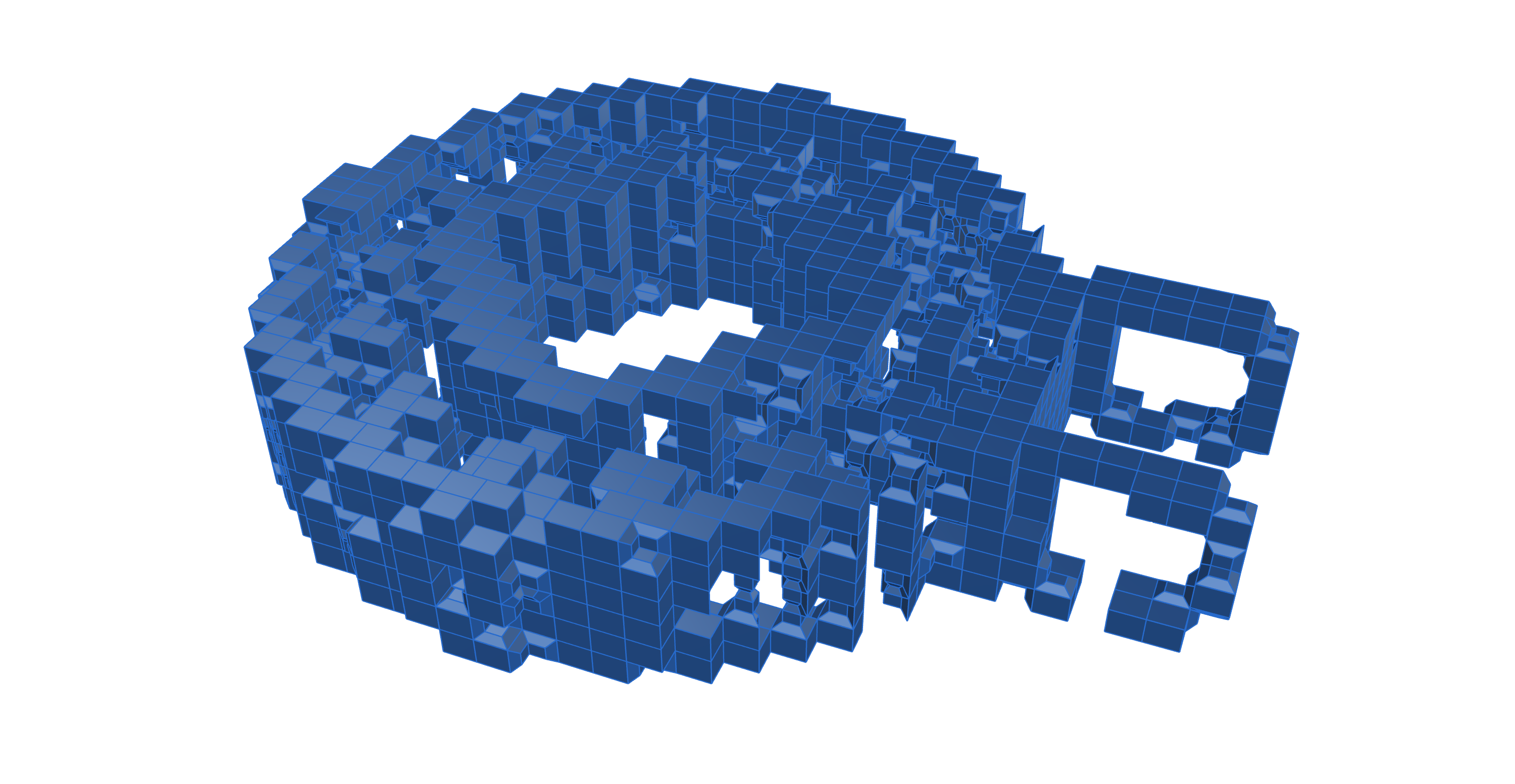}}
\caption{Meshes of bearings are shown.}
\label{fig:bearings}
\end{figure}

\begin{figure}[!tb]
\subfloat[\emph{\hspace{-2pt}The\hspace{-2pt} Bronco\hspace{-2pt} Buster}]
{\includegraphics[trim=29cm 0cm 29cm 0cm, width=0.33\columnwidth, clip]{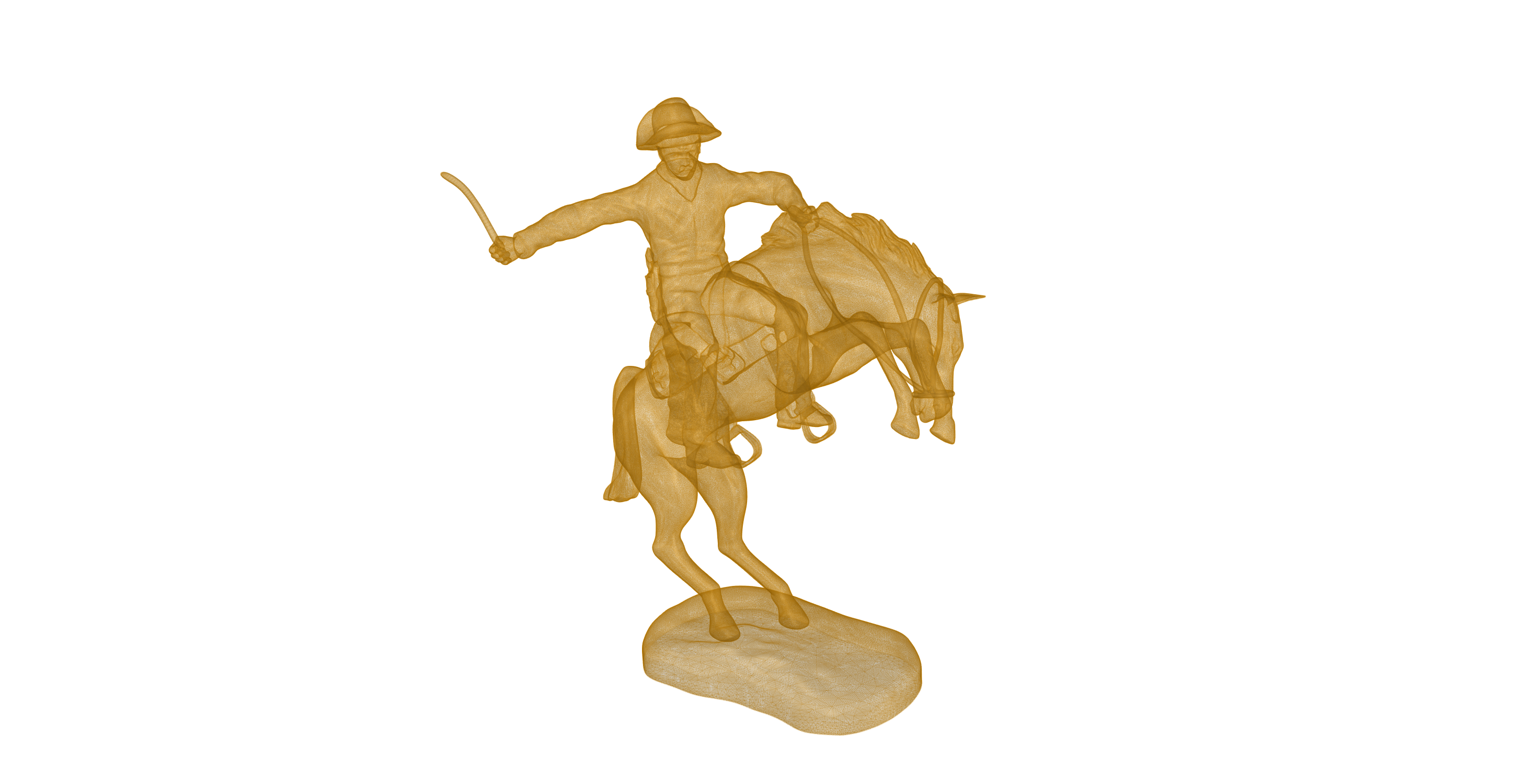}}\hspace{1pt}
\subfloat[\hspace{-2pt}Vol.\hspace{-2pt} Frac.:\hspace{-2pt} 12.5\%]{\includegraphics[trim=25cm 0cm 25cm 0cm, width=0.3\columnwidth, clip]{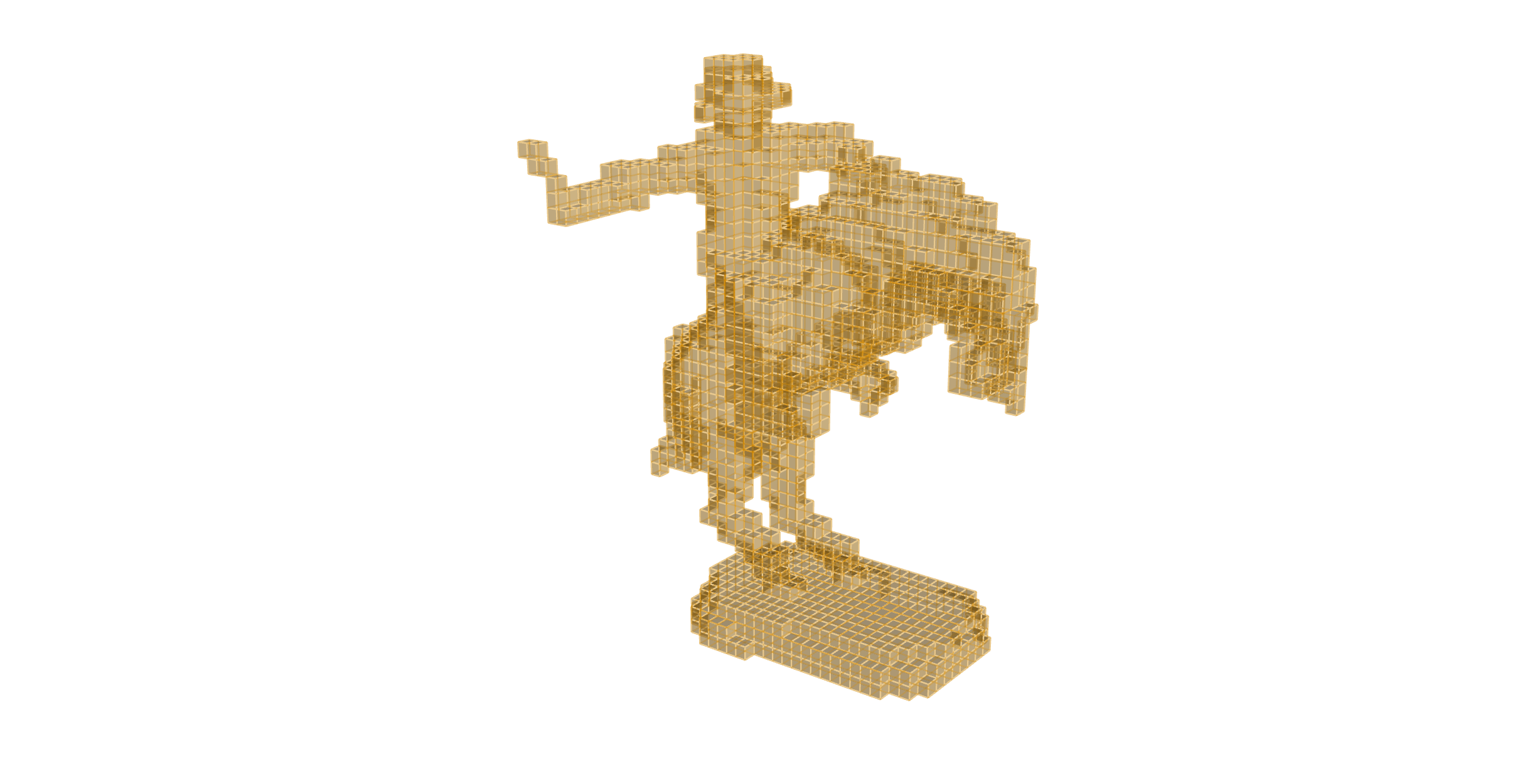}}\hfill
\subfloat[\hspace{-2pt}Vol.\hspace{-2pt} Frac.:\hspace{-2pt} 25\%]{\includegraphics[trim=25cm 0cm 25cm 0cm, width=0.3\columnwidth, clip]{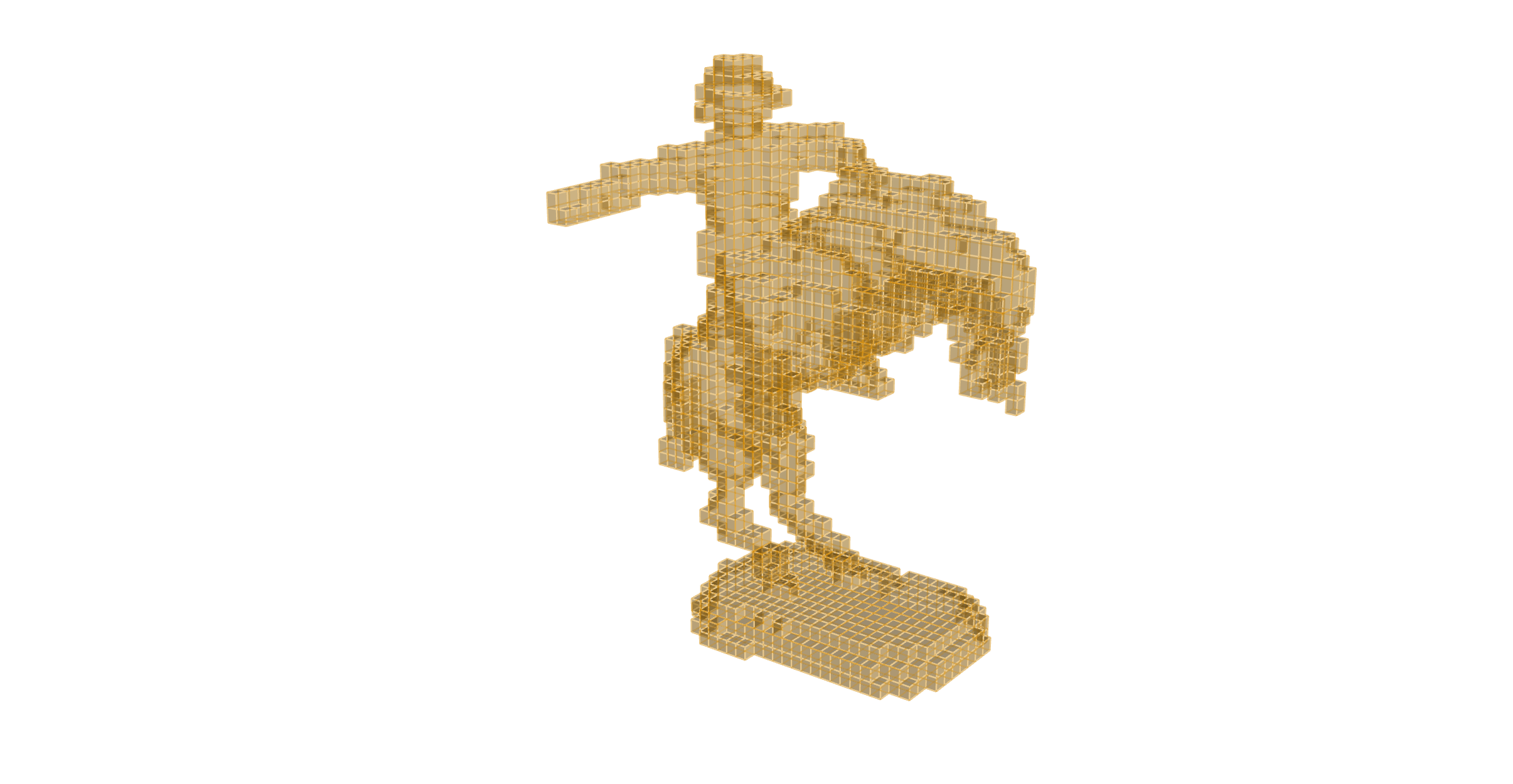}}
\\
\subfloat[\hspace{-2pt}Vol.\hspace{-2pt} Frac.:\hspace{-2pt} 37.5\%]{\includegraphics[trim=25cm 0cm 25cm 0cm, width=0.3\columnwidth, clip]{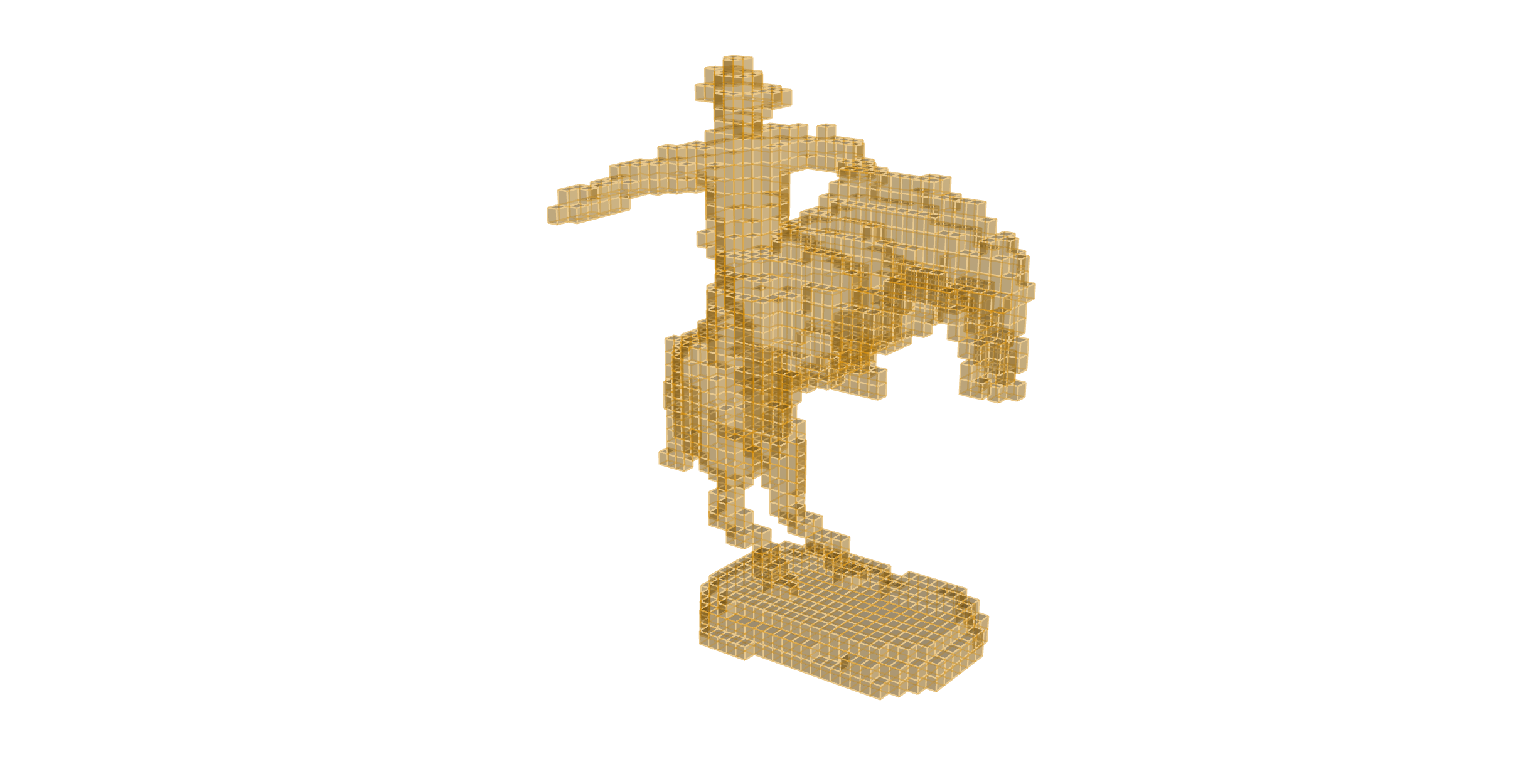}}\hfill
\subfloat[\hspace{-2pt}Vol.\hspace{-2pt} Frac.:\hspace{-2pt} 50\%]{\includegraphics[trim=25cm 0cm 25cm 0cm, width=0.3\columnwidth, clip]{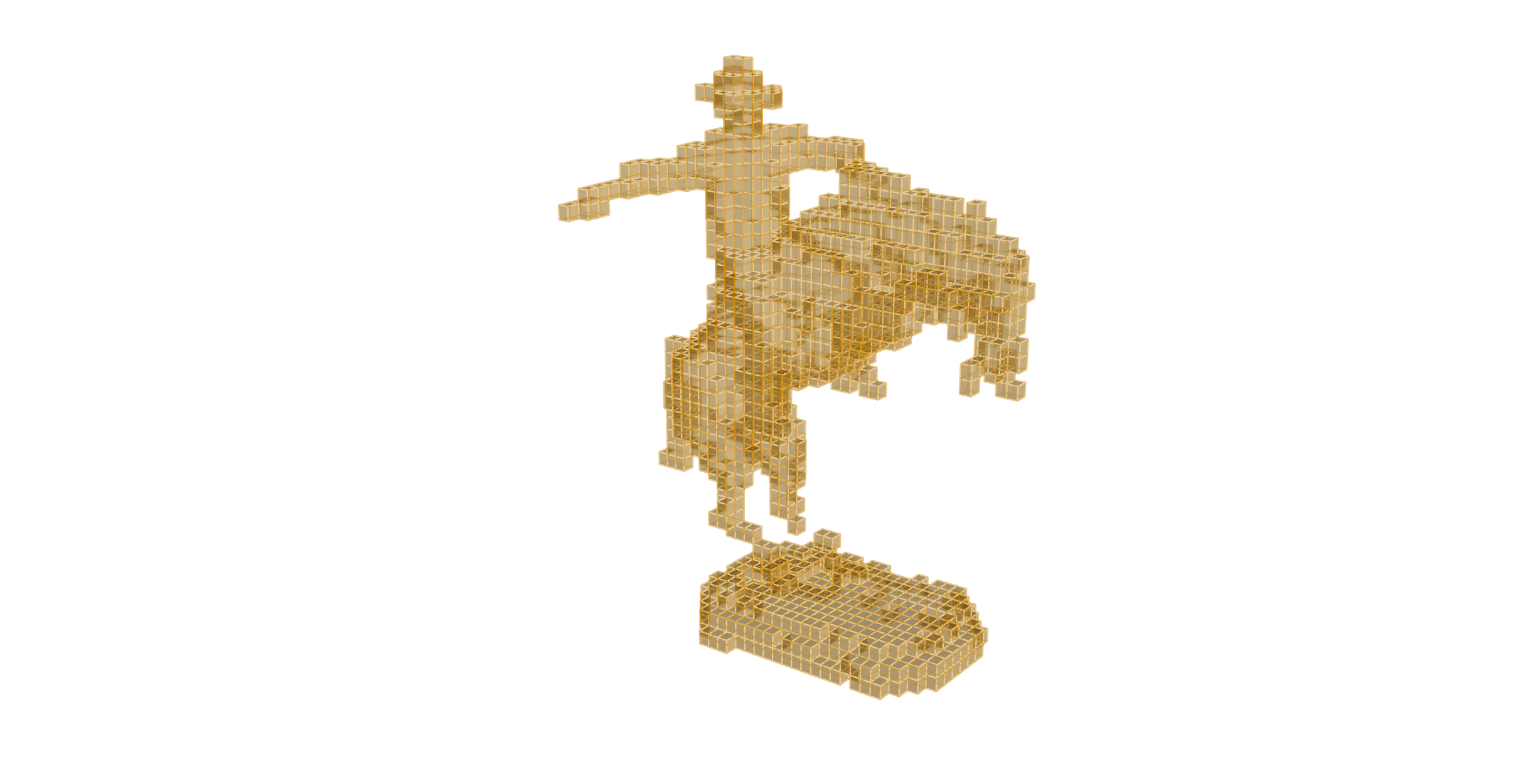}}\hfill
\subfloat[\hspace{-2pt}Vol.\hspace{-2pt} Frac.:\hspace{-2pt} 62.5\%]{\includegraphics[trim=25cm 0cm 25cm 0cm, width=0.3\columnwidth, clip]{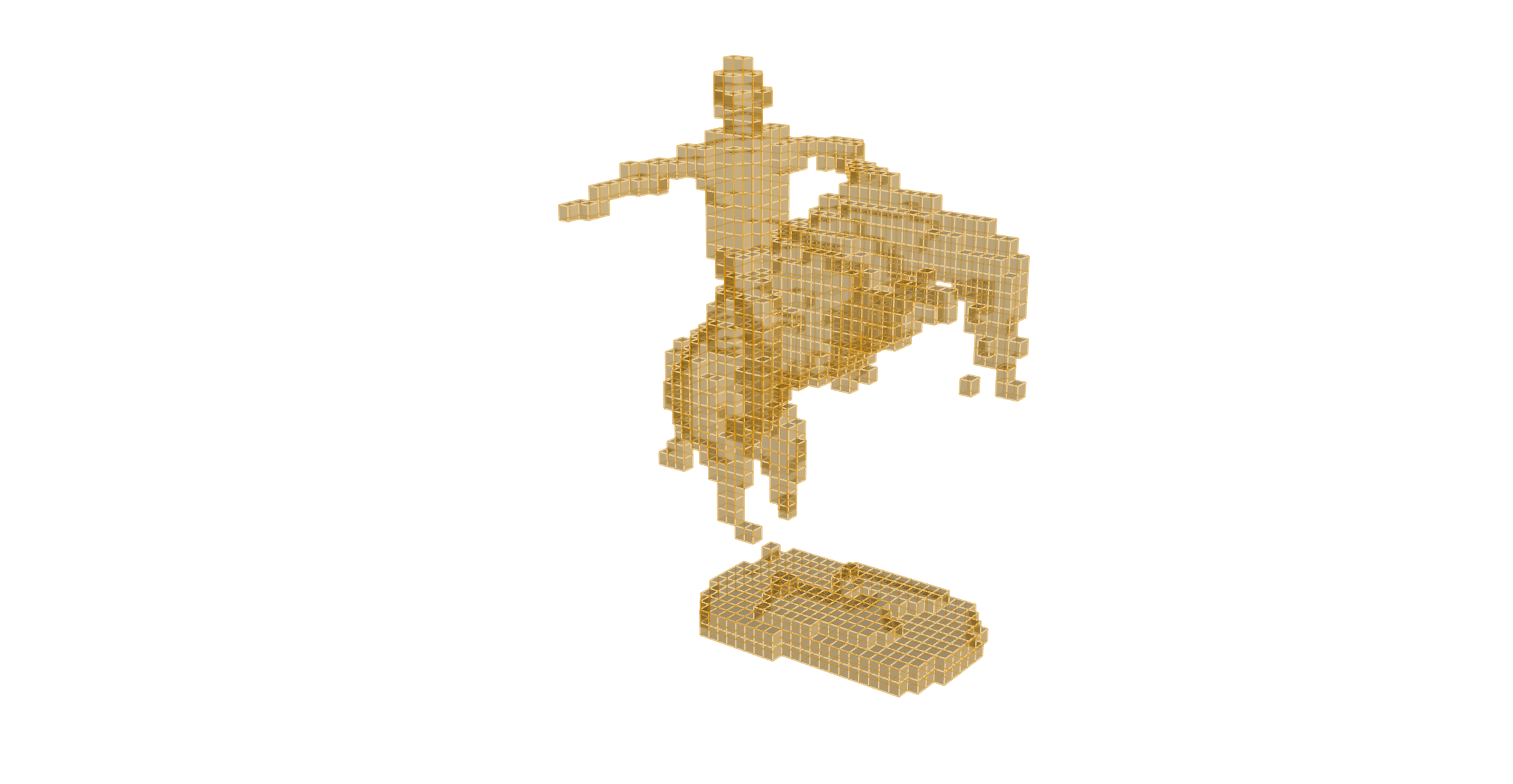}}
\\
\subfloat[\hspace{-2pt}Vol.\hspace{-2pt} Frac.:\hspace{-2pt} 75\%]{\includegraphics[trim=25cm 0cm 25cm 0cm, width=0.3\columnwidth, clip]{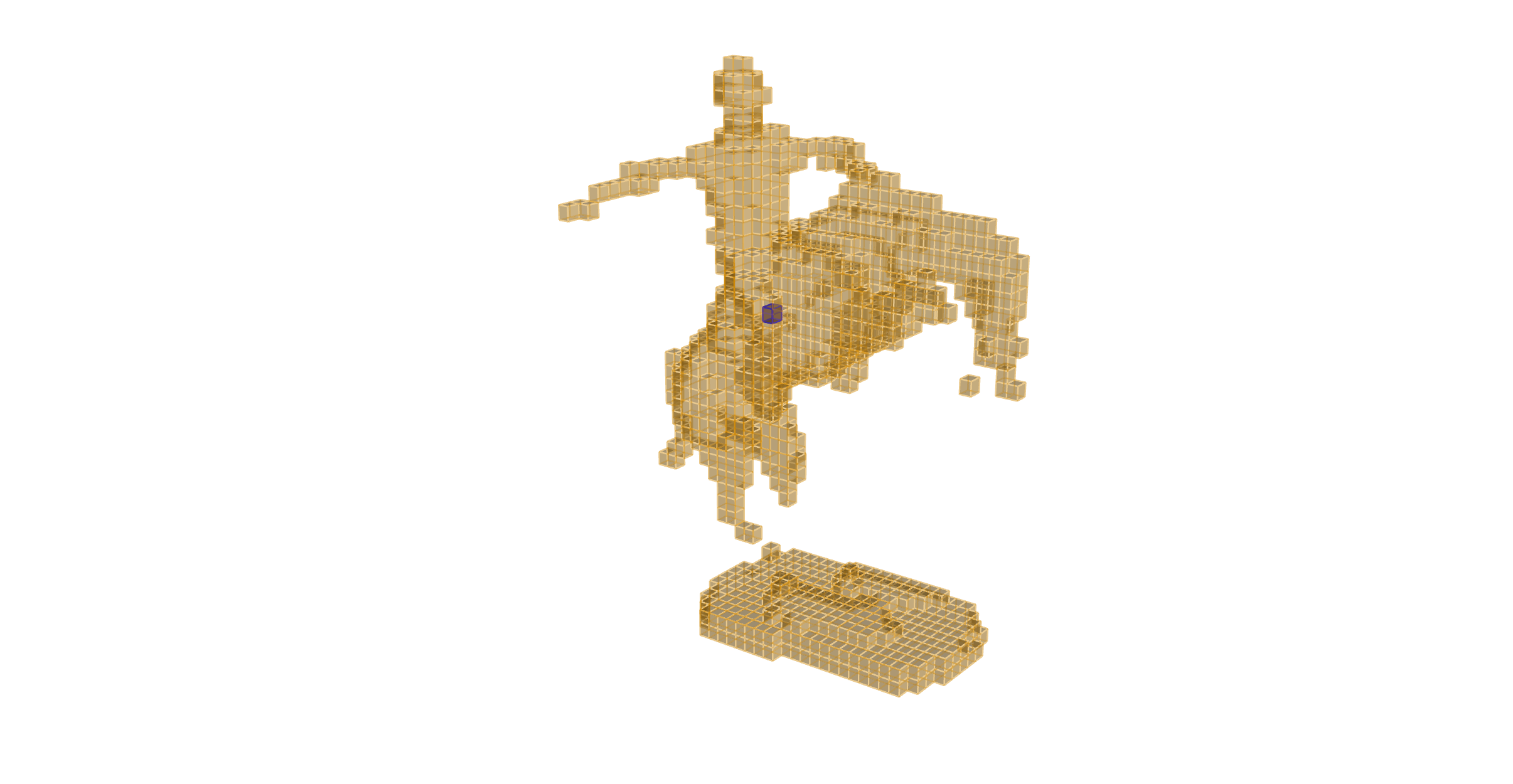}}\hfill
\subfloat[\hspace{-2pt}Vol.\hspace{-2pt} Frac.:\hspace{-2pt} 87.5\%]{\includegraphics[trim=25cm 0cm 25cm 0cm, width=0.3\columnwidth, clip]{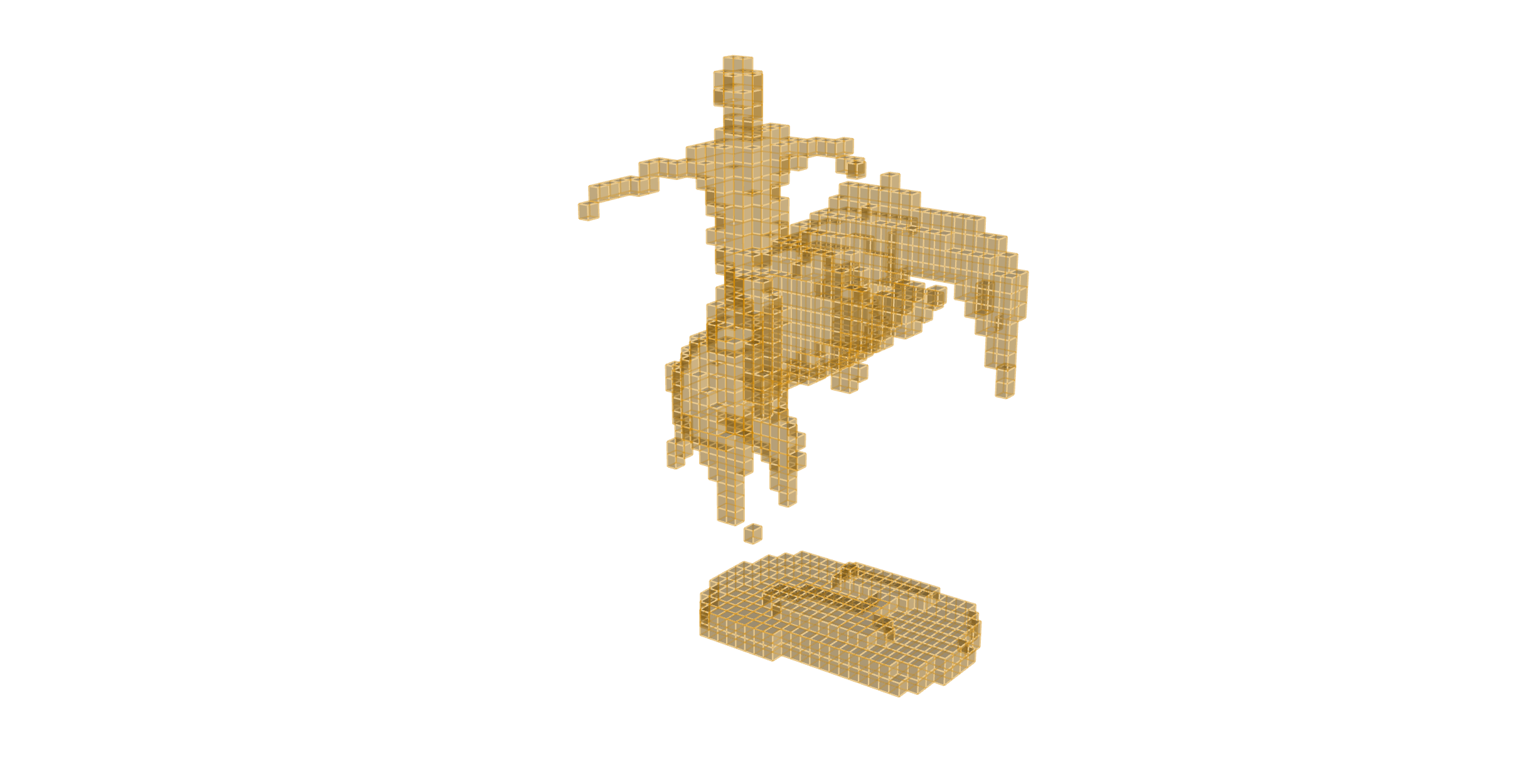}}\hfill
\subfloat[\hspace{-2pt}Vol.\hspace{-2pt} Frac.:\hspace{-2pt} 100\%]{\includegraphics[trim=25cm 0cm 25cm 0cm, width=0.3\columnwidth, clip]{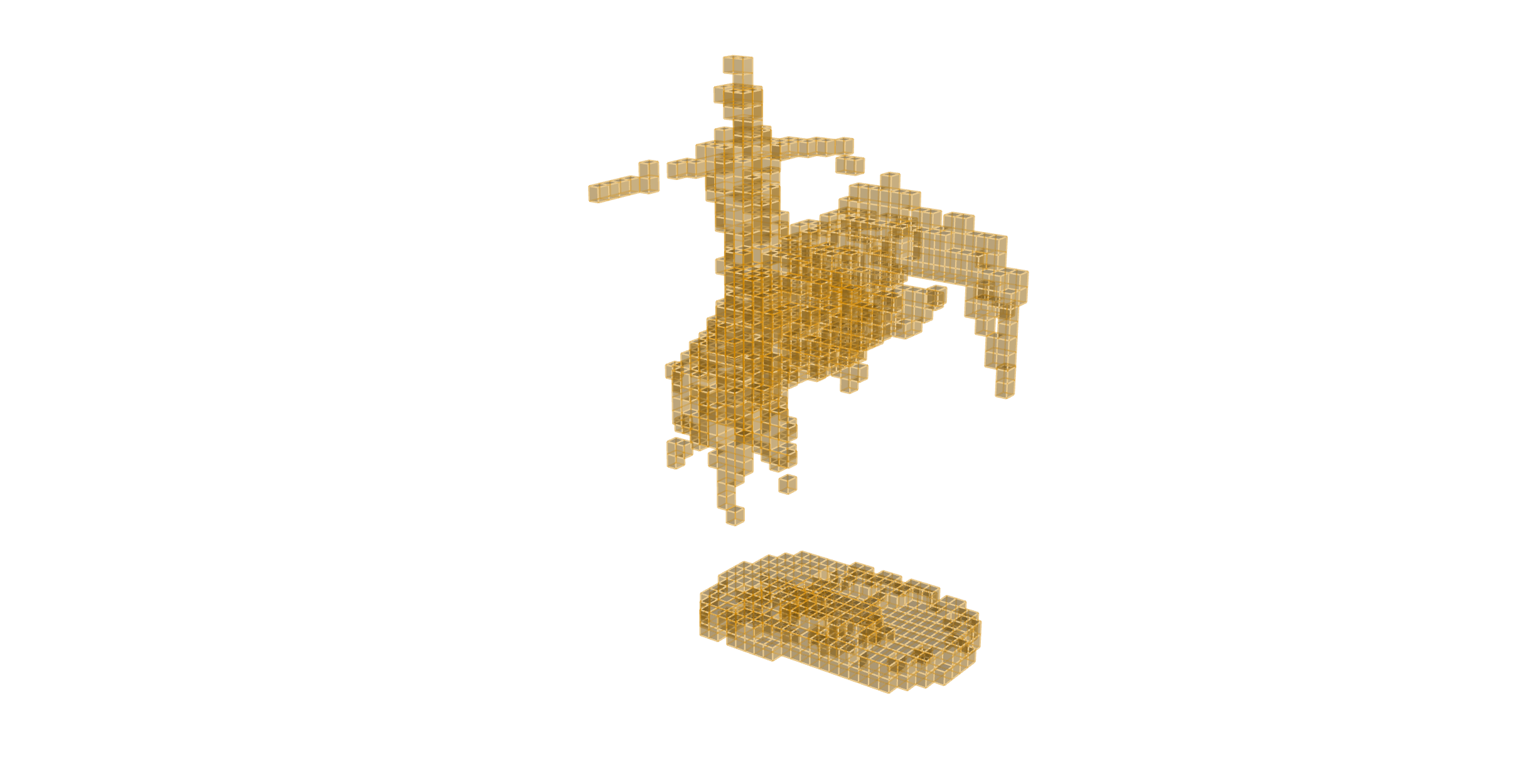}}
\caption{Meshes of \emph{The Bronco Buster} are shown.}
\label{fig:cowboy}
\end{figure}

\section{Results}


 \subsection{Computational Results.}


%

\begin{figure}[!htb]
\centering
\subfloat[Bearings]{
	\includegraphics[width=0.47\columnwidth,trim = 2.5cm 0cm 3cm 1cm, clip]{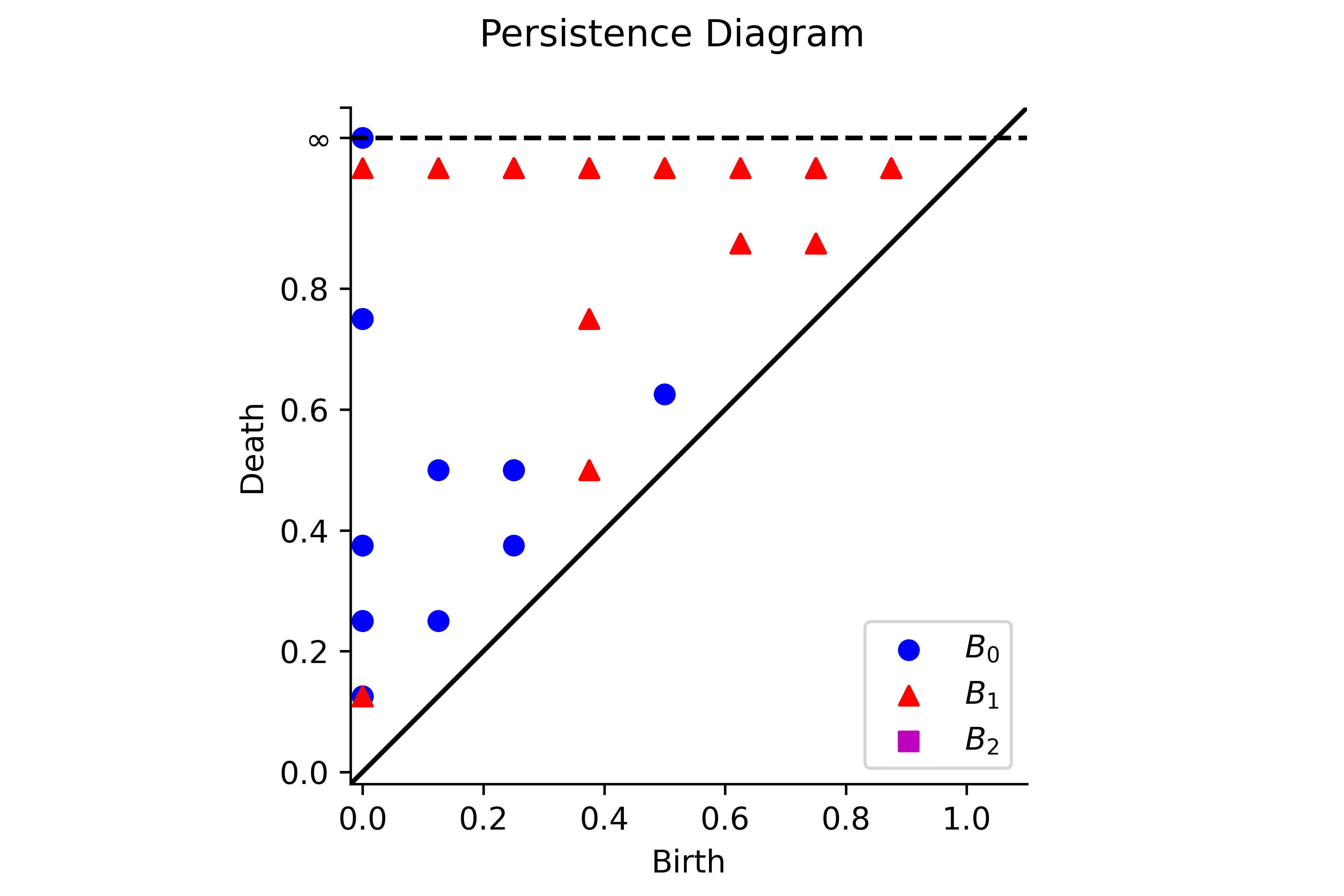}}
\hfill
\subfloat[\emph{The Bronco Buster}]{
	\includegraphics[width=0.47\columnwidth,trim=2.5cm 0cm 3cm 1cm, clip]{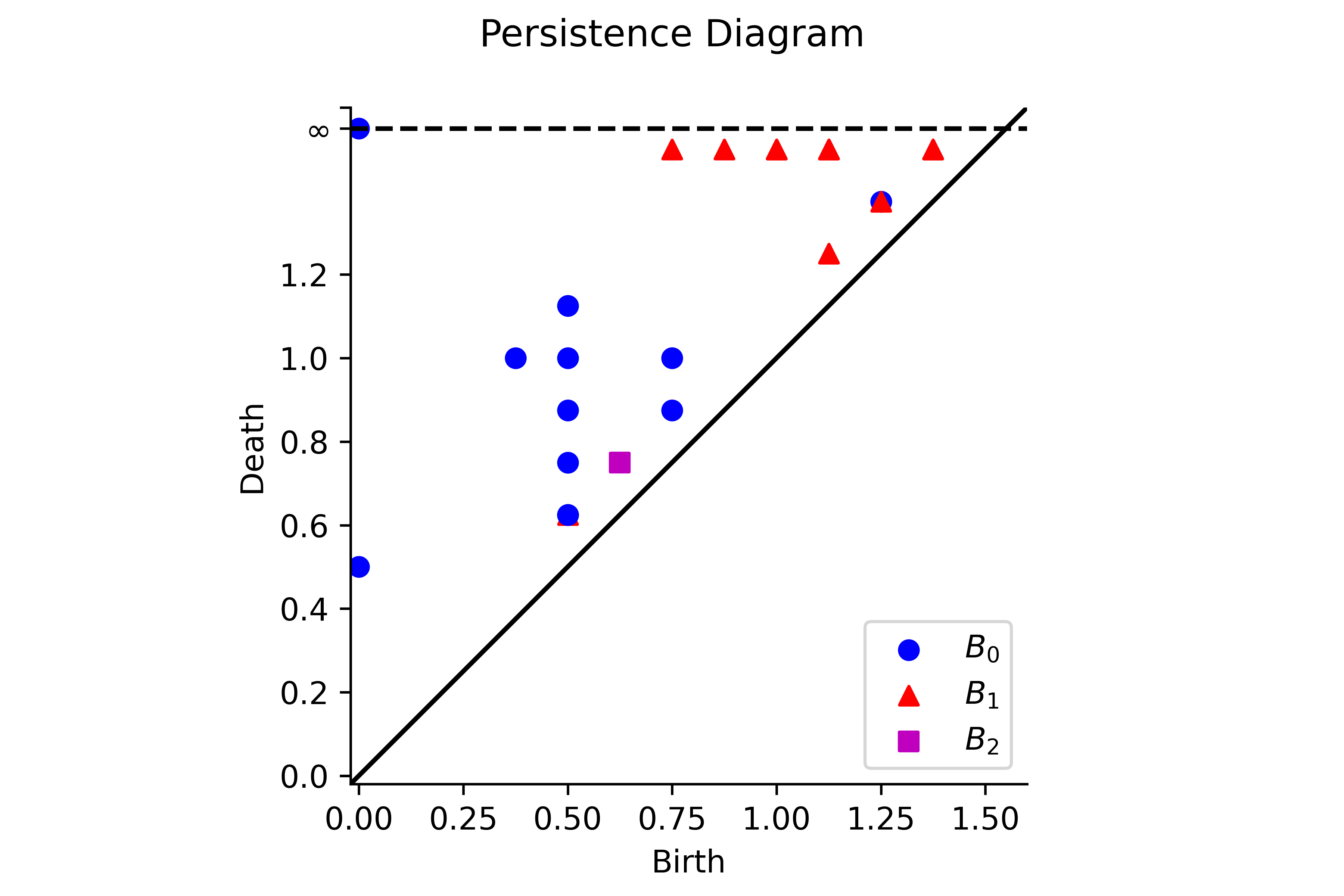}}
\caption{Persistence diagrams for the volumetric meshes are shown, with paramerter 1 minus the volume fraction.}
\label{fig:ph_diagrams}
\end{figure}


\briancode\ was developed and evaluated using a custom plugin to Rhinoceros 3D and Grasshopper.
Winding numbers were computed using libigl \cite{jacobson2013libigl}.
Persistent homology was computed using Aleph \cite{aleph}, which is based on PHAT \cite{Bauer:2017}.

The framework is tested on a 2D model of Chesapeake Bay\footnote{Model derived from \url{https://vecta.io/symbols/281/ecosystems-maps/93/usa-md-va-chesapeake-bay-line-map}} with boundary errors, a 3D model of mechanical bearings \footnote{Model provided at \url{https://ten-thousand-models.appspot.com/detail.html?file_id=1716283}}, and a 3D graphics model of \emph{The Bronco Buster}\footnote{Model provided at \url{https://tinyurl.com/4cmrptev}}.
Snapshots of meshes given computed volume fractions are shown in \cref{fig:volfrac_chesapeake,fig:bearings,fig:cowboy}.
Model errors for the Chesapeake Bay include overlapping edges, repeated/offset edges, and numerous gaps.
Despite the ``interior'' of the bay being ill-defined, the proposed method still captures the intended geographic domain with respect to both the continent and to islands.
A complete view of the homological structure based on varying the volume fraction for the Chesapeake Bay is shown in \cref{fig:persistence_diagram}, while similar persistence diagrams for the volumetric models are shown in \cref{fig:ph_diagrams}.
Results demonstrate that a mesh with the desired homological structure could be extracted from the background grid by selecting the correct threshold. 
These figures are primarily for illustrative purposes: we purposely chose a coarse grid size to generate the topological issues we are addressing. 
In practice, a finer grid would better capture local behavior.

\begin{figure}
\centering
\includegraphics[trim=30cm 0cm 23.5cm 0cm, width=0.48\columnwidth,clip]{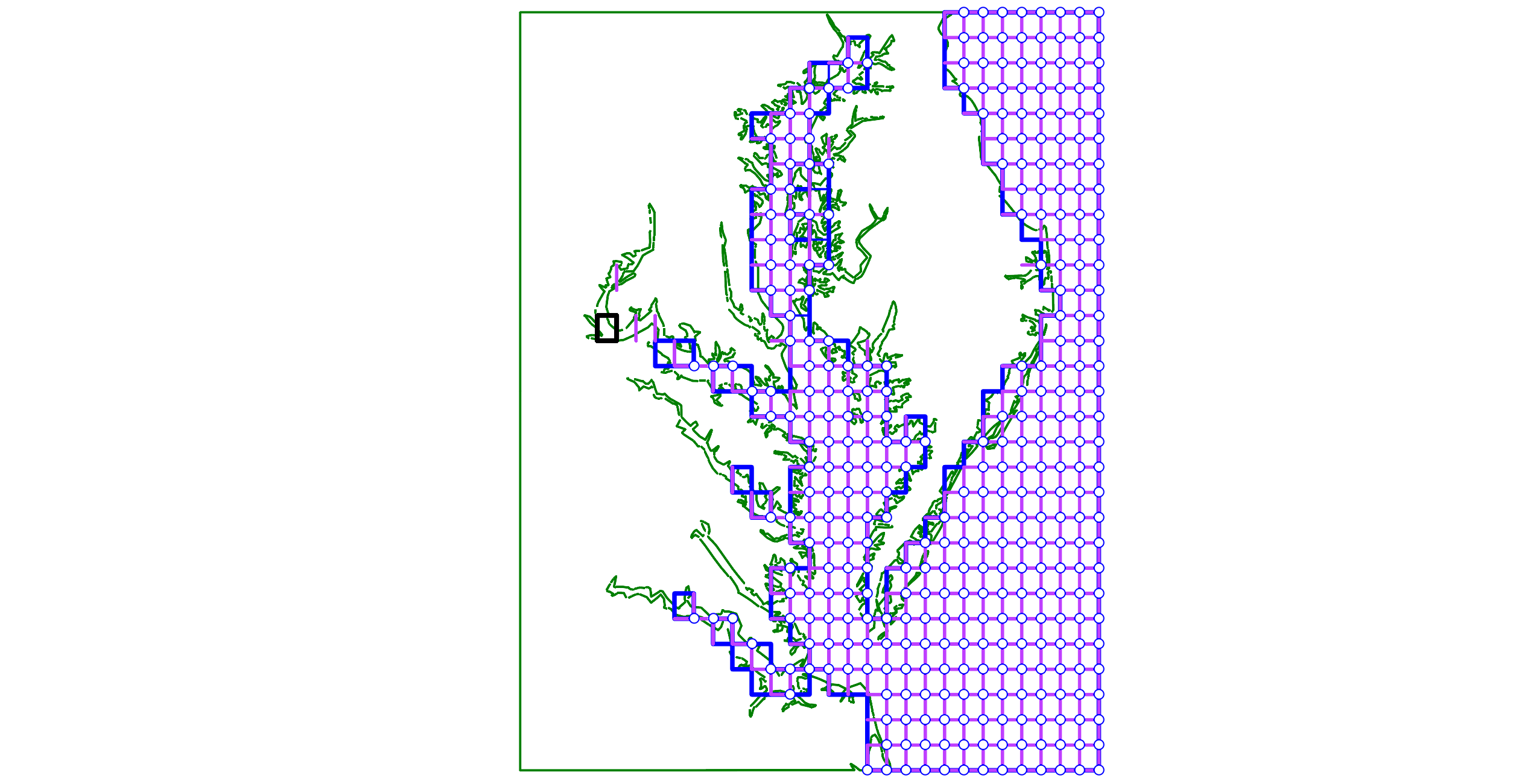}
\hfill
\includegraphics[trim=30cm 0cm 23.5cm 0cm, width=0.48\columnwidth,clip]{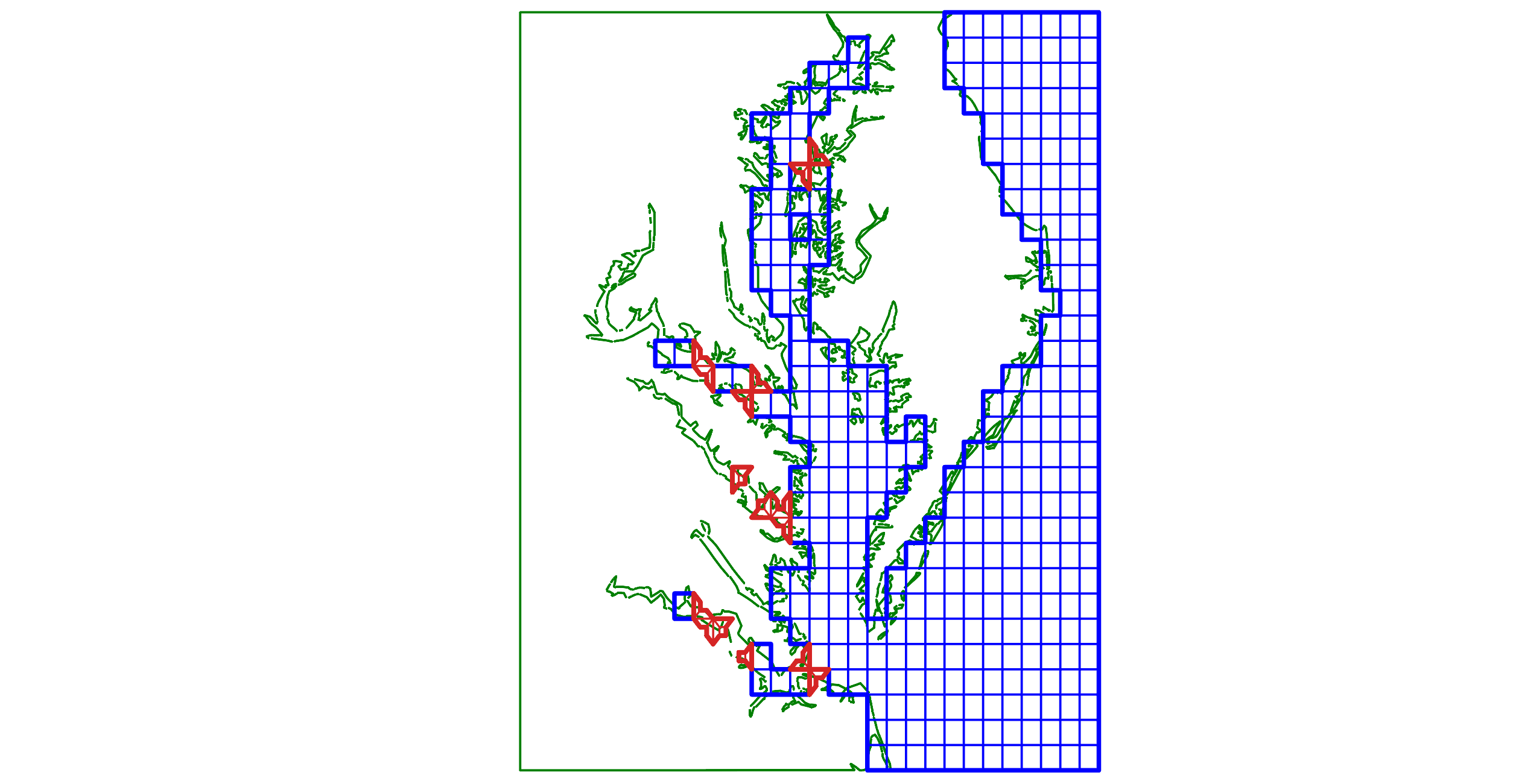}
\caption{Meshes of Chesapeake bay with subcells and anti-aliasing are shown. Interior vertices and edges are shown as blue outlined circles and purple lines respectively, while removed faces are shown in black (left). Connecting and separating templates are in red (right). 
}
\label{fig:antialiasing}
\end{figure}

\begin{figure}
\centering
\subfloat[Time step: 0]{
	\includegraphics[trim=13cm 3cm 4cm 3cm, width=0.3\columnwidth,clip]{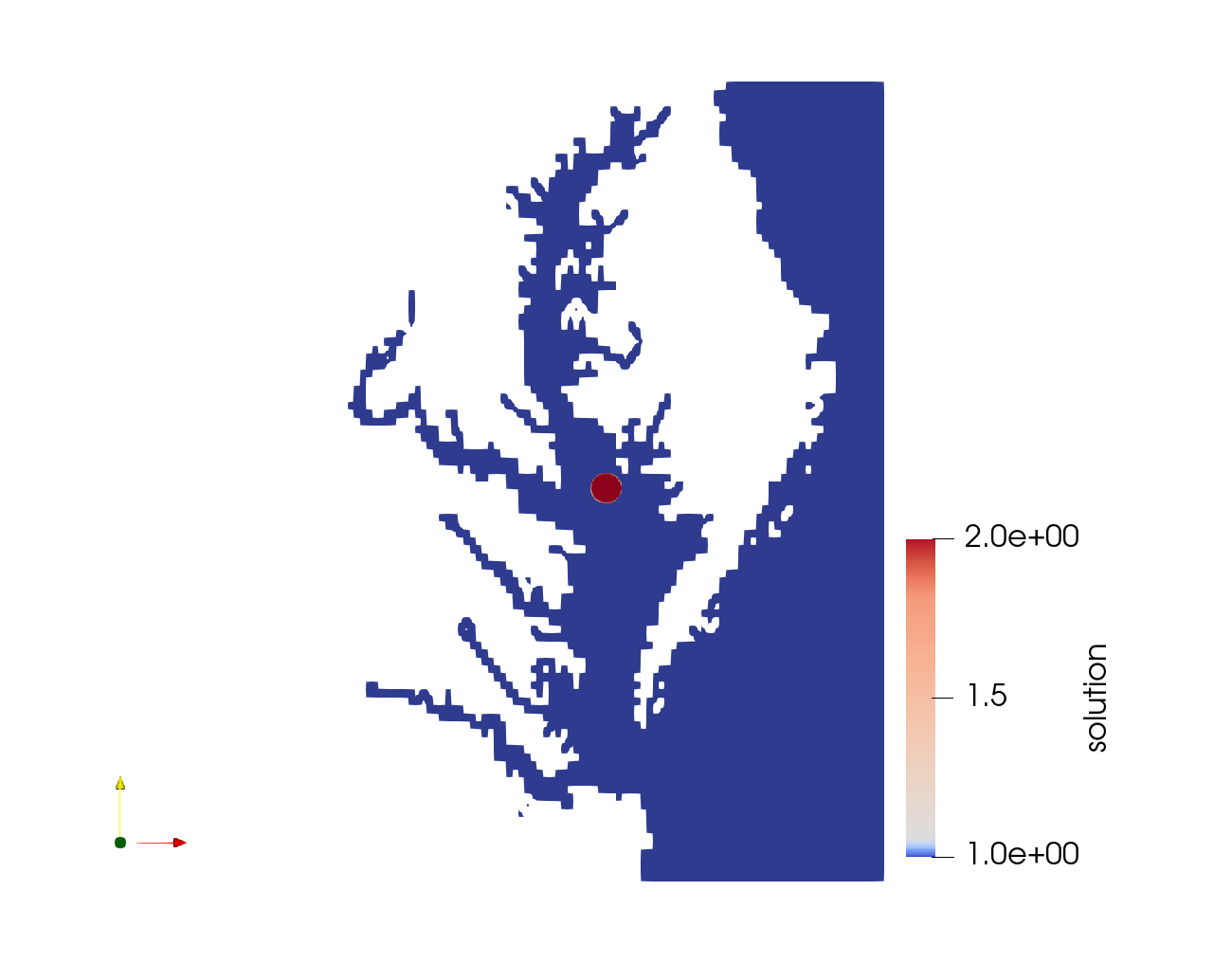}}
\subfloat[Time step: 50]{
	\includegraphics[trim=13cm 3cm 4cm 3cm, width=0.3\columnwidth,clip]{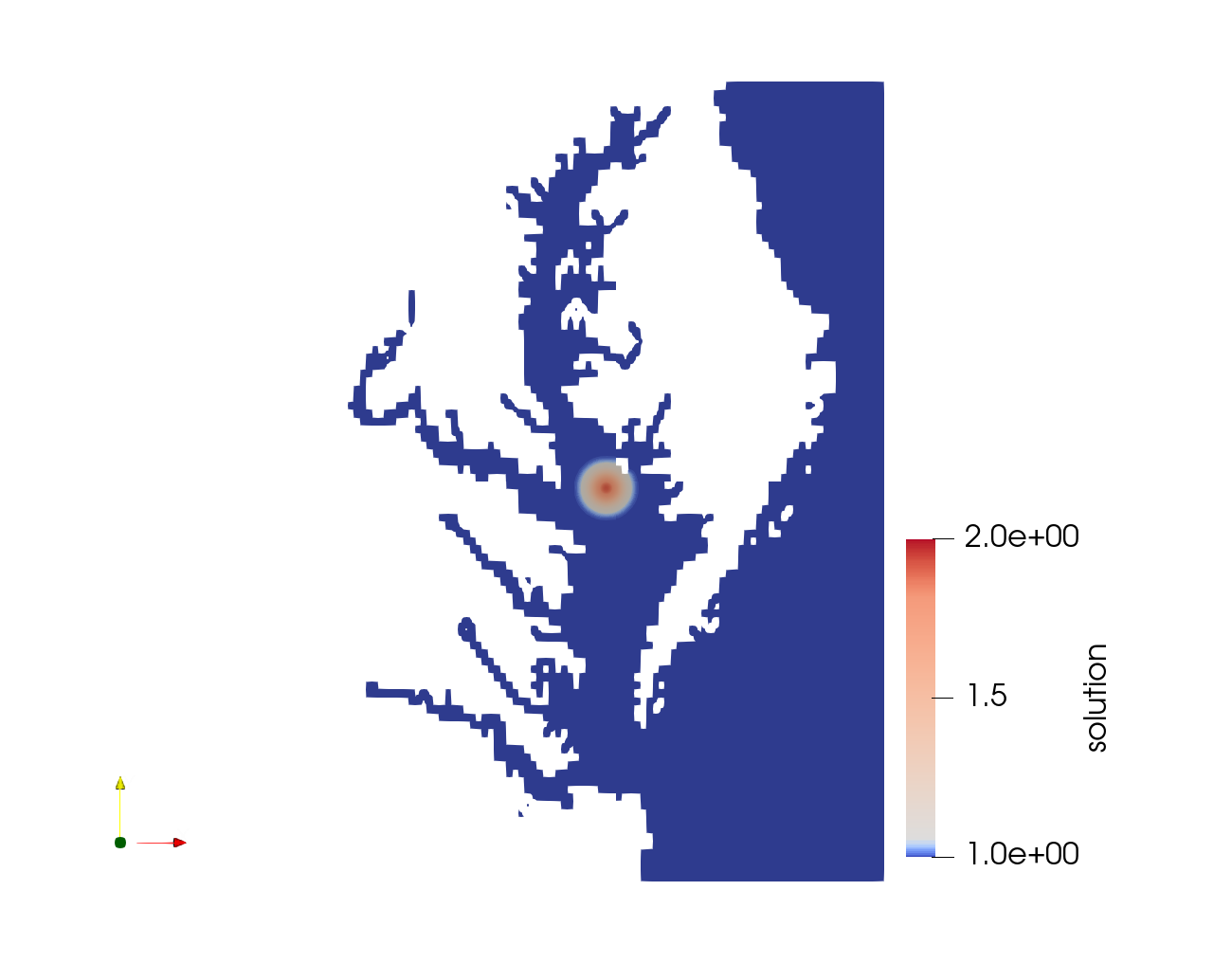}}
\subfloat[Time step: 100]{
	\includegraphics[trim=13cm 3cm 4cm 3cm, width=0.3\columnwidth,clip]{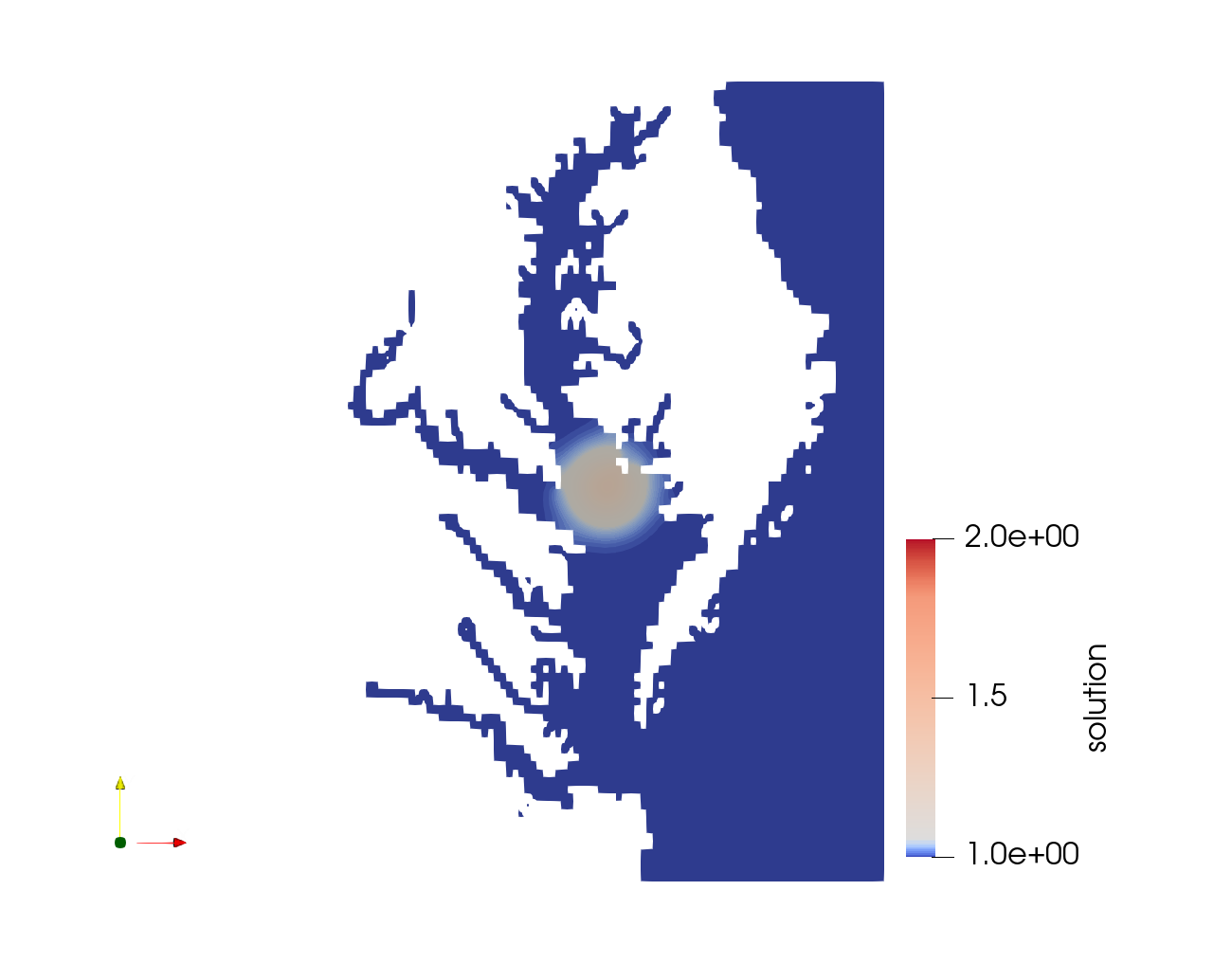}}
\\
\subfloat[Time step: 200]{
	\includegraphics[trim=13cm 3cm 4cm 3cm, width=0.3\columnwidth,clip]{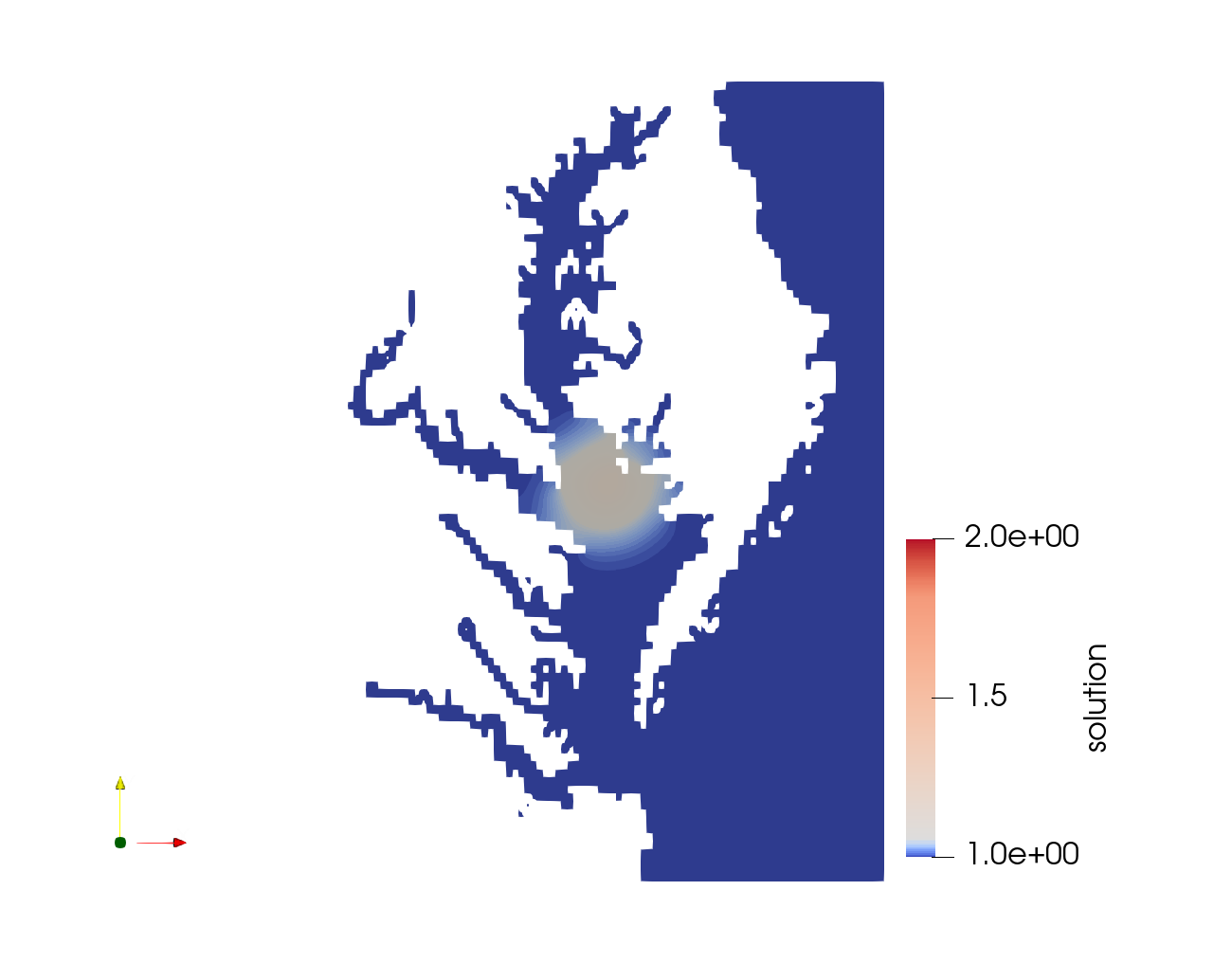}}
\subfloat[Time step: 400]{
	\includegraphics[trim=13cm 3cm 4cm 3cm, width=0.3\columnwidth,clip]{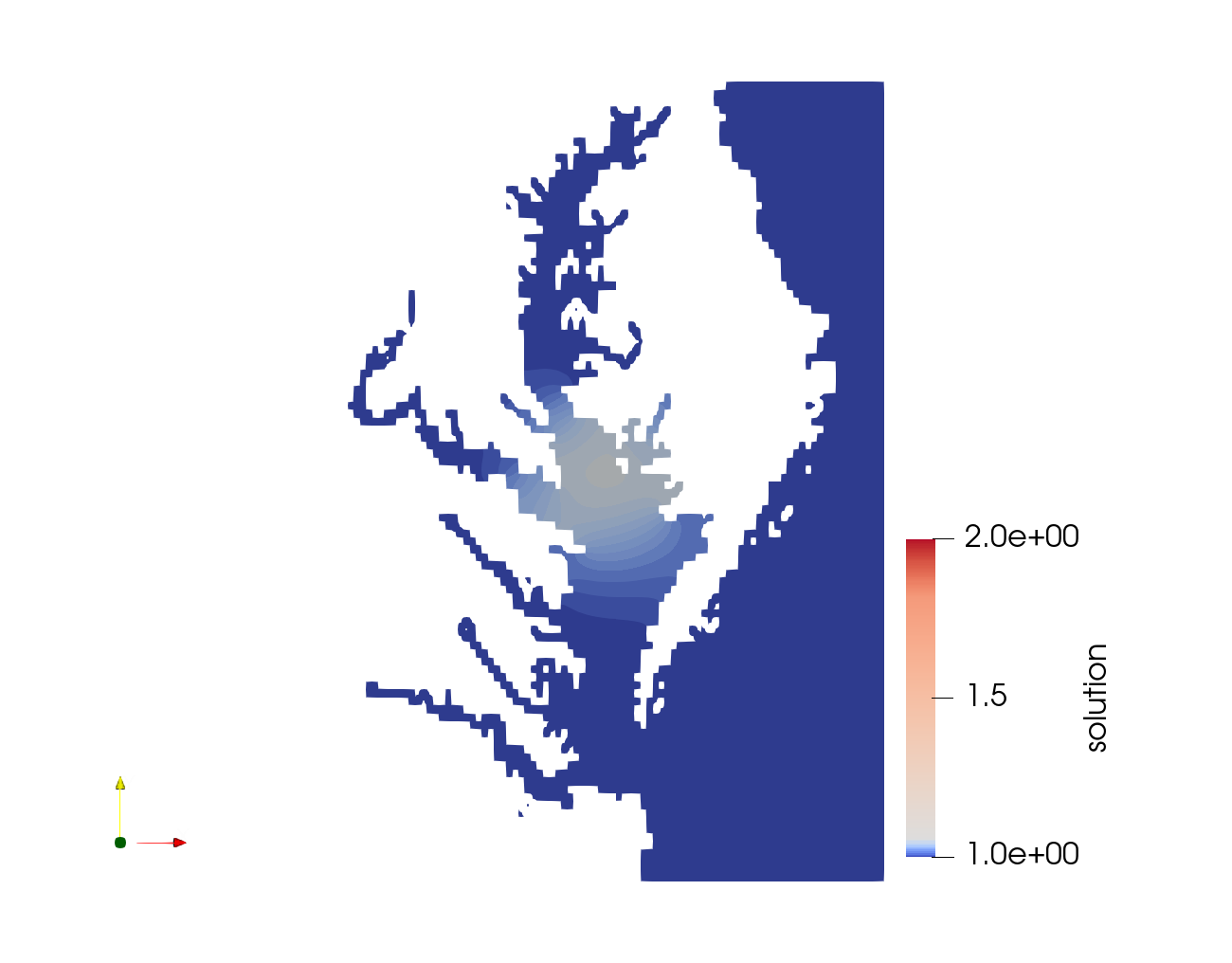}}
\subfloat[Time step: 800]{
	\includegraphics[trim=13cm 3cm 4cm 3cm, width=0.3\columnwidth,clip]{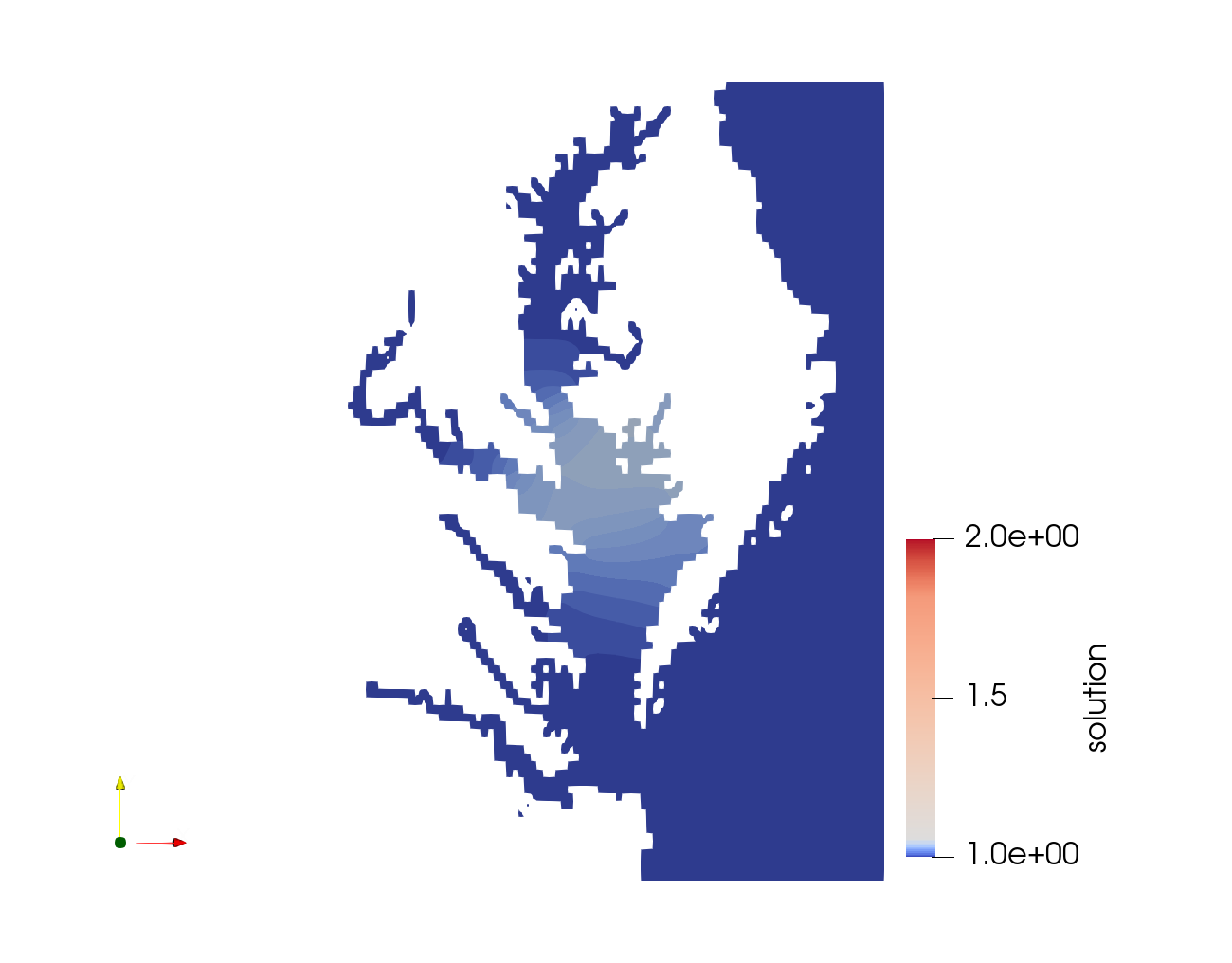}}
\caption{A finite element solution of contaminant propogation is displayed on a \briancode\  mesh of the Chesapeake Bay.}
\label{fig:fea}
\end{figure}

\paragraph{Anti-aliasing.}
\Cref{fig:antialiasing} demonstrates the anti-aliasing technique on a mesh of the Chesapeake Bay to resolve both pinches and archipelagos.
The anti-aliased mesh is analysis suitable, which is illustrated through a simulation of contaminant propagation, e.g. an oil spill, on a refined version of the Chesapeake Bay in \cref{fig:fea}.
Simulations were performed using MFEM \cite{mfem,mfem-web}.

\subsection{Anti-aliasing Guarantees and Limitations.}

As noted in Section \ref{sec:rasterization}, one of the primary difficulties with volume fraction-based meshing methods is mitigating the effects of rasterization (i.e.\ choice of orientation and sample size) through topological anti-aliasing.
The following theoretical result holds regarding the success of the proposed anti-aliasing methods in mitigating topological rasterization.
 A proof of the result can be found in the appendix this document.
\begin{theorem}
Given a rectangular lattice in $\mathbb{R}^2$ with characteristic length $\ell$ overlaying two parallel half-spaces separated by a length of $L$, topological rasterization may occur when $\ell(\sqrt{2}-1)<L\leq\ell$.
For the subgrid sampling scheme proposed in this text, topological rasterization may only occur when $\ell(\sqrt{2}-1)<L<\frac{\sqrt{2}\ell}{2}.$
Finally, topological rasterization due to changes in orientation cannot be resolved for $L$ such that $\ell(\sqrt{2}-1)<L<\frac{\ell}{2}.$
\end{theorem}


 \subsection{Topological Effects of Grid Refinement.}
To conclude the results section, we demonstrate that topological pathologies may arise when naively using persistent homology to find an appropriate grid size to achieve a desired topology.
When features are isolated or globally the same scale, grid refinement has intuitive and predictable topological effects. 
However, general inputs may \emph{not} demonstrate monotonic filtration behavior with grid refinement. 
Counterexamples show non-monotonic filtration behavior by grid size. Discretization by grid cells and their alignment with input features strongly effects topological behavior. Thus algorithm parameters of when to refine the background grid may have unexpected effects on mesh topology. 

\paragraph{Convergence.}
For many inputs, as the grid is refined, topological features of the input are resolved and the output mesh topology becomes stable. 
However, for some inputs, the topology never converges and no filtration is possible.
For some inputs it may be possible to define a filtration, with simplices only appearing, never disappearing. 
If simplices appear and disappear, zig-zag persistence could computationally predict topology.

In \cref{fig:diagonal,fig:jaggies} a feature is inconsistently resolved due to aliasing effects of unaligned grids. 
For the constant-width gap in \cref{fig:diagonal}, refining or coarsening the grid makes the gap resolved consistently as open or closed.
However, for the variable-sized gap in \cref{fig:jaggies}, global uniform refinement may merely move where the problem occurs. 
The example is a wedge of material bounded by two lines meeting at a small angle $\alpha$ at an apex. 
In locations where the grid size is about the same as the local width, whether a cell is included or excluded can change every few grid cells, leading to many separate connected grid components. 
For any small grid size, there will be some portion of the wedge where the lines are about that  size apart, specifically in the range $[\frac{1}{2},1] \cot \alpha$ squares away from the apex.
The geometry is undesirable because the islands move location. 
The grid topology may be constant over refinement, but that topology differs from the topology of the underlying object and is undesirable.

\begin{figure}[!htb]
\centering
\includegraphics[width=0.98\columnwidth]{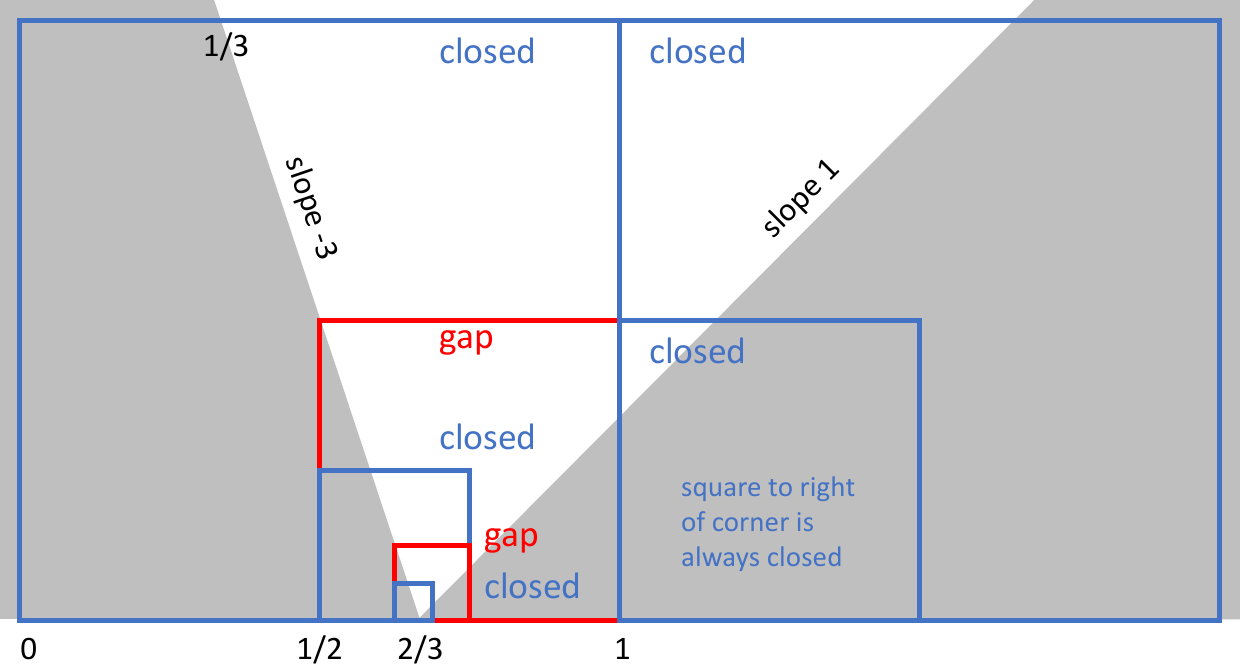}
\includegraphics[width=0.98\columnwidth]{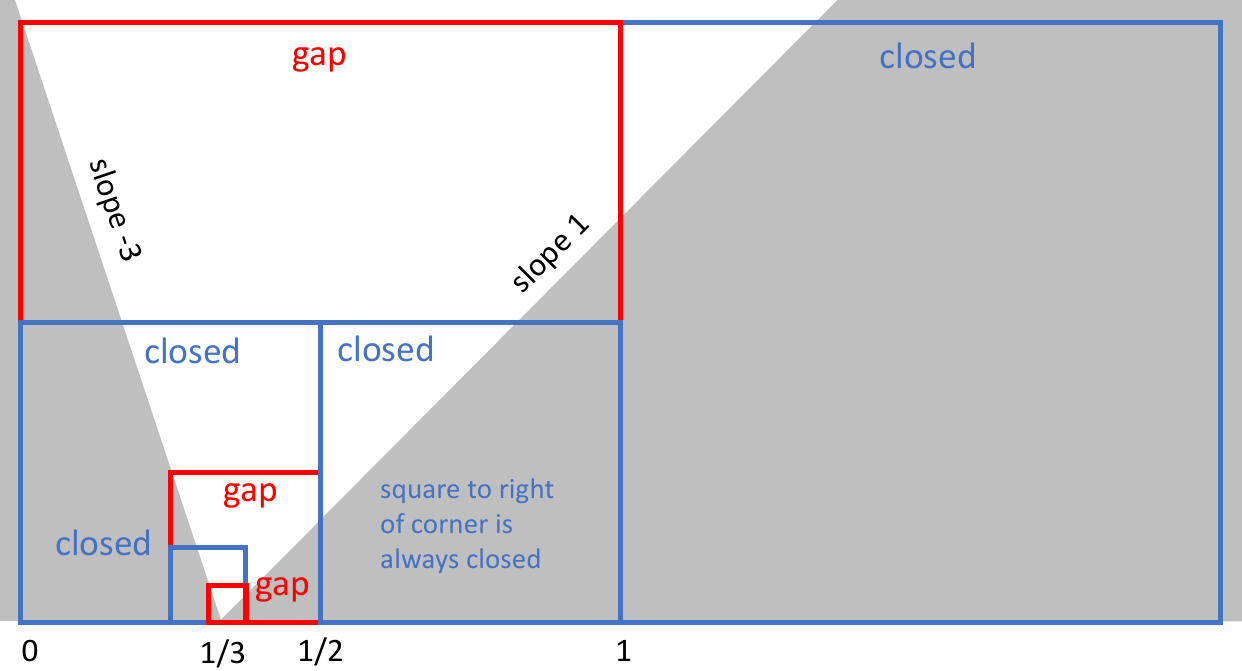}
\caption{In these counterexamples to convergence, the grid topology alternates between one and two connected components ad infinitum under refinement. The alternations of the upper and lower examples are opposite.}\label{fig:two_triangles}
\end{figure}

\paragraph{Non-convergence.}
In each of the examples in \cref{fig:two_triangles} the output mesh topology does not converge under refinement. That is, there is no grid size below which the output mesh topology does not change.
The background grid is uniform, but we only draw some of the relevant cells at each level of refinement. Blue (closure) cells are mostly material and thus included in the output mesh. Red (gap) cells are unfilled and excluded.
Under refinement, the meshes alternate between one and two connected components ad infinitum.
The grid squares containing the corner alternate between filled and open, because of the corner’s relative position inside its square.
The descriptions of the geometries are finite, just two triangular blocks. It is simple, plausible, and without sharp angles. The only unusual feature is the blocks touch at a single pinch point.

The upper and lower examples in \cref{fig:two_triangles} have alternate sizes of when they are open and closed. If an input has both of these pinch features between two material blocks, then exactly one of the pinches will be closed, giving a mesh with the homology of a disk. It is converged in the sense that the homology does not change under refinement, but the local connectivity does change. Hence, even if we were to use zigzag homology it would not distinguish between this case and a single solid block of material.

The analytic description of the geometries in \cref{fig:two_triangles} is two triangular blocks of material with slopes $-3$ and $1$ meeting at a corner. In the upper example, the corner's coordinate is $(\frac{2}{3},0),$ and in the lower it is  $(\frac{1}{3},0).$ 
If we start with a unit grid with a vertex at the origin, then under refinement the grid square containing the corner alternates between having the corner $2/3$ of the way along the bottom edge (blue), and $1/3$ of the way (red). Such blue squares have volume fraction $10/18$ and the red squares $7/18.$ This construction is not tight. The slopes may be different. The corner may lie at some other coordinate, and a grid vertex will never lie on it if its x-coordinate is not $k/2^m$ for some $\{k,m\} \in \mathbb{Z}.$ Thus more complicated sequences than alternating may be constructed.

\section{Conclusion}

In this work, a mesh generation framework is developed to facilitate the creation of meshes on potentially dirty geometry based on user-specified needs through the use of persistent homology.
The framework is built on a volume-fraction based meshing method, similar to Sculpt, in which mesh elements are considered ``in'' or ``out'' of the given geometry depending on whether they have volume fraction above a cutoff range; the most desirable mesh can then be selected based on the  appropriate homological structure induced by this volume fraction.
These volume-fraction based algorithms, including Sculpt, behave quite differently than boundary-fitted algorithms, such as Delaunay Refinement, when the local mesh size is decreased. 
 For boundary-fitted algorithms, reducing the mesh size to the local feature size or less allows the mesh to have good quality and recover the input topology exactly. 
 For the family of volume-fraction codes, the mesh quality is good regardless of the local feature size, but in some cases the topology does not converge as the mesh size decreases by subdividing the background grid. 
 However, for a fixed grid, we may predict the resulting mesh topology, measure how that changes as we vary the volume-fraction threshold using persistent homology, and select an appropriate topology for the mesh's intended purpose. 
When the background grid is about the same size as gaps and thin-region features, the alignment and orientation of the background grid with respect to the features can create aliasing artifacts in the mesh.
There may be spurious pinches and archipelagos.
A gap or thin region may be resolved inconsistently in different locations.
Subgrid sampling provides guidance on whether to connect, separate, or remove components.
These connections and separations can be achieved using templates of small elements. 
We have demonstrated theoretically and experimentally that subgrid sampling can mitigate the effects of aliasing, making connections more consistent.
The software, \briancode{,} demonstrates the potential of the meshing framework in both two and three dimensions, and is planned for open source release.
As a counterpoint we have theoretical analysis showing that for any volume-fraction threshold we choose, there are cases where aliasing artifacts will still arise.

There are a number of avenues for future work to build on and improve this framework.
First, it may be valuable to have adaptive, non-uniform background grids and volume fraction thresholds to mesh more interesting geometries at reduced expense.
However, doing this may require the use of zigzag persistence to accurately capture mesh topology, particularly when refinement does not converge on separation or closure of local mesh features.
Further work should also better utilize fitting algorithms, such as those present in Sculpt, to more accurately fit to the geometric input data that \briancode{} approximates.
To do this more flexibly, additional research should generalize these methods to unstructured cubical complexes.
Finally, it is anticipated that for this framework to be usable in practice, it would need to be redeveloped using C++.
%

%


\section*{Acknowledgements}
\anon{We thank various people for their help (anonymized).\par}
{
We thank Steve Owen for providing Sculpt and explaining its subtleties to us. 
We thank Daniele Panozzo for answering our questions about libigl.
We thank Alec Jacobson and the other authors of libigl, especially the winding number component, for making it freely available to the community. 
We thank Darren Engwirda for discussing coastal climate modeling with us.
We thank Anjul Patney for pointing out the connection between rasterization and volume fraction meshing.\par
}

{\small 
\anon{This work was supported by our institutions and funders (anonymized).\par}
{
This material is based upon work supported by the U.S. Department of Energy, Office of Science, Office of Advanced Scientific Computing Research (ASCR), Applied Mathematics Program. Sandia National Laboratories is a multimission laboratory managed and operated by National Technology \& Engineering Solutions of Sandia, LLC, a wholly owned subsidiary of Honeywell International Inc., for the U.S. Department of Energy’s National Nuclear Security Administration under contract DE-NA0003525.\par
}
}

\appendix
\section{Appendix: Anti-Aliasing Proof}

The following result is proved in this appendix.

\begin{theorem}
Given a rectangular lattice in $\mathbb{R}^2$ with characteristic length $\ell$ overlaying two parallel half-spaces separated by a length of $L$, topological rasterization may occur when $\ell(\sqrt{2}-1)<L\leq\ell$.
For the subgrid sampling scheme proposed in this text, topological rasterization may only occur when $\ell(\sqrt{2}-1)<L<\frac{\sqrt{2}\ell}{2}.$
Finally, topological rasterization due to changes in orientation cannot be resolved for $L$ such that $\ell(\sqrt{2}-1)<L<\frac{\ell}{2}.$
\end{theorem}

Our analysis proceeds in the context of \cref{fig:parallel_axis_aligned_gap,fig:diagonal}.
Assume that two connected components of a 2D domain are parallel to each other and separated by a gap of length $L$.
For the sake of clarity and simplicity, assume both components of the domain are half-spaces, and that as a result, $L$ is the only length of consequence for the geometry to be meshed.
Call the half-spaces $\Omega$ and the complement $\Omega^c$.
Without loss of generality, assume that centerline of $\Omega$ and $\Omega^c$ is the $y$-axis.
Assume a rectangular lattice (``background grid'') tiling the Euclidean plane, rotated by an arbitrary angle and shifted by an arbitrary translation.
Let us assume that lengths of the lattice are equal in both principal directions (e.g. the $x$ and the $y$ direction if no lattice rotation)---else a rotation by 90 degrees will shift biases introduced by anisotropy.
Call this length $\ell$.

By definition of the centroid of a domain, a cell of the lattice, $C$, will intersect $\Omega^c$ maximally if its centroid intersects the centerline of $\Omega^c$.
This is the limiting case for potential rasterization.
As a result, we investigate rotations of the lattice translated such that a cell $C$ intersects $\Omega^c$ on its centerline.
There are three scenarios describing the domain $C\cap \Omega^c$.
\begin{enumerate}
	\item The boundary of $\Omega$ is parallel to $C$, in which case $C \cap \Omega^c$ is a rectangle. 
	\item The boundary of $\Omega$ intersects $C$ at edges on opposite sides of $C,$ in which case $C\cap \Omega^c$ is a parallelogram. 
	\item The boundary of $\Omega$ intersects $C$ at edges that share a common vertex, in which case $C \cap \Omega^c$ is a hexagon.
\end{enumerate}
The first of these cases is a special case of the second, but is particularly important for the analyses and so it has been called out particularly.

\begin{figure}
\centering
	\subfloat[Cell configurations]{
		\includegraphics[width=1\columnwidth]{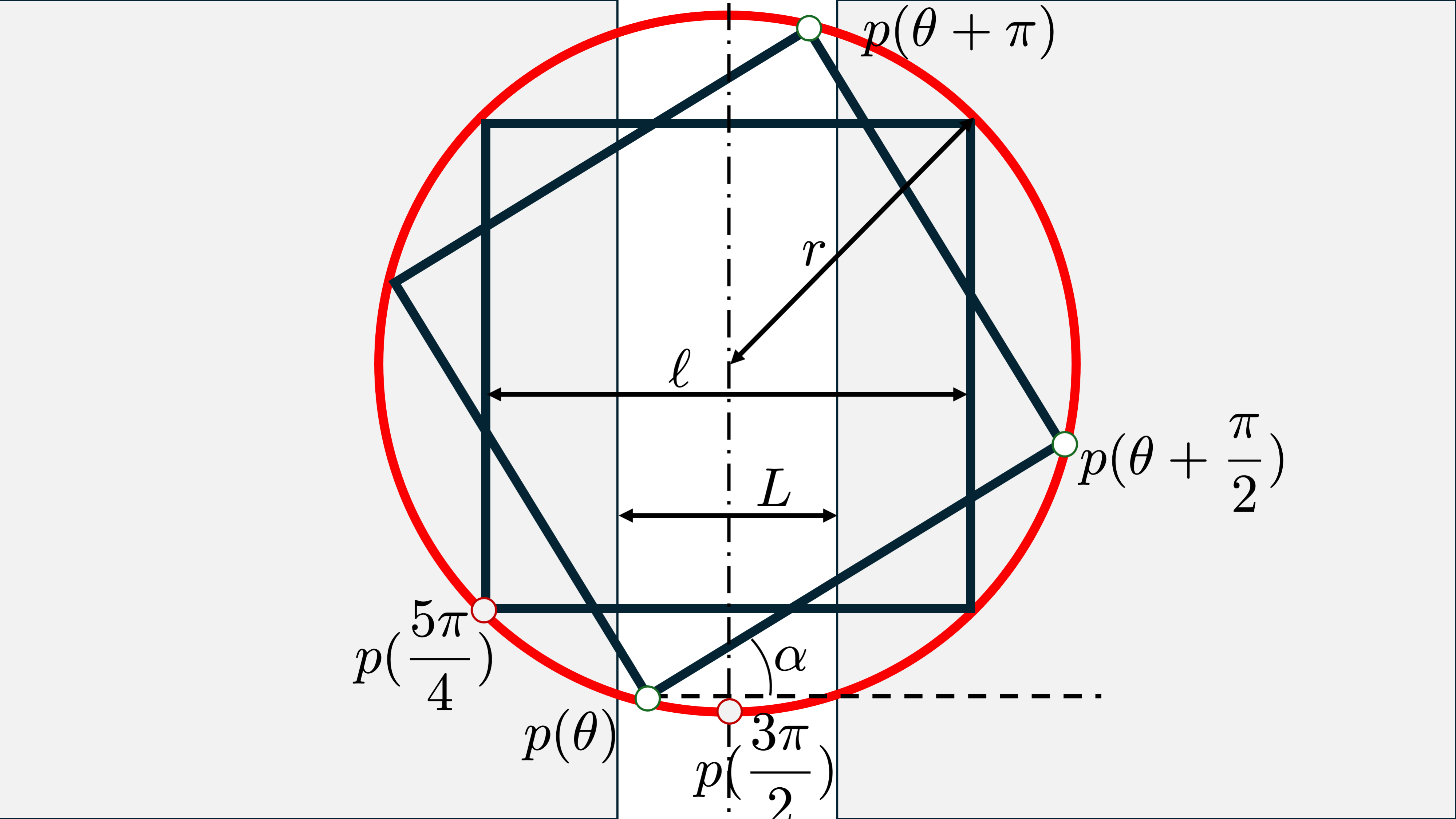}
			\label{fig:proof_pic_setup}
	}
	\qquad
	\subfloat[Area configuration 1]{
		\includegraphics[width=1\columnwidth]{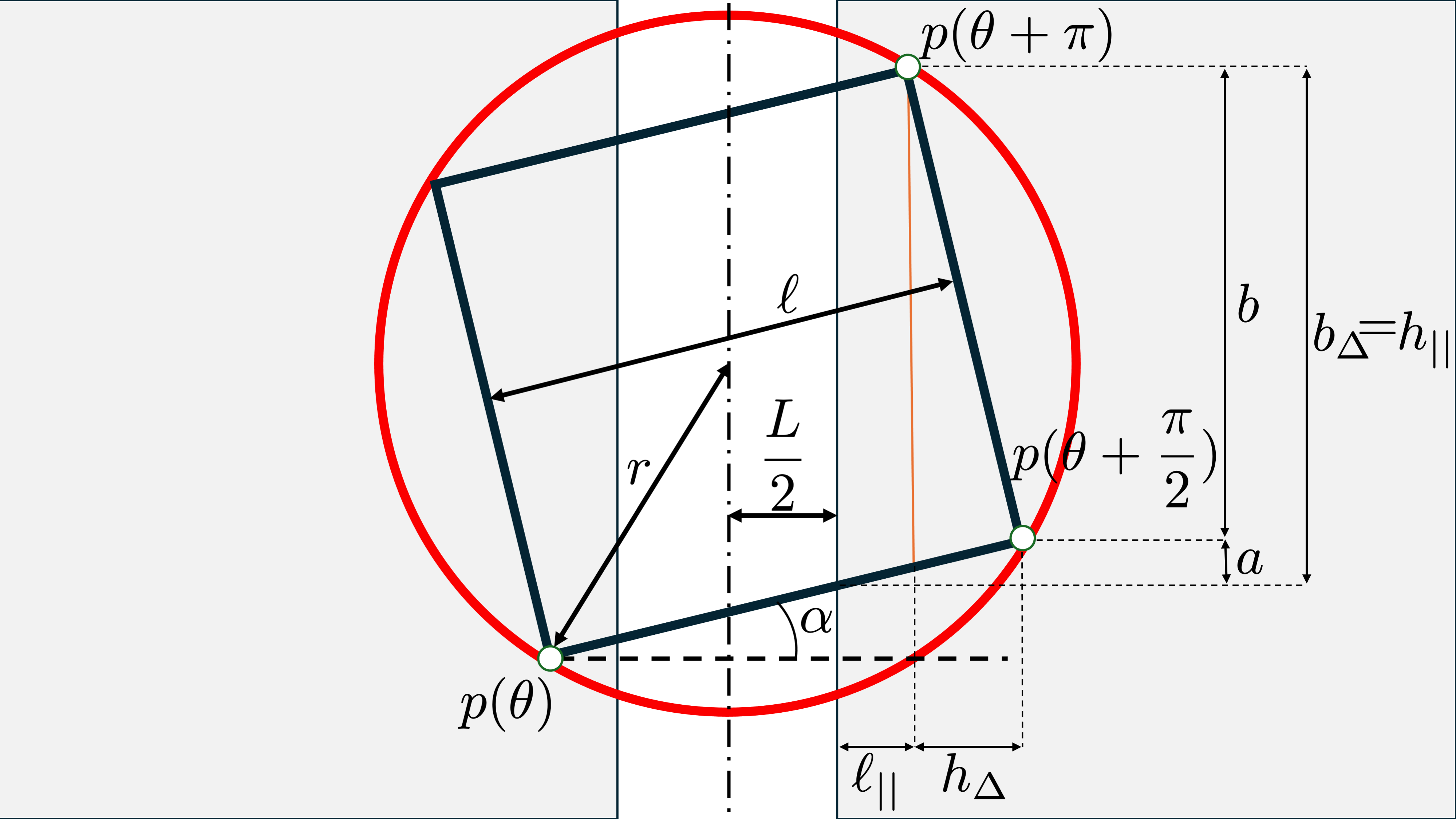}
			\label{fig:proof_pic_parallelogram}
	}
	\qquad
	\subfloat[Area configuration 2]{
		\includegraphics[width=1\columnwidth]{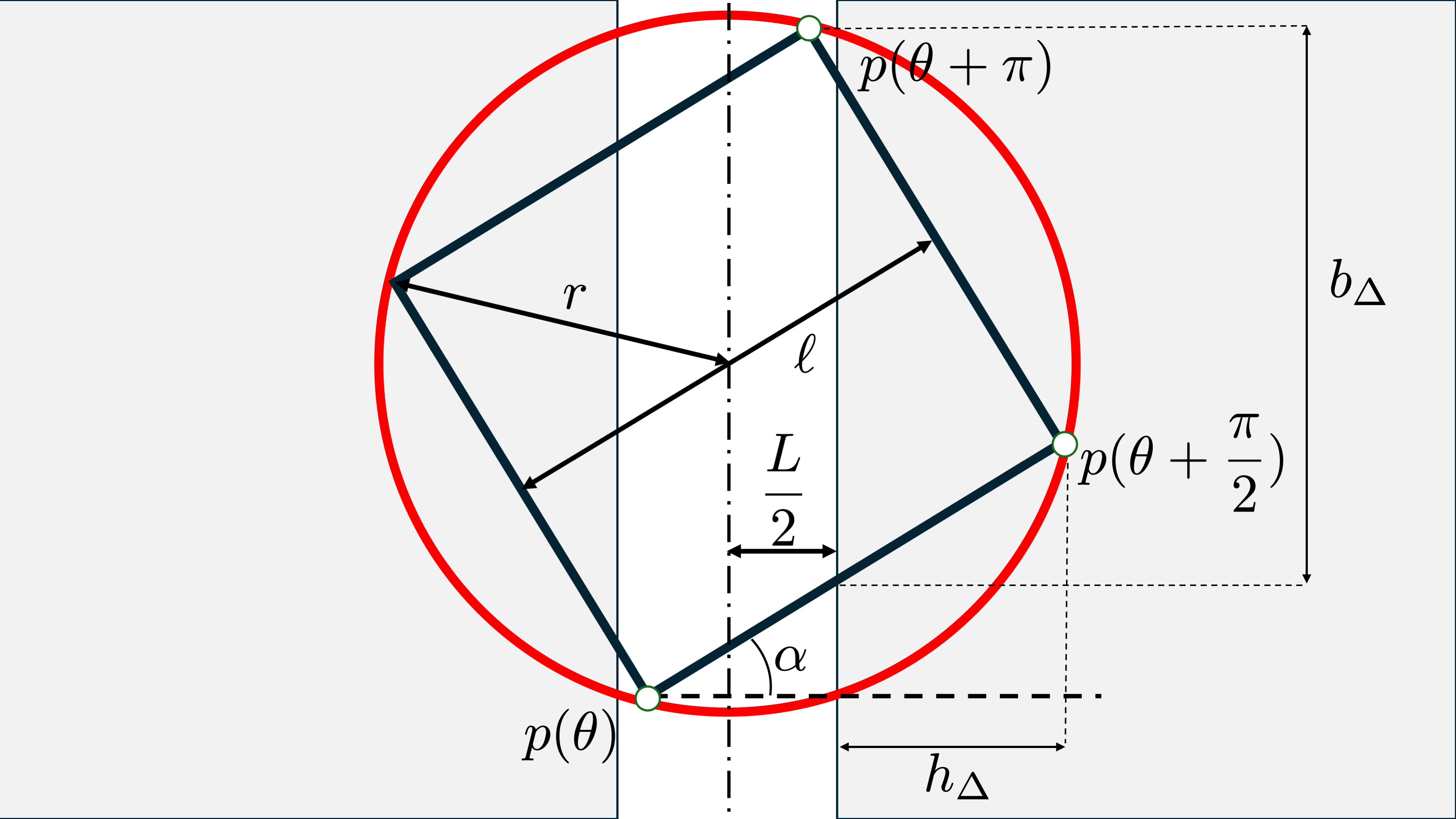}
			\label{fig:proof_pic_triangle}
	}
	\caption{Configurations are shown for maximal intersection of a square with arbitrary orientation centered between two half-spaces.
	Measurements are also depicted which define how to arrive at computed areas.}
	\label{fig:proof_pic}
\end{figure}

Let a circle of radius $r$ be defined as the set of points traced by all potential rotations of $C$ about the centroid of $C$.
By definition
\begin{equation}
	r=\frac{\sqrt{2}\ell}{2}.
\end{equation}
Define the function $p(\theta)$ as the position of a vertex of cell $C$ on this circle at specified radian, $\theta$
\begin{equation}
	p(\theta) = r \big(\cos(\theta),\sin(\theta)\big).
\end{equation}
Due to symmetry of the geometry, it is sufficient to analyze $\theta \in [\frac{5\pi}{4},\frac{3\pi}{2}]$.
Consequently, all following results assume $\theta$ in this range.
This configuration setup is depicted in \cref{fig:proof_pic}.

The equations of the line passing through points $p(\theta)$ and $p(\theta + \frac{\pi}{2})$ can be solved to yield
\begin{equation}
	y(x;\theta) = \frac{r - x \cos(\theta) + x \sin(\theta)}{\cos(\theta) + \sin(\theta)},
\end{equation}
\begin{equation}
	x(y;\theta) = \frac{r - y \cos(\theta) - y \sin(\theta)}{\cos(\theta) - \sin(\theta)}.
\end{equation}
Note that the function $x(y;\theta)$ is ill-defined at $\theta=\frac{5\pi}{4}$.

Using the notation from \cref{fig:proof_pic_parallelogram}, the area of $C$ intersecting one half-space of $\Omega$ in the case where $C\cap\Omega^c$ is a parallelogram is 
\begin{align*}
	A_1 &= \frac{1}{2}h_\Delta b_\Delta + \ell_{||} h_{||}\\
		&= \frac{1}{2}\big(r (\cos(\theta) - \sin(\theta))\big)\frac{-2r}{\cos(\theta)+\sin(\theta)} + \\
		& \qquad\big(-\frac{L}{2}-r\cos(\theta)\big)\frac{-2r}{\cos(\theta)+\sin(\theta)}\\
		&= \frac{r \big(L + r \cos(\theta)  + r\sin(\theta)\big)}{\cos(\theta) + \sin(\theta)} 
\end{align*}
Because $\theta \in [\frac{5\pi}{4},\frac{3\pi}{2}]$, the denominator is always positive.
Similarly, using the notation from \cref{fig:proof_pic_triangle}, the area of $C$ intersecting one half-space of $\Omega$ in the case where $C\cap\Omega^c$ is a hexagon is
\begin{align*}
	A_2 &= \frac{1}{2}h_\Delta b_\Delta \\
	     &= -\big(\frac{L}{2} + r\sin(\theta)\big)^2\sec(2\theta)
\end{align*}
Note that the transition from the first case to the second case occurs when $\ell_{||} h_{||} = 0,$ or in other words when $\frac{L}{2} = -r \cos(\theta) = -\frac{\sqrt{2}\ell}{2} \cos(\theta).$
When $\frac{L}{2}$ is less than this term, the area of intersection will be $A_1$, while when $\frac{L}{2}$ is greater than the term, the area of intersection will be $A_2$.
 
 Taking the derivative of $A_1$ with respect to $\theta$ and setting equal to zero, it is found that $A_1$ is maximal when $L=0$ (i.e. no gap between half-spaces) and when $\theta = \frac{5\pi}{4}$.
 Furthermore, taking the derivative of $A_2$ with respect to $\theta$ and setting equal to zero, critical points of $A_2$ are found to be at $\cos(\theta) = 0, \sec(2\theta) = 0, \frac{L}{2} = -r \sin(\theta), $ and $L = -r \csc(\theta)$.
 The first only occurs when $\theta = \frac{3\pi}{2}$ and the second is never zero and is bounded away from $\theta=\frac{5\pi}{4}$ because $A_2$ can never be the correct area of choice when $\theta=\frac{5\pi}{4}$.
 The third results in $A_2$ equaling zero, and is a minimum area.
 Finally, the fourth introduces a critical point at $\theta = -\csc^{-1}\big(\frac{L}{r}\big)$ provided $\frac{L}{r} = \frac{\sqrt{2}L}{\ell} \geq 1$.
Particularly, $\theta = \frac{3\pi}{2}$ is a local minimum when $L \leq \frac{\sqrt{2}\ell}{2}$ but a local maximum when $L > \frac{\sqrt{2}\ell}{2}$, and $\theta = -\csc^{-1}\big(\frac{L}{r}\big)$ is a local minimum when $L \geq \frac{\sqrt{2}\ell}{2}$.

The above mentioned formulas can be used to show that $A_1\Big|_{\theta = \frac{5\pi}{4}} \geq A_2\Big|_{\theta = \frac{3\pi}{2}}$ for $\frac{L}{2} \leq (2-\sqrt{2})r$ and $A_1\Big|_{\theta = \frac{5\pi}{4}} \leq A_2\Big|_{\theta = \frac{3\pi}{2}}$ for $\frac{L}{2} \geq (2-\sqrt{2})r$.
Particularly, when $\frac{L}{2} < \ell(\sqrt{2}-1)$, the area of $C\cap\Omega$ oriented at $\theta = \frac{5\pi}{4}$ is larger than that when $\theta = \frac{3\pi}{2}$, while the converse is true when $\frac{L}{2} > \ell(\sqrt{2}-1)$.

Now, assume that a persistence parameter, $\phi$ defined by the volume fraction of a square cell, and assume that any cell, $\hat{C},$ with $\phi(\hat{C}) \geq \frac{1}{2}$ is present in the cell complex.
When $L = \ell$ and $\theta = \frac{5\pi}{4}$, depending on the translational positioning of $L$, either all cells $C$ will contain at least 50\% volume fraction (if cell boundaries are aligned to the centerline of $\Omega$) or a vertical line of cells will have volume fraction less that 50\% to potentially 0\%.
For this same orientation, $\theta$, this potential for rasterization will persist until a length $L = \frac{\ell}{2}$, at which point all cells must have at least 50\% volume fraction and will all be contained in the cell complex.
Nonetheless, for any $\theta > \frac{5\pi}{4}$, cells $C$ with centroid on the centerline of $\Omega$  will have volume fraction less than 50\% and thus will not be in the complex.
However, when $L = \ell(\sqrt{2}-1), A_2\Big|_{\theta = \frac{3\pi}{2}} = \frac{\ell^2}{4}$ so all cells $C$ will be contained in the cell complex for all $\theta \in [\frac{5\pi}{4},\frac{3\pi}{2}]$.
Thus, for a gap of length $L$ and a lattice of length $\ell$, the topology of the domain in the lattice cannot be definitively resolved for $\ell(\sqrt{2}-1) < L \leq \ell$.

This above assessment was based on the fact that the maximal area of $C\cap\Omega^c$ would be found when the centroid of $C$ intersected the centerline of $\Omega$.
Similarly, the minimal area of $C\cap\Omega^c$ will be found when the intersection of the centerline of $\Omega$ and $C$ only contains points on the boundary of $C$.
For a fixed $L,$ the maximal intersection will be when $C\cap\Omega^c$ applies when given an orientation of $\theta = \frac{5\pi}{4}$, and the minimal for an orientation of $\theta = \frac{3\pi}{2}$.

Now, consider the proposed topological anti-aliasing technique, where volume fractions for a lattice are defined not only for 2-cells, but also for 1-cells and 0-cells by shifting the lattice by $\frac{\ell}{2}$ in the vertical or horizontal direction for 1-cells and the vertical and horizontal directions for 0-cells.
Assume that $L = \frac{\sqrt{2}\ell}{2} + \epsilon$ for some $\epsilon>0$ whose limit approaches zero.
Assume that $C$ intersects the centroid of $\Omega$.
Then the volume fraction of $C$ is less than 50\% by necessity.
Furthermore, because $C$ intersects the centerline of $\Omega$, a translation of $C$ in one of its principle directions that is also in the positive $y$ direction must also intersect the centerline.
Consequently, it, too, must have volume fraction of less than 50\%.
Proceeding inductively ensures separation of the two half-spaces in the upward direction from $C$, and a similar result can be shown for translations in the downward direction.
By so doing, it is seen that the anti-aliasing scheme correctly separates the two half-spaces.

Finally, for $L = \frac{\sqrt{2}\ell}{2}$, if $C$ intersects the centerline of $\Omega$ and has volume fraction equal to 50\%,  then because of the length of $L$ relative to $\ell$ the orientation of the lattice must be rotated by $\frac{\pi}{4}$ relative to that of the half-spaces.
A translation of $C$ by $\frac{\ell}{2}$ will yield a cell whose centroid lies exactly on the centerline of $\Omega$, and, by earlier analysis, has volume fraction that is less than 50\%.
Call this cell $D$.
Translations of $D$ in both principle directions by $\frac{\ell}{2}$ will yield cells that remain on the centerline of $\Omega$, and thus continue to have volume fractions less than 50\%.
Consequently, the anti-aliasing scheme proposed also separates the two half-spaces.
Rearranging the defined equalities yields the advertised result.

\newpage

\bibliographystyle{siam}
\bibliography{references}

\end{document}